\documentclass[11pt,a4paper]{article}
\usepackage{cite}
\usepackage{graphicx}
\usepackage{amssymb}
\usepackage{amsmath}
\usepackage{dsfont}
\usepackage{mathtools}
\usepackage{array}
\usepackage{rotating}
\usepackage{bbold,amsfonts}

\usepackage[utf8]{inputenc}
\usepackage{bm}
\usepackage{xcolor}
\usepackage{braket}

\usepackage[height=8.8in,width=6.45in]{geometry}
\usepackage[font=small,labelfont=bf]{caption}
\usepackage[hidelinks]{hyperref}
\usepackage{booktabs,float,slashed}
\usepackage{standalone}
\usepackage{caption}
\usepackage{subcaption}
\usepackage[normalem]{ulem}
\usepackage{clipboard}
\usepackage{simplewick}
\usepackage{caption}
\numberwithin{equation}{section}

\newcommand{\beq}{\begin{equation}}
\newcommand{\eeq}{\end{equation}}

\renewcommand{\a}{\alpha}
\renewcommand{\b}{\beta}

\newcommand{\e}{\epsilon}

\renewcommand{\S}{\Sigma}

\DeclareMathOperator{\Tr}{Tr}
\DeclareMathOperator{\tr}{tr}

\newcommand{\ii}{\mathrm{i}}

\makeatletter
\newcommand*{\letterdef@}{}
\newcommand*{\letterdef}[3]{%
	\def\letterdef@##1{\expandafter\newcommand\csname #1\endcsname{#2{##1}}}%
	\@tfor\@tempa :=#3\do{\expandafter\letterdef@\expandafter{\@tempa}}}
\makeatother
\letterdef{c#1} {\mathcal}{ABCDEFGHIJKLMNOPQRSTUVWXYZ} 
\letterdef{rm#1}{\mathrm} {dDeimM} 

\renewcommand{\tt}[1]{{\text{#1}}}
\newcommand{\ttt}[1]{{\text{\texttt{#1}}}}
\newcommand{\cmt}[1]{}
\newcommand{\q}[1]{``#1''} 
\newcommand{\et}[0]{\overset{\theta}{=}}
\newcommand{\tline}[1]{\langle  #1 \rangle}

\newdimen\tableauside\tableauside=1.0ex
\newdimen\tableaurule\tableaurule=0.4pt
\newdimen\tableaustep
\def\phantomhrule#1{\hbox{\vbox to0pt{\hrule height\tableaurule
			width#1\vss}}}
\def\phantomvrule#1{\vbox{\hbox to0pt{\vrule width\tableaurule
			height#1\hss}}}
\def\sqr{\vbox{%
		\phantomhrule\tableaustep
		\hbox{\phantomvrule\tableaustep\kern\tableaustep\phantomvrule\tableaustep}%
		\hbox{\vbox{\phantomhrule\tableauside}\kern-\tableaurule}}}
\def\squares#1{\hbox{\count0=#1\noindent\loop\sqr
		\advance\count0 by-1 \ifnum\count0>0\repeat}}
\def\tableau#1{\vcenter{\offinterlineskip
		\tableaustep=\tableauside\advance\tableaustep by-\tableaurule
		\kern\normallineskip\hbox
		{\kern\normallineskip\vbox
			{\gettableau#1 0 }%
			\kern\normallineskip\kern\tableaurule}%
		\kern\normallineskip\kern\tableaurule}}
\def\gettableau#1 {\ifnum#1=0\let\next=\null\else
	\squares{#1}\let\next=\gettableau\fi\next}
\allowdisplaybreaks
\makeatletter
\g@addto@macro\bfseries{\boldmath}
\makeatother
\graphicspath{{./figures/},{./figures/generic/},{./figures/3vv/},{./figures/chiralantichiral/},{./figures/nonscalar/},{./figures/marco-temp/},{./figures/displayingdiagrams/}}
\begin{document}
\begin{titlepage}
\vspace*{10mm}
\begin{center}
{\LARGE \bf 
        $\theta$-diagram technique for $\cN=1$, $d=4$ superfields
}

\vspace*{15mm}

{\Large D. Bason${}^{\,a,b}$, M. Bill\`o${}^{\,c,d}$}

\vspace*{8mm}
	
${}^a$ Universit\`a degli Studi di Trieste, Dipertimento di Fisica,\\
			Via Alfonso Valerio, 2, 34127 Trieste, Italy			
			\vskip 0.3cm

${}^b$ I.N.F.N. - sezione di Trieste,\\
			Via Alfonso Valerio, 2, 34127 Trieste, Italy    
			\vskip 0.3cm

${}^c$ Universit\`a degli Studi di Torino, Dipartimento di Fisica,\\
			Via P. Giuria 1, I-10125 Torino, Italy
			\vskip 0.3cm
			
${}^d$   I.N.F.N. - sezione di Torino,\\
			Via P. Giuria 1, I-10125 Torino, Italy

\vskip 0.8cm
	{\small
		E-mail:
		\texttt{marco.billo@unito.it, davide.bason@phd.units.it}
	}
\vspace*{0.8cm}
\end{center}

\begin{abstract}
We describe a diagrammatic procedure to carry out the Grassmann integration in super-Feynman diagrams of 4d theories expressed in terms of $\mathcal{N}=1$ superfields. This method is alternative to the well known $D$-algebra approach. We develop it in detail for theories containing vector, chiral and anti-chiral superfields, with the type of interactions which occur in $\mathcal{N}=2$ SYM theories with massless matter, but it would be possible to extend it to other cases. The main advantage is that this method is algorithmic; we implemented it as a  Mathematica program that, given the description of a super Feynman diagram in momentum space, returns directly the polynomial in the momenta produced by the Grassmann integration. 
\end{abstract}
\vskip 0.5cm
	{
		Keywords: {$\mathcal{N}=1$ susy theories, superdiagrams, Grassmann integration}
	}
\end{titlepage}
\setcounter{tocdepth}{2}
\tableofcontents
\vspace{1cm}

\section{Introduction}
\label{sec:intro}
Supersymmetric field theories have been intensely investigated since their discovery \cite{Gervais:1971ji,Golfand:1971iw,Volkov:1972jx,Wess:1973kz,Wess:1974tw}, for various important reasons related to their possible phenomenological implications, to their r\^ole as simplified, highly symmetric theoretical scenarios in which to shed light on deep properties of QFTs and their enticing connections with advanced mathematical structures. 
In all of these directions, the fact that supersymmetry simplifies the perturbative expansion exploiting partial cancellations between bosonic and fermionic loops has always played an important part.

Supersymmetric theories can be described in terms of superfields \cite{Salam:1974jj}, which collect the bosonic and fermionic fields that sit in a given representation of the supersymmetry algebra (a multiplet) within an expansion in terms of Grassmann variables, usually dubbed with the letter $\theta$. The perturbative expansion when organized in terms of superfields of the correlation functions is expressed in terms of super Feynman diagrams. These encompass the contributions of all ordinary Feynman diagrams for the component fields of the various multiplets. The evaluation of super Feynman diagrams requires the integration of their internal vertices over superspace, i.e. over both their space-time positions and their Grassmann $\theta$ coordinates. The Grassmann integration is essentially an algebraic operation, and poses no conceptual problem. However, from the practical point of view, it can rapidly become complicated when the order in perturbation theory and/or the number of external points increases. It is therefore important to develop efficient strategies to tackle it. Often, the space-time dependence of the diagrams is mapped via Fourier transform to momentum space. In this setting, the result of the Grassmann integration is a polynomial in the external and the loop momenta. 

In this work we focus on four space-time dimensions. The description of theories with $\cN=1$ supersymmetry, i.e., with four supercharges, in terms of $\cN=1$ superfields is rather simple. In fact $\cN=1$ superfields are often used also for theories with higher supersymmetry, such as $\cN=2$ theories; in this case, a single $\cN=2$ supermultiplet encompasses more than one $\cN=1$ superfield. 
In particular we consider theories that, when decomposed into $\cN=1$ superfields, contain chiral/anti-chiral and vector superfields. The superpropagators for these superfields and the vertices for typical interactions are standard, see for instance \cite{Bagger:2020}. 

A well-known strategy to carry out Grassmann integrations in superdiagrams of this kind was proposed long time ago in \cite{Grisaru:1979} and it is referred to as ``$D$-algebra approach''. Roughly, one eliminates all $\theta$ integrations but one. The remaining integrand comprises a string of spinorial covariant derivatives (usually denoted with the symbol $D$) which have to be simplified using the algebraic relations satisfied by the latter.

Here we illustrate another strategy in which one remains with multiple Grassmann integrations of rather simple integrands which can be described in diagrammatical terms and performed using a few general rules. This $\theta$-diagrammatic procedure is algorithmic and can be implemented in a computer code. 

A partial version of this approach was already proposed in \cite{Billo:2019} as an instrument to ease the computation of certain\cmt{scalar} superdiagrams occurring in $\cN=2$ SYM theories with massless matter. The diagrammatic rules devise in \cite{Billo:2019}, however, did not allow to treat algorithmically diagrams with self-interaction vertices of the vector superfield. 
In this work, we generalize the $\theta$-diagrammatic method so that also applies to such superdiagrams. We also show how to treat 
diagrams in which all possible fields, bosonic and fermionic, appear in the external legs. 

Our method is quite general. To describe it explicitly, however, we focus on the kind of fields and interactions which typically occur in the context of 
$\cN=2$ conformal SYM theories. Thus we consider massless superfields with the type of interactions that occur in such theories; for instance, we do not consider vertices with spinorial covariant derivatives acting on the chiral/anti-chiral superfields. Also, given that in the $\cN=2$ SYM context it is often convenient to study the so called ``difference theory'' with respect to $\cN=4$\cmt{I think one should add some references to this} SYM and in this difference ghost contributions typically cancel out, we do not consider explicitly diagrams which contain the ghosts. However, the propagator of the ghost superfield has exactly the same expression as that of chiral superfields, see for instance \cite{Kovacs:1999}, and their interaction vertices are of the type that can be described with our method. They can therefore be included without problems, provided that one takes into account in the loops their flipped statistics.

Our diagrammatic method is most straightforward if we don't impose the Wess-Zumino gauge, at the price of having to include higher and higher interaction vertices when we increase  the perturbative order. In principle, all such interactions are manageable within our approach.  

This paper is organized as follows. Section 2 describes the type of $\cN=1$ superdiagrams we consider, describing the relevant Feynman rules and in particular their Grassmann content. Section 3 introduces the basics of the $\theta$-diagram approach to the Grassmann integrations in superdiagrams. In section 4 we analyze in detail an example. Section 5 generalizes the method to all possible types of fields in the external legs. In section 6 we describe an implementation of our approach in a Mathematica code, which is provided together with this work.

\section{$\cN=1$ superdiagrams for $\cN=2$ theories}
\label{sec:N1super}
We consider supersymmetric theories in $d=4$ whose content can be arranged into chiral (or anti-chiral) and vector $\cN=1$ superfields. In our conventions, see Appendix \ref{app:notations}, the expansion of a chiral superfield has the form  
\begin{align}
	\label{e:superfieldchir}
	\Phi(x,\theta,\bar{\theta})
	= \phi(x)+\sqrt{2}\theta\psi(x) + \ii(\theta\sigma^\mu\bar{\theta})\partial_\mu\phi(x)-\theta^2 	F(x)-\frac{\ii}{\sqrt{2}}\theta^2\partial_\mu\psi(x)\sigma^\mu\bar{\theta}
	-\frac{1}{4}\theta^2\bar{\theta}^2\square\phi(x)~,
\end{align}	
where $\phi$ is a complex scalar, $\psi$ its fermionic partner (a chiral Weyl spinor) and $F$ an auxiliary field. The antichiral superfield contains $\bar{\phi}$ and the antichiral spinor $\bar{\psi}$.  
For a vector superfield we have
\begin{align}
	\label{e:superfieldvect}
	V(x,\theta,\bar{\theta})
	& =  C(x) + \ii\theta\chi(x) - \ii\bar{\theta}\bar{\chi}(x) + (\theta\sigma^\mu\bar{\theta})v_\mu(x) + \frac{\ii}{2}\theta^2(M(x) + \ii N(x))
	-\frac{\ii}{2}\theta^2(M(x)-\ii N(x))\notag \\
	& + \ii \theta^2\bar{\theta}(\bar{\lambda}(x) + \frac{i}{2}\bar{\sigma}^\mu\partial_\mu\chi(x))
	- \ii\bar{\theta}^2\theta(\lambda(x) + \frac{\ii}{2}\sigma^\mu\partial_\mu\bar{\chi}(x)) + \frac{1}{2}\theta^2\bar{\theta}^2(D(x) - \frac{1}{2}\square C(x)~,
\end{align}
where $v_\mu$ is the connection for a gauge group $G$, $\lambda$ and $\bar{\lambda}$ are the the (anti)chiral parts of the gaugino and $D$ a real auxiliary field; they all carry indices in the adjoint of $G$. The remaining fields can be fixed exploiting the supersymmetric  gauge transformation
\begin{align}
	\label{sgt}
	V \to V + \Phi_{\rm gauge} + \bar{\Phi}_{\rm gauge}~,
\end{align} 
where the parameter $\Phi_{\rm gauge}$ is itself a chiral superfield. A typical choice is the Wess-Zumino gauge, that exploits this invariance to set $C,M,N$ and $\chi$ to zero, while retaining the usual gauge freedom with parameter $2\mathrm{Im}~\phi_{\rm gauge}$. The Wess-Zumino gauge choice does not commute with supersymmetry transformations, though, and we will not impose it; thus we retain supersymmetry explicit, at the price of keeping the full fledged expression (\ref{e:superfieldvect}) of the vector superfield and having more terms in the action. 

We consider theories in which the group $G$ is in general non Abelian, and the chiral superfields are charged, i.e. transform in some representation $R$ -- in general reducible --  of $G$. We also include the possibility of cubic chiral or anti-chiral vertices. For the sake of simplicity, we do not include other types of interactions among the matter fields, but the method we propose can be generalized in this sense.

\subsection{$\cN=2$ SYM theories}
\label{subsec:N2}
As we already remarked in the introduction, the initial idea of the method we describe here was put forward in the context of $\cN=2$ SYM theories with matter \cite{Billo:2019}. The fields of these theories can be organized, with respect to an $\cN=1$ supersymmetry subalgebra, into vector and (anti)-chiral superfields with the type of interactions we mentioned above. Indeed the degrees of freedom of the $\cN=2$ vector multiplet can be encoded in a vector superfield $V$ plus a chiral superfield $\Phi$, both in the adjoint of the gauge group $G$. The components of an $\cN=2$ matter hypermultiplet can instead be arranged in two chiral superfields, $Q$ and $\tilde Q$, transforming in conjugate representations of $G$. 

\begin{figure}
	\begin{center}
		\includegraphics[width = 0.85\textwidth]{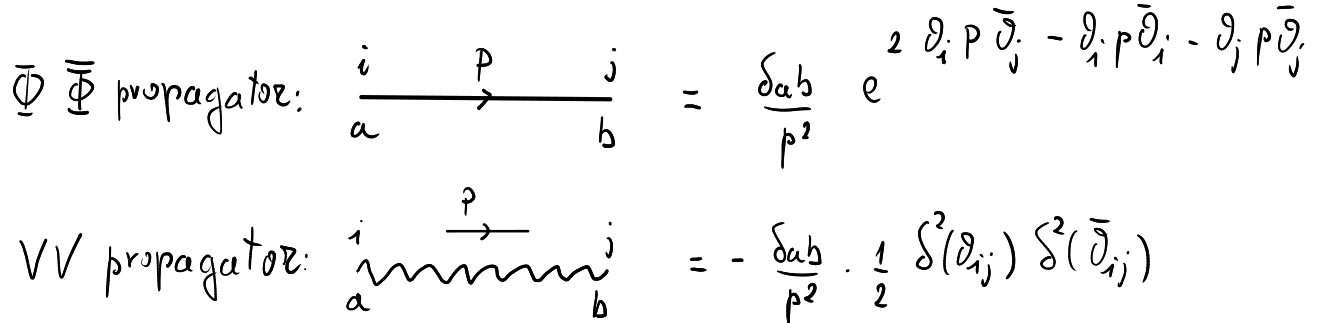}
		\caption{Superpropagators for the $\cN=2$ gauge action.}
	\label{fig:vecprop}
\end{center}
\end{figure}

\begin{figure}
\begin{center}
	\includegraphics[width = 0.85\textwidth]{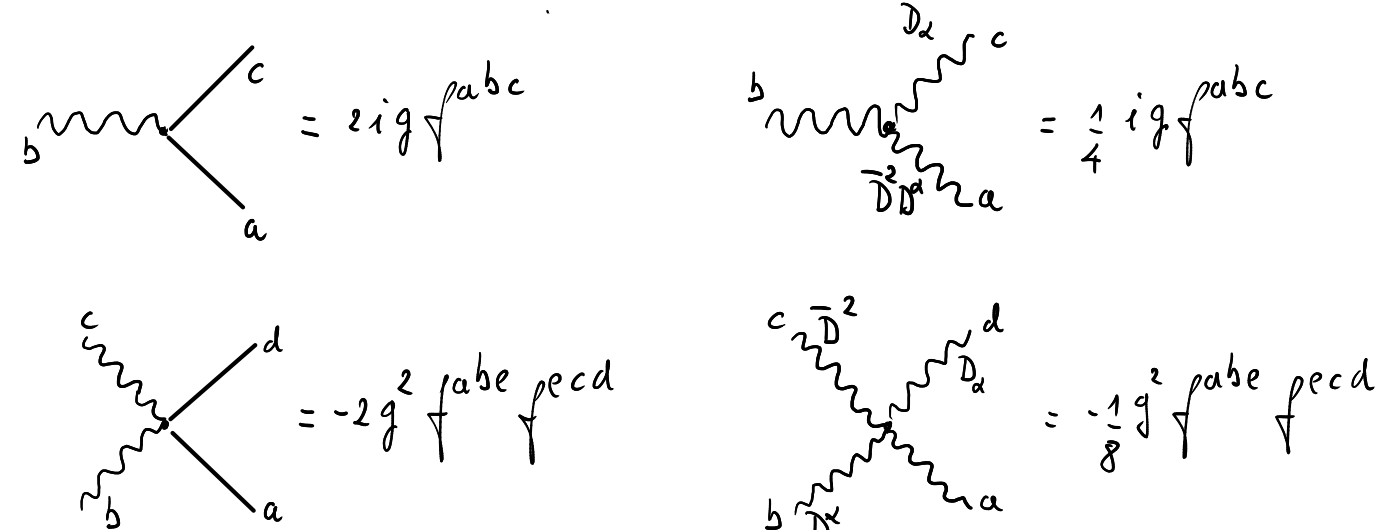}
	\caption{Vertex rules for the $\cN=2$ gauge action, up to order $g^2$.}
\label{fig:vecvert}
\end{center}
\end{figure}

The pure gauge action can be written in the $\cN=1$ superspace as follows:
\begin{align}
	\label{SN2gauge}
	S_{\rm gauge} = \int\! d^4x\, d^4\theta 
	\ \text{Tr}\left.\left[\frac{1}{8g^2}\left(W_\alpha 	W^\alpha\delta^2(\bar{\theta})+\text{h.c.}\right)
	- \frac{\xi}{4}D^2(V)\bar{D}^2(V)\right|_{\xi=1}+2\rme^{-2gV}\bar{\Phi}\rme^{2gV}\Phi\right]\text{,}
\end{align}
where $g$ is the bare gauge coupling and  
\begin{align}
	\label{e:gluinosuperfield}
	W_\alpha = -\frac{1}{4}\bar{D}^2(\rme^{-2gV}D_\alpha(\rme^{2gV}))~,~~~ 	
	\bar{W}_{\dot{\alpha}} = -\frac{1}{4}D^2(\rme^{2gV}\bar{D}_{\dot{\alpha}}(\rme^{-2gV}))\text{.}
\end{align}
This action is given in the Fermi-Feynman gauge ($\xi=1$), but we did not write the ghost-antighost part since it is not needed in order to explain our algorithm\footnote{Their propagators can be inserted with the function \texttt{CP} and their vertices with the function \texttt{GV}, as shall be explained in section \ref{sec:description}}.       
Expanding up to order $g^2$ and taking the colour trace we have
\begin{align}
	\label{e:SYMnoghosts}
	S_{\rm gauge} & = \int\! d^4x\, d^4\theta\  \bigl[-V^a\square V^a +\bar{\Phi}^{a}\Phi^a
	+g f^{abc}(\frac{\ii}{4}(\bar{D}^2(D^\alpha V^a))V^b(D_\alpha V^c) + 	2\ii\bar{\Phi}^{a}V^b\Phi^c)+\notag\\
	& + g^2f^{abe}f^{ecd}(-\frac{1}{8}V^a (D^\alpha V^b)(\bar{D}^2 V^c)(D_\alpha V^d)-2\bar{\Phi}^{a} V^b 	V^c \Phi^d) + O(g^3)\bigr]~.
\end{align}
From this action one can derive Feynman rules in terms of superpropagators and supervertices. Note that every vertex carries a superspace integration; one can as usual trade the vertex positions for the momenta running in the diagram, and remain with a $\int d^4\theta_i$ integral in the $i$-th vertex. The Feynman rules in momentum space, understanding these Grassmann integrals at each vertex, are displayed in figures \ref{fig:vecprop} and \ref{fig:vecvert}.  
Note that the cubic and quartic vector vertices contain spinorial derivatives; see Appendix \ref{subapp:spinconv}, in particular eq. (\ref{Dmom}),
for our conventions and notations. Thus in figure \ref{fig:vecvert} a covariant derivative placed on one leg in which a momentum $p$ flows out of the vertex is to be understood as
$D_\alpha= \partial_\alpha - p_{\alpha\dot{\alpha}}\bar{\theta}^{\dot{\alpha}}$. These covariant derivatives are among the chief sources of algebraic complications in carrying out the Grassmann integrations in the superdiagrams. Our approach, that will be discussed in section \ref{subsec:diavect}, associates these covariant derivatives to the vector superpropagators connecting the vertices rather than to the vertices themselves. For this reason, in the right hand sides of figure \ref{fig:vecvert} we have not written them.     

\begin{figure}
\begin{center}
\includegraphics[width = 0.85\textwidth]{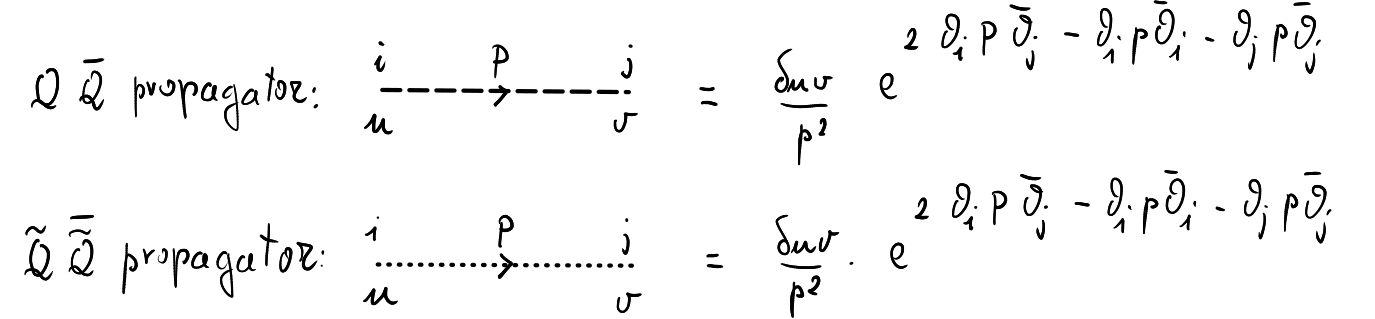}
\caption{Superpropagators for the $\cN=2$ hypermultiplets.}
\label{fig:hyperprop}
\end{center}
\end{figure}

\begin{figure}
\begin{center}
\includegraphics[width = 0.85\textwidth]{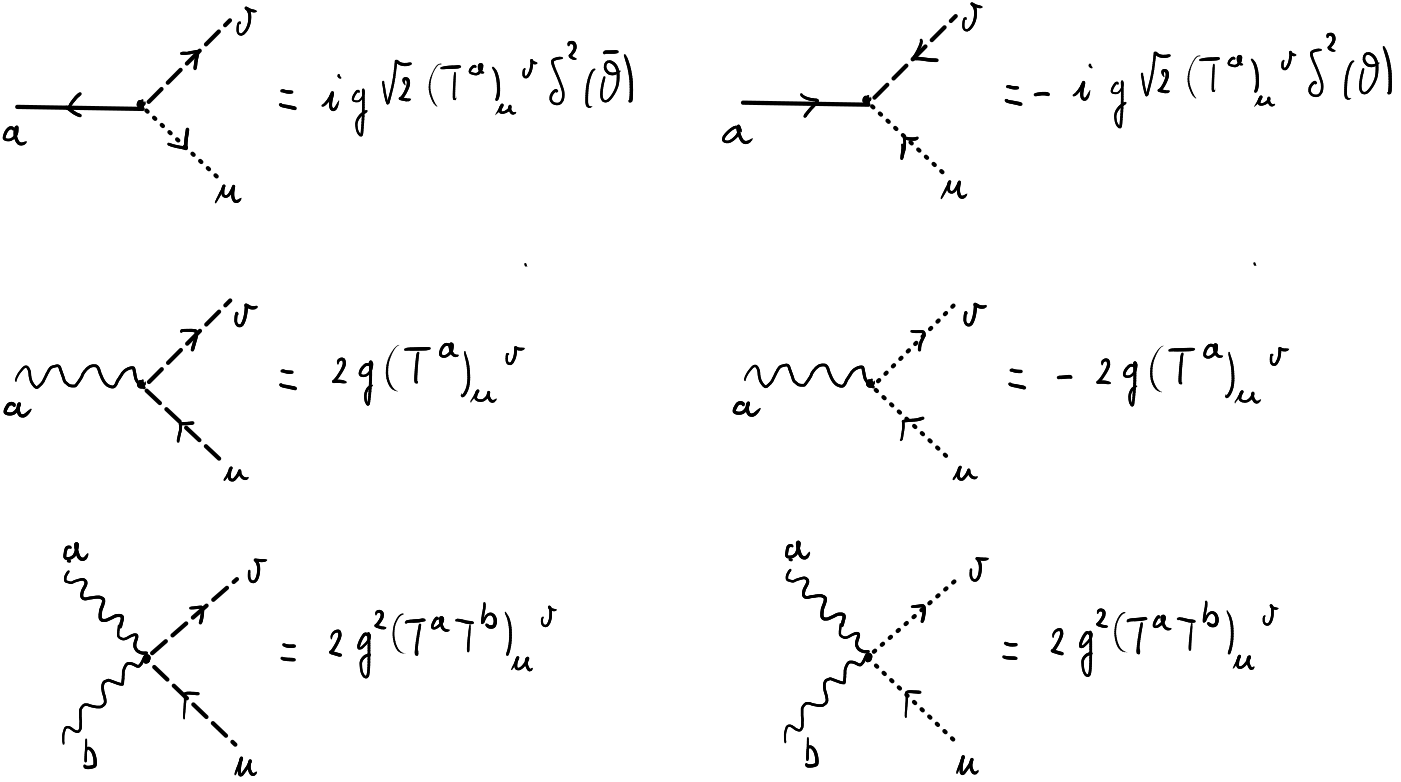}
\caption{Vertex rules for the $\cN=2$ hypermultiplet action, up to order $g^2$.}
\label{fig:hypervert}
\end{center}
\end{figure}

The hypermultiplet part of the action is given by
\begin{align}
\label{Sgauge2}
S_{\rm hyper} 
& = \int\! d^4 x\, d^4\theta\ \bar{Q} \rme^{2gV} Q + \tilde{Q}\rme^{-2gV} \bar{\tilde{Q}} 
+ \ii\sqrt{2}g \tilde{Q}\Phi Q\delta^2(\bar{\theta})
- \ii\sqrt{2}g\bar{Q}\bar{\Phi}\bar{\tilde{Q}}\delta^2(\theta)~.
\end{align}
Expanding up to order $g^2$ and making explicit the colour indices%
\footnote{The chiral superfield $Q$ transforms in a representation $R$ whose indices we denote by $u,v,\ldots$; by $(T^a)^u_{~v}$ we denote the hermitian generators of the gauge group in this representation. The superfield $\tilde Q$ transforms in the conjugate representation.} we get
\begin{align}
\label{Shyper2}
S_h & =\int d^4x d^2\theta d^2\bar{\theta} \ 
\biggl[\bar{Q}^u Q_u + \tilde{Q}^u\bar{\tilde{Q}}_u 
+g \bigl(2\bar{Q}^uV^a(T^a_R)^v_uQ_v -2\tilde{Q}^u V^a(T^a_R)^v_u\bar{\tilde{Q}}_v 
\notag\\
& + \ii\sqrt{2}\tilde{Q}^u\Phi^a(T^a_R)_u^v Q_v\delta^2(\bar{\theta})-i\sqrt{2}\bar{Q}^u\bar{\Phi}^a(T^a_R)_u^v\bar{\tilde{Q}}_v\delta^2(\theta)
\bigr)
\notag\\
& + g^2 \bigl(2\bar{Q}^uV^aV^b(T^a_RT^b_R)^v_u Q_v+2\tilde{Q}^uV^aV^b(T^a_RT^b_R)^v_u \bar{\tilde{Q}}_v\bigr)
+ O(g^3)\biggr]\text{.}
\end{align}
The corresponding superpropagators are displayed in figure \ref{fig:hyperprop} and the vertices in figure \ref{fig:hypervert}.

\subsection{Factorizing the Grassmann part}
\label{subsec:fact_grass}
We are interested in evaluating superdiagrams, constructed with the Feynman rules described above, that contributes to a correlator with fixed external states. Such a superdiagram  depends on the set of external momenta, which we label collectively by $q$, and carries colour structure and Lorentz indices associated to the external states. 
We write it in the following form%
\footnote{In the the $\cN=2$ theory all fields are massless. If we would apply this decomposition in presence of massive fields, of course we should use massive propagators in the appropriate legs.}:  
\begin{align}
\label{gen-diag}
\cW_{\rm colour}^{\rm Lorentz}(q) = 
\cN\times \cT_{\rm colour}\times \!\int \!\prod_s \frac{d^d k_s}{(2\pi)^d}
~\delta^{(d)}(\mathrm{cons}) ~\frac{\cZ^{\rm Lorentz}(k)}{\prod_s k_s^2}~.
\end{align}
Here $\cN$ is the product of the symmetry factor of the diagram and all the factors (like 
the powers of the coupling constants) appearing in the vertices - except for the color 
factors which give rise to the tensor $\cT_{\rm colour}$. By the index $s$ we enumerate the internal lines of the superdiagram. The scalar integral over the internal momenta $k_s$ is typically performed using dimensional regularization, setting 
$d=4-2\varepsilon$\cmt{necessary here?}. The momenta are subject to the appropriate
conservation relations enforced by the $\delta$-functions $\delta^{(d)}(\mathrm{cons})$. 
Beside the denominator coming from the massless propagators, the integrand 
contains also a numerator $\cZ^{\rm Lorentz}(k)$ which is the result of the integration over all the 
Grassmann variables of the $\theta$-dependent expressions present in the superdiagram and is the object of this paper.

Since the difference between the chiral multiplets $\Phi$, $Q$ and $\tilde Q$ resides only in their colour properties and numerical factors in the vertices, which are factored out in the decomposition (\ref{gen-diag}), 
the Grassmann part of propagators and vertices are the same independently of the specific chiral multiplets involved. Thus, denoting any chiral field simply by a solid line, and introducing the symbol $\et$ such that $\cW \et \cZ$ means ``the Grassmann part of $\cW$ is $\cZ$'', we have the following rules. For the chiral propagator between two points $i$ and $j$ we have
\begin{align}
\label{det1}
\parbox[c]{.45\textwidth}{\includegraphics[width = .45\textwidth]{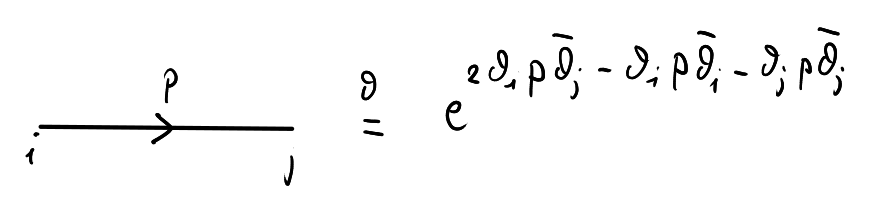}}~,
\end{align}	
For the chiral and anti-chiral cubic vertices we have	
\begin{align}
\label{det2}
\parbox[c]{.7\textwidth}{\includegraphics[width = .7\textwidth]{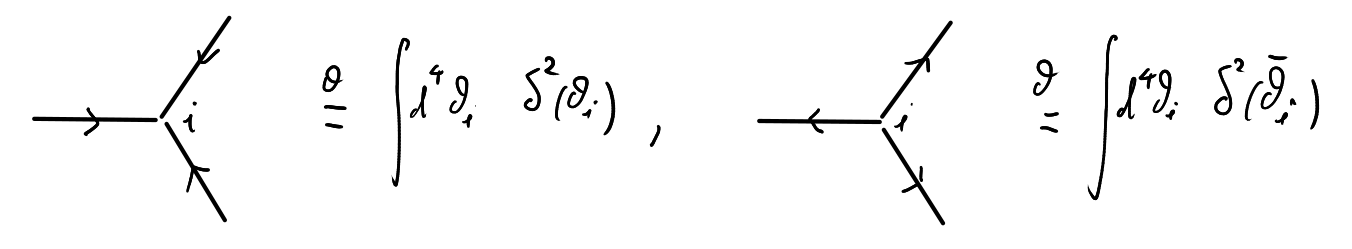}}~,
\end{align}	
where we reinstated the integration over the variables $\theta_i$ and $\bar{\theta}_i$ pertaining to the vertex which was understood in figure \ref{fig:hypervert}. For the vector propagator we have
\begin{align}
\label{det3}
\parbox[c]{.35\textwidth}{\includegraphics[width = .35\textwidth]{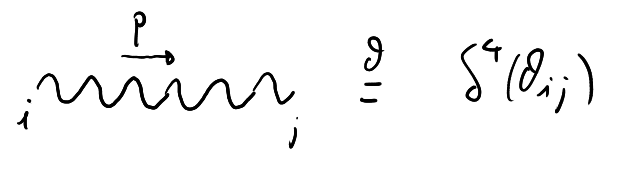}}
\end{align}	
and for the vertices involving the vectors
\begin{align}
\label{det4}
\parbox[c]{.8\textwidth}{\includegraphics[width = .8\textwidth]{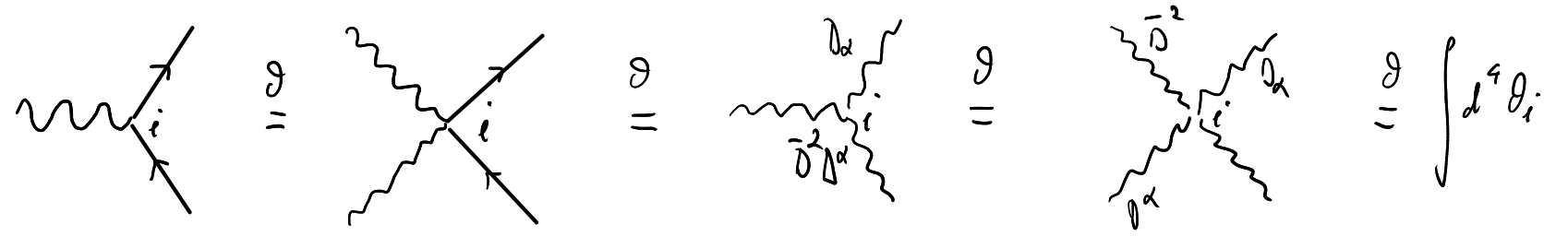}}~,
\end{align}	
where, as discussed above, we do not include in the r.h.s. the spinorial covariant derivatives because we will associate them with the vector lines connected to the vertices. 

\subsection{External lines in superdiagrams}
\label{subsec:ext_lines}
Often we are interested in correlators of specific fields rather than of superfields. In this case, super Feynman diagrams contain field-superfield propagators in the external lines. Such propagators, that arise from the Wick contraction of an external field with a superfield appearing in an internal vertex, are perfectly well defined. For instance, considering a generic chiral superfield and omitting internal symmetry indexes, we have
\begin{align}
\braket{\phi(x_1)\bar{\Phi}(x_2,\theta_2,\bar{\theta}_2)}_0
& = \braket{\phi(x_1)(\bar{\phi}(x_2)+\sqrt{2}\bar{\theta}_2\bar{\psi}(x_2)+...)}_0=
\\ \notag 
& = \braket{\phi(x_1)\bar{\phi}(x_2)}_0 + \sqrt{2}\bar{\theta}_{2,\dot{\alpha}} \braket{\phi(x_1)\bar{\psi}^{\dot{\alpha}}(x_2)}_0 + ...~.
\end{align}
We can also write them in terms of the super-propagators, and this will be useful for us in the following. Indeed one can see, by eq. (\ref{e:superfieldchir}), that for the chiral content we have:
\begin{align}
\label{phitoPhi}
&\phi(x)=\left.\Phi(x,\theta,\bar{\theta})\right|_{\theta=\bar{\theta}=0}~,\\
&\psi_\alpha(x)=\left.\frac{1}{\sqrt{2}}\partial_\alpha\Phi(x,\theta,\bar{\theta})\right|_{\theta=\bar{\theta}=0}~.
\end{align}
Thus, considering the above relations for a generic chiral superfield and the similar ones for the anti-chiral content, we have
\begin{align}
\braket{\phi(x_i)\bar{\Phi}(x_j,\theta_j,\bar{\theta}_j)}_0
& = \left.\braket{\Phi(x_i,\theta_i,\bar{\theta}_i)\bar{\Phi}(x_j, \theta_j,\bar{\theta}_j)}_0\right|_{\theta_i=\bar{\theta}_i=0}~,
\label{e:scalarexternal1}\\ 
\braket{\Phi(x_i,\theta_i,\bar{\theta}_i)\bar{\phi}(x_j)}_0
& = \left.\braket{\Phi(x_i,\theta_i,\bar{\theta}_i)\bar{\Phi}(x_j, \theta_j,\bar{\theta}_j)}_0\right|_{\theta_j=\bar{\theta}_j=0}~,
\label{e:anti-scalarexternal1}\\
\braket{\psi_\alpha(x_i)\bar{\Phi}(x_j,\theta_j,\bar{\theta}_j)}_0
& = \frac{1}{\sqrt{2}}\left.\partial_{i,\alpha} \braket{\Phi(x_i,\theta_i,\bar{\theta}_i)
\bar{\Phi}(x_j,\theta_j,\bar{\theta}_j)}_0\right|_{\theta_i=\bar{\theta}_i=0}~,
\label{e:spinorchiralexternal1}\\
\braket{\Phi(x_i,\theta_i,\bar{\theta}_i)\bar{\psi}^{\dot{\alpha}}(x_j)}_0
& = \frac{1}{\sqrt{2}}\left.\bar{\partial}^{\dot{\alpha}}_j
\braket{\Phi^a(x_i,\theta_i,\bar{\theta}_i) \bar{\Phi}^b(x_j,\theta_j,\bar{\theta}_j)}_0\right|_{\theta_j=\bar{\theta}_j=0}~.
\label{e:anti-spinoranti-chiralexternal1}
\end{align}
Writing the above relations in momenta space we obtain:
\begin{align}
\parbox[c]{.17\textwidth}{\includegraphics[width = .17\textwidth]{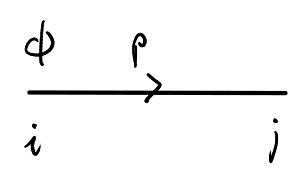}}	
& =\frac{1}{p^2} \rme^{-\theta_jp\bar{\theta}_j}
\et \rme^{-\theta_jp\bar{\theta}_j}~,
\label{e:se2}\\ 
\parbox[c]{.17\textwidth}{\includegraphics[width = .17\textwidth]{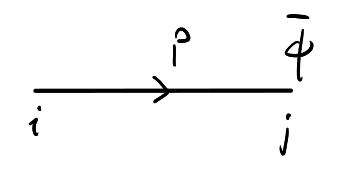}}	
& =\frac{1}{p^2} \rme^{-\theta_ip\bar{\theta}_i}
\et \rme^{-\theta_ip\bar{\theta}_i}~,
\label{e:ase2}\\
\parbox[c]{.17\textwidth}{\includegraphics[width = .17\textwidth]{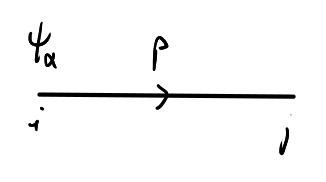}}
&=\frac{\sqrt{2}}{p^2} p_{\alpha\dot{\beta}}\bar{\theta}_j^{\dot{\beta}}
\rme^{-\theta_jp\bar{\theta}_j}\et p_{\alpha\dot{\beta}}\bar{\theta}_j^{\dot{\beta}}
\rme^{-\theta_jp\bar{\theta}_j}~,
\label{e:sce2}\\
\parbox[c]{.17\textwidth}{\includegraphics[width = .17\textwidth]{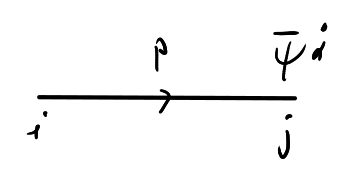}}
& =-\frac{\sqrt{2}}{p^2}\bar{p}^{\dot{\alpha}\beta}\theta_{i,\beta}
\rme^{-\theta_ip\bar{\theta}_i} \et \bar{p}^{\dot{\alpha}\beta}\theta_{i,\beta}
\rme^{-\theta_ip\bar{\theta}_i}~.
\label{e:asce2}
\end{align}

We can relate the fields in the vector multiplet to the vector superfield in a similar way to what we did for the chiral supermultiplet, see eq. (\ref{phitoPhi}), namely using derivatives and the setting some variables to zero. Indeed one can see, omitting colour indexes, that
\begin{align}
\label{vmuV1}
v^\mu(x)=\frac{1}{2}\left.\sigma^\mu_{\alpha\dot{\beta}}\partial^\alpha\bar{\partial}^{\dot{\beta}}V(x,\theta,\bar{\theta})\right|_{\theta=\bar{\theta}=0}
\end{align}
from which we can obtain:
\begin{align}
\braket{v^\mu(x_i)V(x_j,\theta_j,\bar{\theta}_j)}_0=\left.\frac{1}{2}\sigma^\mu_{\alpha\dot{\beta}}\partial^\alpha\bar{\partial}^{\dot{\beta}}\braket{V(x_i,\theta_i,\bar{\theta}_i)V(x_j,\theta_j,\bar{\theta}_j)}_0\right|_{\theta_i=\bar{\theta}_i=0}~.
\end{align}
Then in momentum space we have:
\begin{align}
\parbox[c]{.17\textwidth}{\includegraphics[width = .17\textwidth]{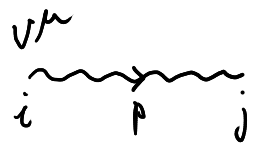}}
& =\frac{1}{p^2}(\theta_j p\bar{\theta}_j) \et 2(\theta_j p\bar{\theta}_j)~.
\end{align}
Also these external lines can be treated through the diagrammatical method that we propose in this paper, but a bit of care is needed. Suppose that the external vector field be connected to a vector superfield on which a certain expression $f(D,\bar{D})$, built of spinorial covariant derivatives, acts. This happens, for instance, for some of the vector Super Yang-Mills vertices -- see fig. \ref{fig:vecvert}. Then the Grassmann part of  $\braket{v(x_i)f(D_j,\bar{D_j})V(x_j,\theta_j,\bar{\theta}_j)}_0$ would be given by 
\begin{align}
\label{vmuf}
\braket{v^\mu f(D_j,\bar{D_j})V(\theta_j,\bar{\theta}_j)}_0 \et \sigma^\mu_{\alpha,\dot{\beta}}\partial^\alpha\bar{\partial}^{\dot{\beta}}f(D_j,\bar{D}_j)\delta^4(\theta_{ij})~.
\end{align}
Such an expression is not easily handled in our diagrammatical method, but there is a way around. Instead of expressing $v^\mu$ in terms of the superfield $V$ by means of Grassmann derivatives, we can exploit Grassmann integrations and write
\begin{align}
\label{e:vectorvectorexternal1}
v^\mu(x)&=-2\int\! d^2\theta d^2\bar{\theta} \ (\theta\sigma^\mu\bar{\theta})V(x,\theta,\bar{\theta})~,
\end{align}	
Then we can also replace eq. (\ref{vmuf}) with
\begin{align}
\label{vmufbis}
\braket{v^{\mu}(x_i) f(D_j,\bar{D}_j)V(x_j,\theta_j,\bar{\theta}_j)}_0
&= -2\! \int\! d^ \theta_i d^2\bar{\theta}_i\ (\theta_i\sigma^\mu\bar{\theta}_i)
\braket{V(x_i,\theta_i,\bar{\theta}_i) f(D_j,\bar{D}_j) V(x_j,\theta_j,\bar{\theta}_j)}_0~.
\end{align}
In momentum space this leads to
\begin{align} 
\parbox[c]{.17\textwidth}{\includegraphics[width = .17\textwidth]{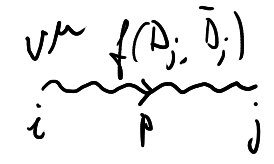}}
& \et \int\! d^2 \theta_i d^2 \bar{\theta}_i\ 2(\theta_i \sigma^\mu\bar{\theta}_i) 
f(D_j,\bar{D}_j) \left[\delta^4(\theta_{ij})\right]~,
\label{e:vmue}
\end{align}
In this way%
\footnote{Note that it is also possible to use field to superfield relations based on Grassmann integration, instead of derivations as in eq. (\ref{phitoPhi}), also for the fields in the chiral multiplet. Again, this could in principle be useful for theories containing vertices with  covariant spinorial derivatives acting on (anti-)chiral superfields. For theories that do not contain such vertices, as is the case for instance for the $\mathcal{N}=2$ SYM theory, this is not needed and it is more efficient to use the formulas (\ref{e:scalarexternal1}-\ref{e:anti-spinoranti-chiralexternal1}) which do not introduce integration of auxiliary variables. 
}
we only have to deal with Grassmann integrals and covariant derivatives acting on $\delta^4(\theta_{ij})$, which can be treated within our diagrammatic formalism, as we will discuss in section \ref{subsec:diavect}.

We can deal similarly with the gaugino, even if the expression is slightly more complicated and contains two terms%
\footnote{This is due to the fact that we do not impose the Wess-Zumino gauge and thus we have to disentangle the $\chi$ spinor from the gaugino.}:
\begin{align}
\braket{\lambda_\alpha(x_i) V(x_j,\theta_j,\bar{\theta}_j)}_0 
& =\int\! d^4\theta_i\ \left(-2\ii\theta_{i,\alpha} - \sigma^\mu_{\alpha\dot{\beta}}\bar{\theta}_i^\beta\bar{\theta}_i^2\, \partial/\partial x_{i,\mu}\right)
\braket{V(x_i,\theta_i,\bar{\theta}_i) V(x_j,\theta_j,\bar{\theta}_j)}_0~,
\label{e:spinorvector1}\\
\braket{V(x_i,\theta_i,\bar{\theta}_i)\bar{\lambda}^{\dot{\alpha}}(x_j)}_0
& =\int\! d^4\theta_j\ \left(2\ii\bar{\theta}_j^{\dot{\alpha}} + \bar{\sigma}^{\mu,\dot{\alpha},\beta}\theta_{j,\beta}\theta_j^2\, \partial/\partial x_{j,\mu}\right)
\braket{V(x_i,\theta_i,\bar{\theta}_i)V(x_j,\theta_j,\bar{\theta}_j)}_0~.
\label{e:anti-spinorvector1}
\end{align} 	
In momentum space and focusing only on the Grassmann part we obtain:
\begin{align}
\parbox[c]{.17\textwidth}{\includegraphics[width = .17\textwidth]{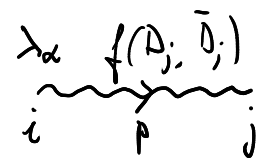}}
& \et \int\! d^2\theta_i d^2\bar{\theta}_i\ \left(\theta_{i,\alpha}+\frac{p_{\alpha\dot{\beta}}}{2}\bar{\theta}^{\dot{\beta}}_i\theta^2_i\right) f(D_j,\bar{D}_j)\left[\delta^4(\theta_{ij})\right]~,
\label{e:ge}\\
\parbox[c]{.17\textwidth}{\includegraphics[width = .17\textwidth]{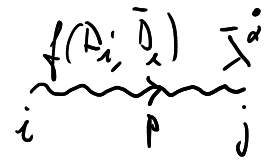}}
& \et \int\! d^2\theta_j d^2\bar{\theta}_j\ \left(\bar{\theta}_j^{\dot{\alpha}} - \frac{\bar{p}^{\dot{\alpha}\beta}}{2}\theta_{j,\beta}\bar{\theta}^2_j\right)
f(D_i,\bar{D}_i)\left[\delta^4(\theta_{ij})\right]	~.
\label{e:age}
\end{align}
In making explicit the spinorial derivatives in the expressions above one has to take into account that eq. (\ref{Dmom})  is written with the momentum flowing out of the vertex $j$ (so in some of the cases above it is actually the momentum $-p$). 

\section{Grassmann integration in superdiagrams}
\label{sec:Grassmann}
As already stated, the focus of this paper is on the Grassmann integration occurring in superdiagrams, namely on the computation of the quantity $\cZ^{\rm Lorentz}$ defined in eq. (\ref{gen-diag}). Of course, methods to perform such integration are well known, and in particular the so-called $D$-algebra approach of \cite{Grisaru:1979} is widely used. Here we propose a method that we find  efficient and that is very algorithmic, so that it can be implemented in a symbolic language code. This method is based on the construction of what we call $\theta$-diagrams.

\subsection{A simple example}
\label{subsec:simpleexample}
To fix the ideas, let us consider a very simple example, the one-loop correction to the propagator of the scalar in an adjoint chiral multiplet with matter chiral multiplets running in the loop. In this case the Lorentz structure is trivial, and the colour factor follows easily from the colour part of the vertices and propagators given in figures \ref{fig:hypervert} and \ref{fig:hyperprop}, while the normalization factor turns out to be just $2g^2$. With respect to eq. (\ref{gen-diag}) it is immediate in this example to exploit the momentum conservation, obtaining 
	\begin{equation}
		\label{1loopQ}
		\parbox[c]{.35\textwidth}{\includegraphics[width 
			= .35\textwidth]{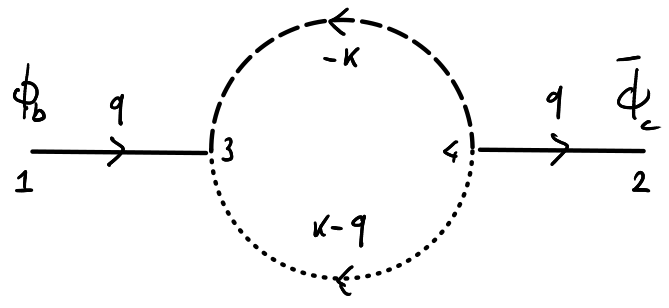}}
		= 2 g^2 \times \Tr_\cR(T^b T^c)\, \times
		\int \frac{d^dk}{(2\pi)^d} \frac{1}{(q^2)^2}\frac{1}{k^2(k-q)^2}\, \cZ(k,q)~.
	\end{equation}
	The factor $\cZ(k,q)$ is the result of the integration over the Grassmann variables at each 
	internal vertex.
	According to the rules in figures \ref{fig:hyperprop} and \ref{fig:hypervert} and eq.s (\ref{e:se2}) and (\ref{e:ase2})  it reads%
	\begin{align}
		\label{Z1loopQ}
		\cZ(k,q) = \int d^4\theta_3 \,d^4\theta_4\, (\theta_3)^2 ({\bar\theta}_4)^2\,
		\rme^{2 \theta_4 (k-q) \bar\theta_3 - \theta_4 (k-q) \bar\theta_4 - \theta_3 (k-q) \bar\theta_3 }
		\rme^{-2 \theta_4 k \bar\theta_3 + \theta_4 k \bar\theta_4 + \theta_3 k \bar\theta_3 }~.
	\end{align}
	This integral is straightforward, as we will describe shortly. 
	In general, the computation of $\cZ^{\rm Lorentz}(k)$ through Grassmann integrations
	remains essentially algebraic and presents in principle no conceptual difficulty for any superdiagram. Nevertheless, it typically becomes very involved as the complexity of the superdiagram increases.
	
	\subsection{$D$-algebra}
	\label{s:Dalgebra}
	The key idea of this approach is to make the Grassmann integration local, namely to reduce it to the integral over a single  pair of Grassmann variables.
	This can be achieved because the Grassmann variables to be integrated over are associated to the vertices in the superdiagram. The vertices are connected by superpropagators. The Grassmann part of the vector superpropagator directly contains $\delta$ functions, see eq. (\ref{det3}), 
	that identify the $\theta$-variables in the vertices $i$ and $j$ it connects. Also the Grassmann part of the chiral-antichiral propagator, given in eq. (\ref{det1}), can be rewritten so as to display such $\delta$ functions; indeed one can show that%
	\begin{align}
		\label{propchirdelta}
		\rme^{2\theta_i p \bar\theta_j - \theta_i p \bar\theta_i - \theta_j p \bar\theta_j }
		= \frac{1}{16} \bar D^2_{i,p}\, D^2_{i,p}
		\delta^2(\theta_{ij})\delta^2(\bar{\theta}_{ij})~,
	\end{align}	
	where we have remarked that the spinorial covariant derivatives contain the momentum $p$.
	Exploiting appropriately the $\delta$ functions present in the propagators, one can then eliminate the integrals over all but one of the pair of Grassmann variables. The remaining integrand, however, contains in general spinor covariant derivatives arising both from the vector vertices, see figure \ref{fig:vecvert}, and from the chiral propagators rewritten using eq. (\ref{propchirdelta}). This expression needs to be evaluated using the algebraic properties of the spinorial covariant derivatives $D$ and $\bar D$, i.e., carrying out the $D$-algebra.  
	
	For the example of eq. (\ref{Z1loopQ}), this procedure amounts to rewrite $\cZ(k,q)$ as 
	\begin{align}
		\label{Z1loopQ0}
		\cZ(k,q) & = \frac{1}{16^2}\int d^4\bar\theta_3 \,d^4\theta_4\,\theta_3^2\,\bar\theta_4^2
		\left[D^2_{4,-k}\, \bar D^2_{4,-k} \delta^4(\theta_{34})\right]\,
		\left[D^2_{4,k-q}\, \bar D^2_{4,k-q}\delta^4(\theta_{34})\right]
	\end{align}
	and integrate it by parts%
	\footnote{The rules for Grassmannian integrations by part are collected in Appendix \ref{s:integrationbyparts}.} to obtain
	\begin{align}
		\label{Z1loopQ2}
		\cZ(k,q) & = \frac{1}{16^2} \int d^4\bar\theta_3 \,d^4\theta_4\, 
		\left[D^2_{4,q-k}\, \bar D^2_ {4,q-k}
		\left[ \theta_3^2\,\bar\theta_4^2
		\bar D^2_{4,-k}\, D^2_{4,-k} \delta^4(\theta_{34})\right]\right]\, \delta^4(\theta_{34})~.
	\end{align}
	Using the fact that $D_4^2 $ and $\bar D_4^2$ commute with $\theta_3^2$ and exploiting the $\delta$-function we 
	can write
	\begin{align}
		\label{Z1loopQ3}
		\cZ(k,q) & = \frac{1}{16^2} \int \,d^4\bar\theta_4\, \theta_4^2 
		\left[D^2_{4,q-k}\, \bar D^2_{4,q-k}
		\left[\bar\theta_4^2
		\bar D^2_{4,-k}\, D^2_{4,-k} \delta^4(\theta_{34})\right]\right]_{3\to 4}~,
	\end{align}
	where in the string of covariant derivatives we have to identify the variables in the node $3$ with those in the node $4$ \emph{after} having carried out the derivatives. In this way we have recast $\cZ$ in term of a Grassmannian integration at a single node. As argued in \cite{Grisaru:1979}, this can be done for any loop within a superdiagram and, iterating the procedure, for any irreducible superdiagram. 
	
	The explicit evaluation of $\cZ$ written in the form  (\ref{Z1loopQ3}) can be carried out exploiting the algebraic properties of the covariant derivatives
	and the final outcome is that 
	\begin{align}
		\label{Z1loopQbis}
		\cZ(k,q) = 
		- q^2~.
	\end{align}
	
	\subsection{Introducing the $\theta$-diagrams}
	\label{subsec:intrthetad}
	In this example, the $D$-algebra strategy turns out to be rather more involute than carrying out the integration over the Grassmann variables in both nodes with the propagators in exponential form, as in eq.s (\ref{Z1loopQ},\ref{Z1loopQbis}). In fact, the latter procedure can be described in a diagrammatic, algorithmic form which turns out to be generalizable to a large class of superdiagrams and can be implemented in a computer program. Let us now introduce this strategy starting from the example at hand. 
	
	The basic ingredient are the exponentials appearing in the chiral propagators. We introduce the following diagrammatic notation:
	\begin{equation}
		\label{td1}
		\parbox[c]{.8\textwidth}{\includegraphics[width = .8\textwidth]{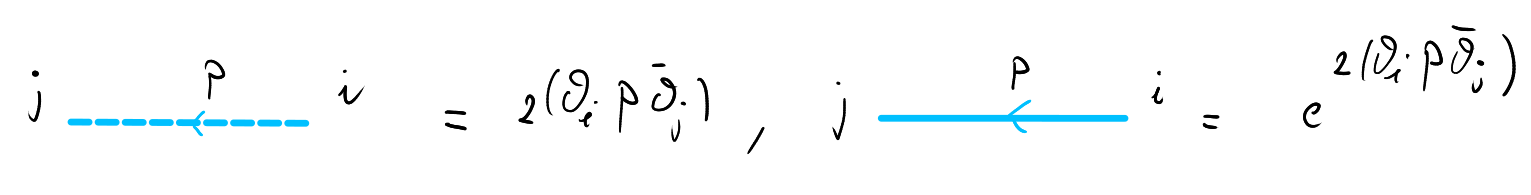}}~.
	\end{equation}
	Due to the Grassmannian nature of the $\theta$ and $\bar\theta$ variables, the exponentials are in fact polynomial:
	\begin{align}
		\label{expexp}	
		\rme^{2(\theta_i p\bar{\theta}_j)}=
		1 + 2(\theta_i p\bar{\theta}_j) + \frac{1}{2}\times \left(2(\theta_i p\bar{\theta}_j)\right)^2~.
	\end{align}
	In diagrammatic notation, this reads
	\begin{equation}
		\label{td2}
		\parbox[c]{.8\textwidth}{\includegraphics[width = .8\textwidth]{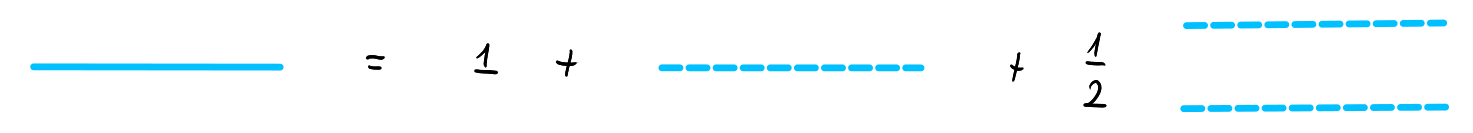}}~,
	\end{equation}
	where the momentum and the end-point labels are the same in all lines and are understood. 
	
	The superdiagrams contain integrations over Grassmann variables $\theta_i$ and $\bar\theta_i$ at the various sites. These we represent graphically as black or white dots:
	\begin{equation}
		\label{td3}
		\parbox[c]{.4\textwidth}{\includegraphics[width = .4\textwidth]{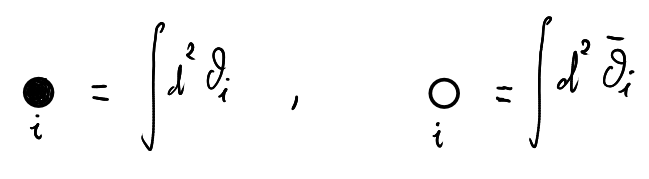}}~.
	\end{equation}
	
	The cubic (anti)-chiral super-vertices bring in factors of Grassmannian delta-functions:
	\begin{align}
		\label{thetadelta}
		\theta_i^2 = \delta^2(\theta_i)~,~~~ 
		\bar\theta_i^2 = \delta^2(\bar\theta_i)~.
	\end{align}  
	
	Taking into account all this, we can rewrite diagrammatically eq. (\ref{Z1loopQ}) as follows:
	\begin{equation}
		\label{td4}
		\cZ(k,q) = \parbox[c]{.5\textwidth}{\includegraphics[width = .5\textwidth]{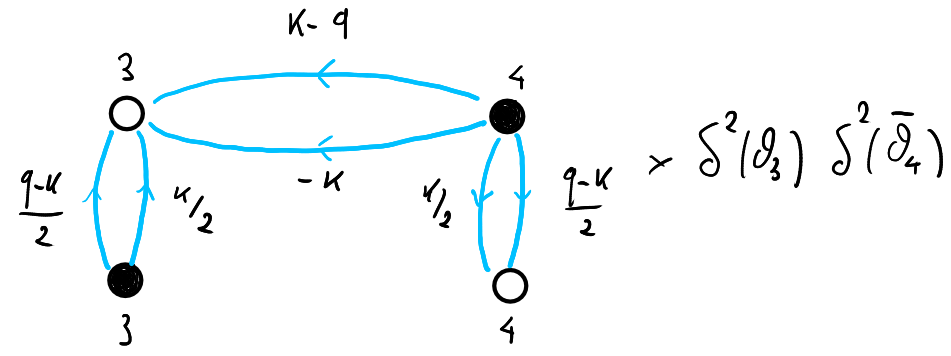}}~.
	\end{equation}
	
	This expression can be simplified in some obvious ways. The Grassmannian delta-function $\delta^2(\theta_3)$ soaks up the corresponding integration and sets to zero all occurrences of $\theta_3$. On the diagram, this correspond simply to remove the black dot labeled by $3$ and the two solid lines connecting to it, since the associated exponentials reduce to 1. Similarly, the $\delta^2(\theta_4)$ removes the white dot labeled by 4 and the lines attached to it. We remain thus with
	\begin{equation}
		\label{td5}
		\cZ(k,q) = \parbox[c]{.22\textwidth}{\includegraphics[width = .22\textwidth]{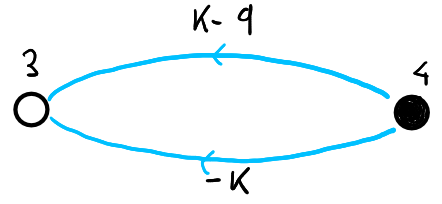}}~.
	\end{equation}
	Note that the labels $3$ and $4$ attached to the dots just assign a name to the integration variables, and they are thus actually redundant.
	
	Moreover, we can take into account the following general property: the product of two solid lines with the same endpoints gives a single solid line carrying the total momentum, namely 
	\begin{equation}
		\label{td6}
		\parbox[c]{.44\textwidth}{\includegraphics[width = .44\textwidth]{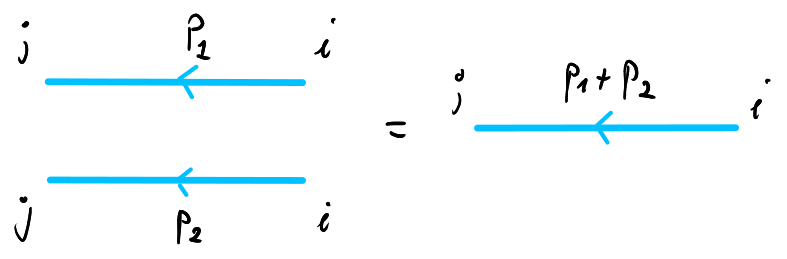}}~.
	\end{equation}
	This just corresponds to the product rule for the corresponding two exponentials, see eq. (\ref{td1}). Thus we arrive at
	\begin{equation}
		\label{td7}
		\cZ(k,q) = \parbox[c]{.2\textwidth}{\includegraphics[width = .2\textwidth]{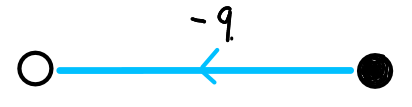}}~.
	\end{equation}
	We can expand the solid $\theta$-lines into dashed lines according to eq. (\ref{td2}). Then we have to take into account the presence of white and black dots, that correspond to Grassmann integrals over two-component spinors. According to eq. (\ref{td1}), each dashed line leaving the vertex $i$ carries one $\theta_i$ variable, and each dashed line entering the vertex $j$ carries one $\bar\theta_j$. 
	The Grassmann integrals $\int\! d\theta^2$ or  $\int\! d\bar{\theta}^2$ vanish unless they act on expression quadratic in $\theta_{\alpha}$  or $\bar{\theta}^{\dot{\alpha}}$. Graphically this means that a $\theta$-diagram is non-null if and only if exactly two dashed lines are attached to each dot. This is a very simple but crucial property of the $\theta$-diagrams. In particular, we have 
	\begin{equation}
		\label{td8}
		\parbox[c]{.48\textwidth}{\includegraphics[width = .48\textwidth]{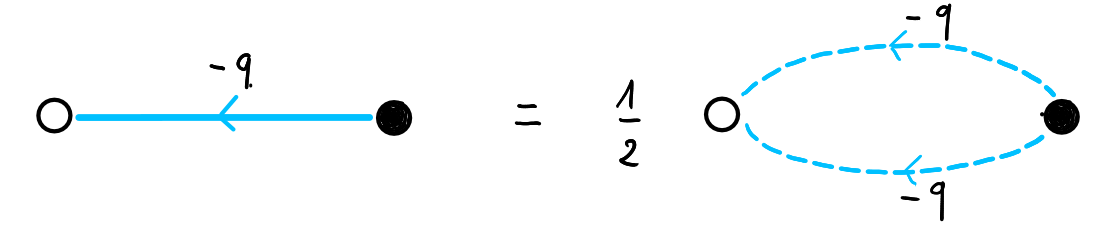}}~.
	\end{equation}
	Finally, the dashed loop on the r.h.s. above is very easily evaluated. One has
	\begin{align}
		\label{l2int}	
		\parbox[c]{.22\textwidth}{\includegraphics[width = .22\textwidth]{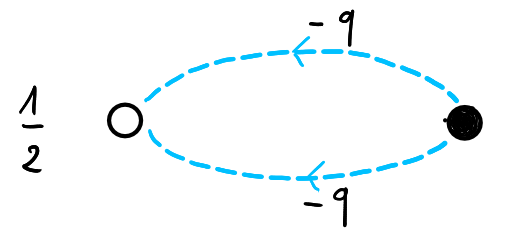}}	
		= \frac 12 \int d^2\theta_4 \, d^2\bar{\theta}_3\,  2(\theta_4 q \bar{\theta}_3) 2(\theta_4 q \bar{\theta}_3)
		= \frac 12 \tr (q\bar q) = -  q^2~,
	\end{align}
	where
	we made use of a Fierz rearrangement, see eq.s (\ref{fierzchirug},\ref{fierzachirug}), and carried out the integrations according to the basic rules in eq.s (\ref{intcomp}--\ref{intt2}). We obtain therefore
	\begin{align}
		\label{Ztdis}
		\cZ(k,q) = - q^2~,
	\end{align}
	in agreement with the result of the $D$-algebra manipulations.
	
	Of course, the Grassmann integrations in this $\cZ(k,q)$ are very easy. It may seem that our diagrammatic notation unnecessarily complicates it. However, it readily generalizes and is algorithmic. As a first step, we show how the Grassmann part $\cZ$ of all superdiagrams involving only (anti)-chiral superfields can be managed using  $\theta$-diagrams. 
	
	\subsection{Diagrams with (anti)chiral superfields}
	\label{subsec:dcac}
	This is the simplest case. Let us describe how to manage it.
	
	\paragraph{Deriving the $\theta$-diagram}
	The Grassmann part of all chiral-superchiral propagators is given by a momentum dependent exponential of the type considered above, see eq. (\ref{det1}), thus we can say that
	\begin{equation}
		\label{td10}
		\parbox[c]{.9\textwidth}{\includegraphics[width = .9\textwidth]{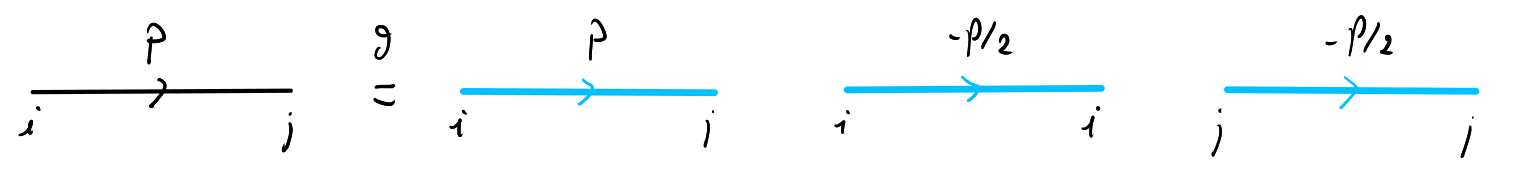}}~.
	\end{equation}
	Note that in eq. (\ref{td10}) and in the following, the cyan, thick lines that appear on the r.h.s. of the $\et$ symbol have a completely different meaning from the black ones on the left: they are the solid $\theta$-lines defined in eq. (\ref{td1}). 
	
	For propagators connected to external scalars, given in eq.s (\ref{e:se2}) and (\ref{e:ase2}), we have simply  
	
	\begin{equation}
		\label{td13}
		\parbox[c]{.51\textwidth}{\includegraphics[width = .51\textwidth]{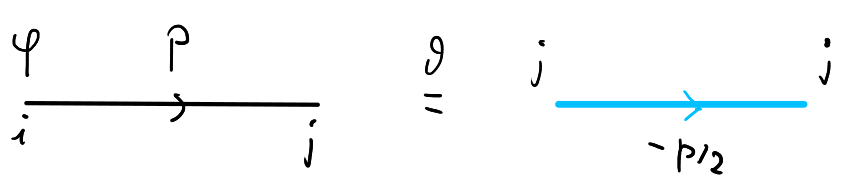}}
	\end{equation}
	and
	\begin{equation}
		\label{td14}
		\parbox[c]{.51\textwidth}{\includegraphics[width = .51\textwidth]{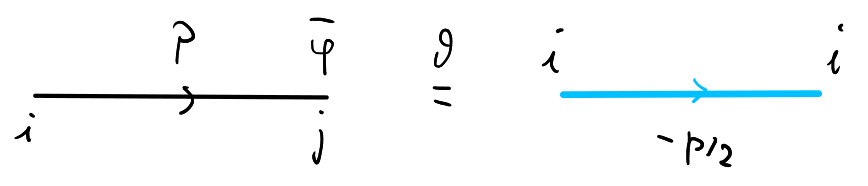}}
	\end{equation}
	
	Also the contribution of cubic vertices involving chiral superfields to the Grassmann part of the diagram are very easy to describe: 
	\begin{equation}
		\label{td11}
		\parbox[c]{.48\textwidth}{\includegraphics[width = .48\textwidth]{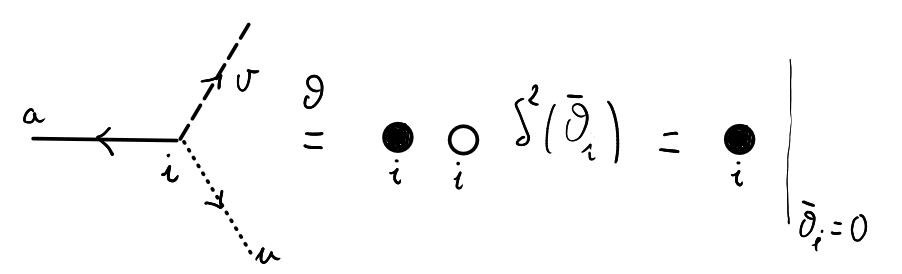}}~.
	\end{equation}
	Similarly, for the anti-chiral vertices one has
	\begin{equation}
		\label{td12}
		\parbox[c]{.48\textwidth}{\includegraphics[width = .48\textwidth]{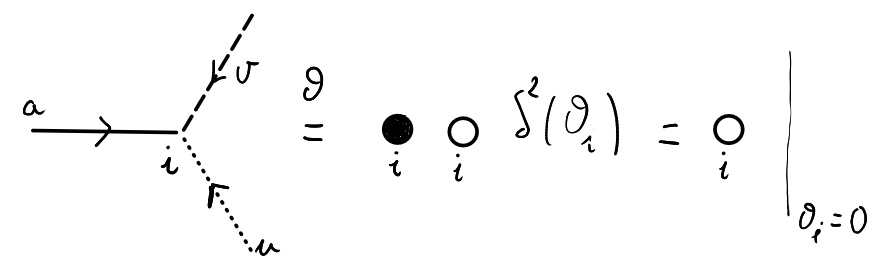}}~.
	\end{equation}
	
	It follows from eq.s (\ref{td13},\ref{td14}) that a chiral propagator connecting an external scalar to a cubic vertex gives a trivial contribution to the $\theta$-diagram. Indeed, suppose for instance that the propagator 
	(\ref{td13}) be attached to a vertex of the type (\ref{td12})  in position $j$; then its $\theta$-line $\exp(-\theta_j p\bar\theta_j)$ reduces to $1$ since the vertex sets all occurrences of $\theta_j$ to zero. 
	
	The properties we have listed are sufficient to determine the $\theta$-diagrammatic form of the Grassmann factor $\cZ$ associated to any superdiagram $\cW_{\rm colour}$ involving only (anti) chiral superfields with scalar external states; we will consider later the possibility that the external states correspond to other components of the multiplet. 
	
	\paragraph{An example}
	Let us illustrate this by a specific example, namely
	\begin{equation}
		\label{td15}
		\cW_{ab}(q) = \parbox[c]{.4\textwidth}{\includegraphics[width = .4\textwidth]{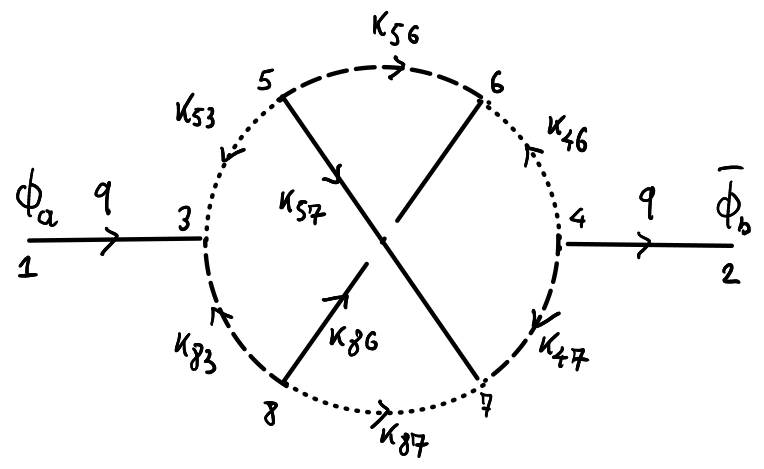}}~.
	\end{equation}
	This diagram represents a three-loop contribution to the adjoint chiral scalar propagator, and has been considered for instance in \cite{Billo:2019}. Using the rules from (\ref{td10}) to (\ref{td12}) for all the propagators and vertices and making use of the $\delta$-functions that they contain we obtain    
	\begin{align}
		\label{WabtoZ3l}
		\cW_{ab} \et \cZ = \parbox[c]{.25\textwidth}{\includegraphics[width = .25\textwidth]{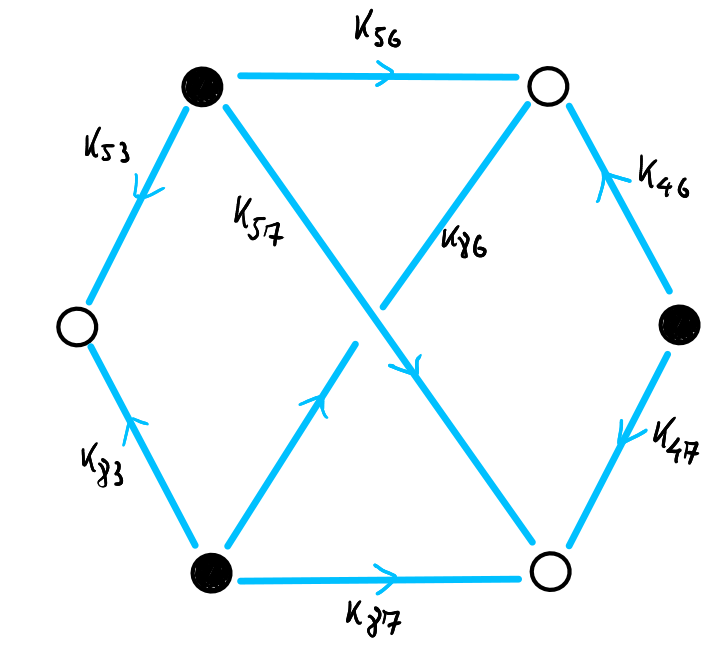}}~.
	\end{align}
	This is perfectly analogous to the form we obtained in eq. (\ref{td5}) for the one-loop example. 
	It is easy to convince oneself that in fact in all cases involving only chiral/antichiral superfields the structure of vertices and lines of the $\theta$-diagram mimics the one of the original superdiagram, except for dropping the external lines attached to scalar components, and associating a white dot to every chiral cubic vertex and a black dot to every anti-chiral one. 
	
	\paragraph{Expanding the $\theta$-diagram}
	To compute a $\theta$-diagram of the type just described, one can expand its solid lines according to eq. (\ref{td2}).
	Having done this, we can exploit the crucial observation we already stated  just before eq. (\ref{td8}): the only non zero contributions are those in which exactly two dashed lines enter or leave every dot. As a consequence, such non-zero contributions consist of products of loops of dashed lines%
	\footnote{Each loop consisting of only two lines, though, comes with a factor $1/2$ because it arises from the expansion of a single solid line as in eq. (\ref{td8}).} going through non-overlapping subsets of dots, leaving no dot isolated.  
	
	Let us illustrate this with another example. Consider a $\theta$-diagram with the following structure:
	\begin{align}
		\label{td17}
		\cZ = \parbox[c]{.17\textwidth}{\includegraphics[width = .17\textwidth]{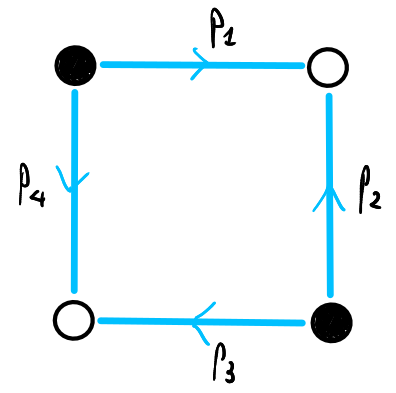}}~.
	\end{align}
	This can arise, for instance, from a one-loop contribution to a four-point function of chiral/antichiral scalars. According to the rules just stated, it is given by
	\begin{align}
		\label{td18}
		\cZ = \parbox[c]{.62\textwidth}{\includegraphics[width = .62\textwidth]{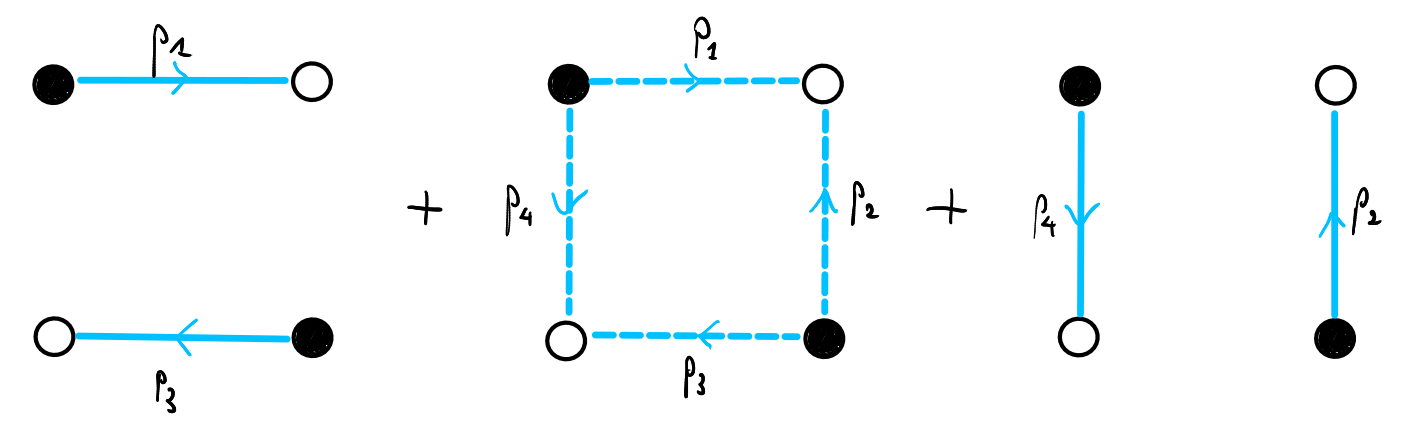}}~.
	\end{align}
	The terms in the right hand side represent precisely the possible decompositions of $\cZ$ into loops of dashed lines; note that we have drawn the loops involving two dots as single solid lines, see eq. (\ref{td8}).
	
\begin{figure}
		\begin{center}
			\includegraphics[width= \textwidth]{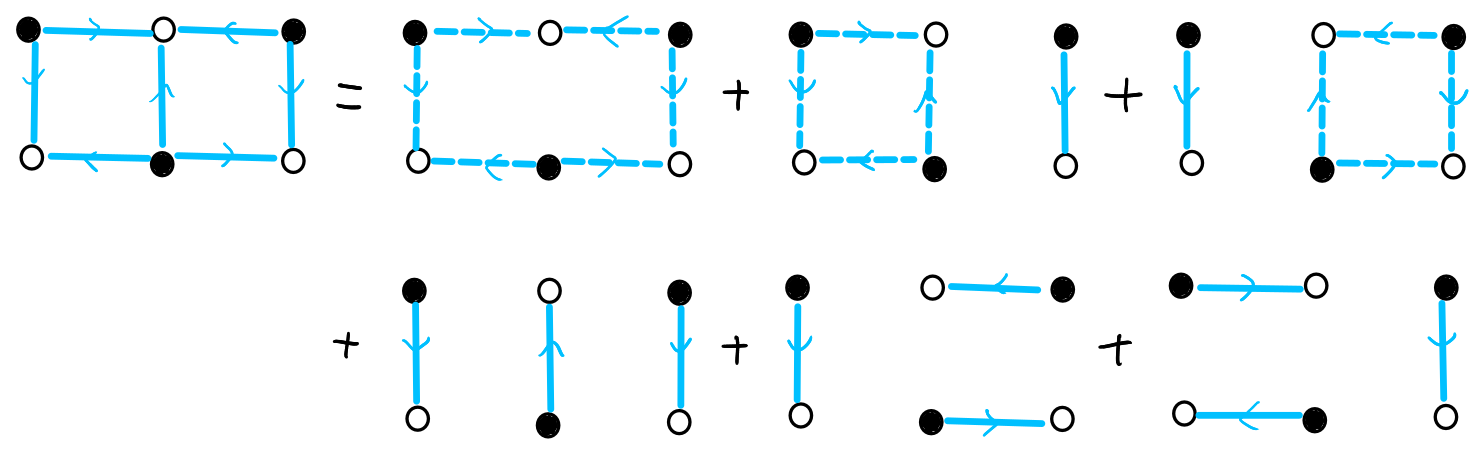}
		\end{center}
		\caption{An example of the decomposition of a $\theta$-diagram into its non-vanishing contributions. For simplicity we do not label the vertex and the lines.}
		\label{fig:doublesquare}
	\end{figure}

	Another example of the decomposition of a $\theta$-diagram is given in figure \ref{fig:doublesquare}. 
	
	The decomposition into loops can be realized by means of a path-finding algorithm and implemented in a computer code; this was done in \cite{Armato:2019}. In the code presented here we follow a different strategy, based on expanding in turn all lines and taking into account at each steps the simplifications owing to the rule that selects exactly two dashed lines in each node. This approach turns out to be computationally more efficient.  
	
	\paragraph{Getting the final result}
	The next ingredient is the following: each loop of dashed lines can be evaluated explicitly and the result can be given in a general form. Using repeatedly the Fierz rearrangements in eq.s (\ref{fierzchir}--\ref{fierzachirug}) and the integration rules (\ref{intcomp},\ref{intt1},\ref{intt2}) one easily shows that 
	\begin{align}
		\label{td20}
		\parbox[c]{.3\textwidth}{\includegraphics[width = .3\textwidth]{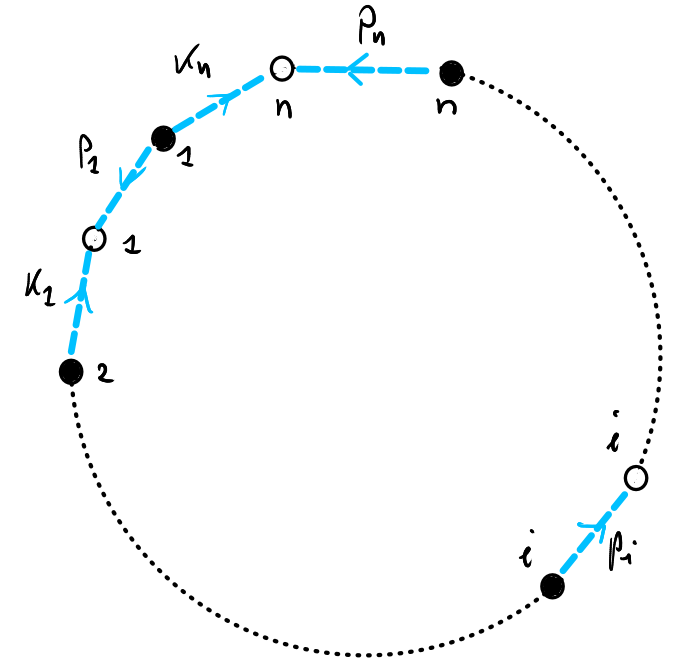}}
		& = \int d^2\theta_1\, d^2\bar{\theta}_1 \ldots \ldots d^2\theta_n\,d^2\bar{\theta}_n\,
		2 \big(\theta_1 p_1\bar{\theta}_1\big)2\big(\theta_2 k_1 \bar{\theta}_1\big)
		\ldots
		\nonumber\\[-15mm]
		& \times 2\big(\theta_n p_n \bar{\theta}_n\big)2\big(\theta_1 k_n \bar{\theta}_n\big)
		= (-1)^{n+1} \, \tr\big(p_1\,\bar{k}_1 \ldots p_n\,\bar{k}_n\big)
		\\
		\nonumber
	\end{align}
	The traces appearing above are of the form
	\begin{align}
		\text{tr}(p_1\bar{k}_1...p_n\bar{k}_n)=p_{1,\mu_1}k_{1,\nu_1}...p_{n,\mu_n}k_{n,\nu_n}\text{tr}(\sigma^{\mu_1}\bar{\sigma}^{\nu_1}...\sigma^{\mu_n}\bar{\sigma}^{\nu_n})
	\end{align}
	They can be recursively computed using the properties of the matrices $\sigma^\mu$ and $\bar{\sigma}^\mu$, see Appendix \ref{subapp:spinconv}.
	
	For instance for eq. (\ref{td18}) we obtain
	\begin{align}
		\label{z4res}
		\cZ =  p_1^2\, p_3^2 - 2(p_1\cdot p_2)(p_3\cdot p_4) + 2(p_1\cdot p_3)(p_2\cdot p_4)
		- 2 (p_1\cdot p_4)(p_2 \cdot p_3) + p_2^2\, p_4^2+p_{1,\mu}p_{2,\nu}p_{3,\rho}p_{4,\sigma}\epsilon^{\mu\nu\rho\sigma}~.
	\end{align}
	
	\paragraph{Perspective} Let us summarize. Drawing the $\theta$-diagrams and evaluating them according to the rules we just described, we can compute explicitly the Grassmannian integrations for all superdiagrams involving only chiral and antichiral superfields, with scalar external states. All the three main steps of the procedure, namely writing down the $\theta$-diagram, finding its decomposition into loops of dashed lines and evaluating the latter, are completely algorithmic and can be implemented in a computer code.
	
	This procedure was already described in \cite{Billo:2019}, without however detailing the algorithm for determining all non-zero contributions and without providing a computer code. In the present work we aim at extending this approach to a much larger class of superdiagrams, maintaining its main advantage, namely that of being algorithmic and readily implemented in a code. The basic idea is to cast the $\theta$-dependent part of the superdiagram in terms of Grassmannian delta-functions and exponentials, which we can then evaluate with the $\theta$-diagrammatic method described in the present section. 
	
	\subsection{Diagrams with (internal) vector superfields}
	\label{subsec:diavect}
	The Grassmann part of the propagator of the vector superfields is very simple: 
	\begin{equation}
		\label{td21}
		\parbox[c]{.45\textwidth}{\includegraphics[width = .45\textwidth]{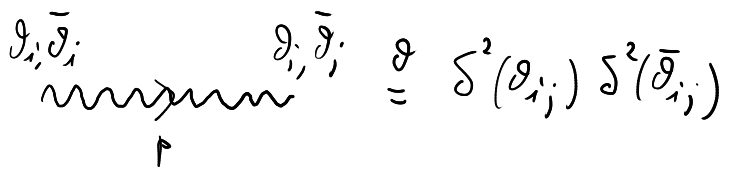}}~.
	\end{equation}
	
	As was mentioned in the introduction, we restrict in this paper to theories that can be formulated in terms of  $\mathcal{N}=1$ chiral and vectorial superfields, with interactions that have covariant derivatives that act only on vectorial superfields. This class of theories contain in particular $\cN=2$ SYM theories with matter hypermultiplets. In this case, beside the cubic vertices of the type considered in eq.s (\ref{td11},\ref{td12}), we have other interactions that involve both vectors and chiral superfields. Up to order $g^2$, they are elementary from the point of view of their Grassmann structure:
	\begin{equation}
		\label{td22}
		\parbox[c]{0.45\textwidth}{\includegraphics[width = 0.45\textwidth]{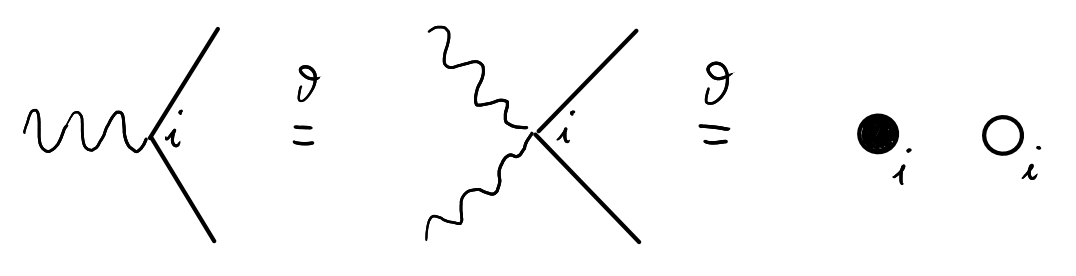}}~.
	\end{equation}
	
	From the action (\ref{e:SYMnoghosts}) we see that there are also interaction vertices, involving vector superfields only, which include spinor covariant derivatives. Connecting vertices of this kind to other vertices by means of the vector superpropagator we effectively obtain expressions in which such covariant derivatives act on the vector lines. The possible cases, and their Grassmann structure, can be worked out with a little bit of algebra. For instance, as in eq.(\ref{s4_3}), there might be a case where we have a single covariant derivative acting on a vector propagator. This implies a contribution like $D_{i,\alpha}\delta^4(\theta_{ij})$. It is straightforward to see that the following two expression coincide:
	\begin{align}
		D_{i,\alpha}\delta^4(\theta_{ij})=2\theta_{ij,\alpha}\delta^2(\bar{\theta}_{ij})e^{\frac{(\theta_i p_{ij}\bar{\theta}_i)-(\theta_j p_{ij}\bar{\theta}_i)}{2}}\text{.}
	\end{align}
	Graphically we have:
	\begin{equation}
		\label{vl1}
		\parbox[c]{0.7\textwidth}{\includegraphics[width = 0.7\textwidth]{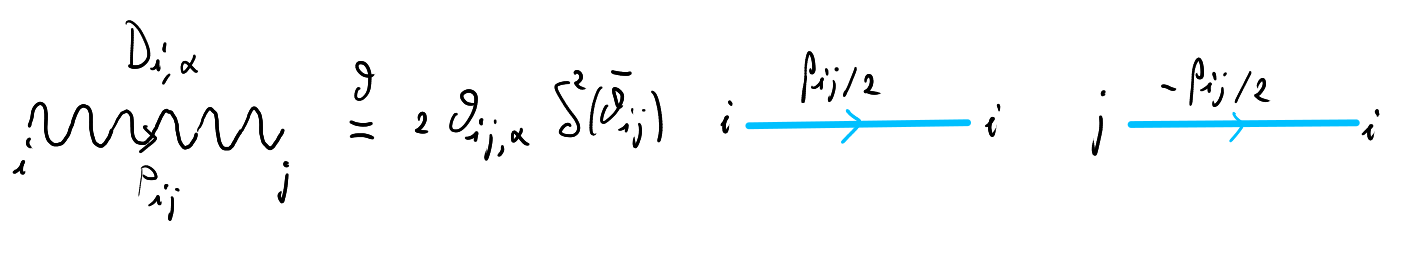}}\text{.}
	\end{equation}
	This can be extended with further derivatives analogously. In two cases we might have two spinor derivatives:
	\begin{align}
		\label{vl2a}
		\parbox[c]{0.7\textwidth}{\includegraphics[width = 0.7\textwidth]{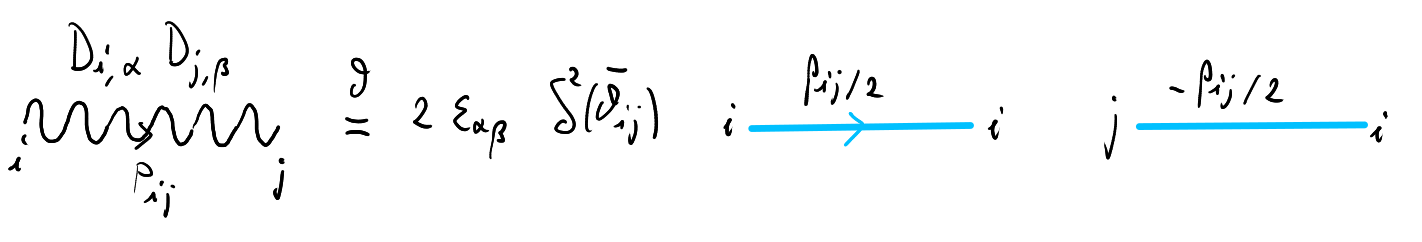}}~,\\
		\label{vl2b}	
		\parbox[c]{0.7\textwidth}{\includegraphics[width = 0.7\textwidth]{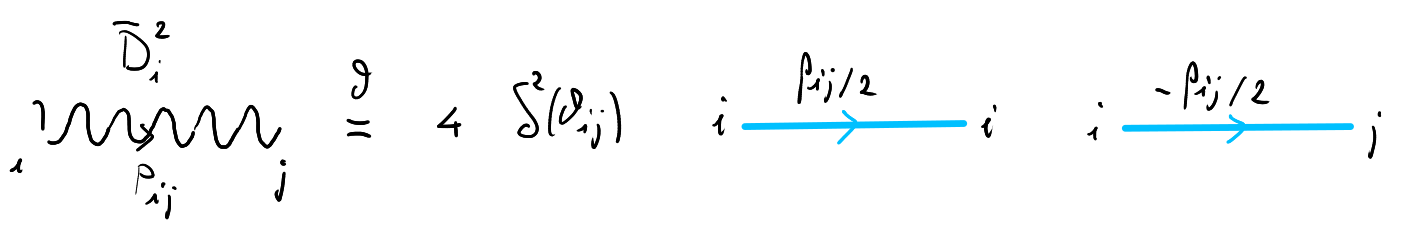}}~,\\
	\end{align}	
	In eq. (\ref{vl2a}) we have took into account that a line might have a derivatives coming from both vertices it attaches to, see figure \ref{fig:double3vertexdecomposition}.
	
	We now proceed with the numeration. we have two contributions with three spinor derivatives:
	\begin{align}
		\label{vl3a}
		& \parbox[c]{0.82\textwidth}{\includegraphics[width = 0.82\textwidth]{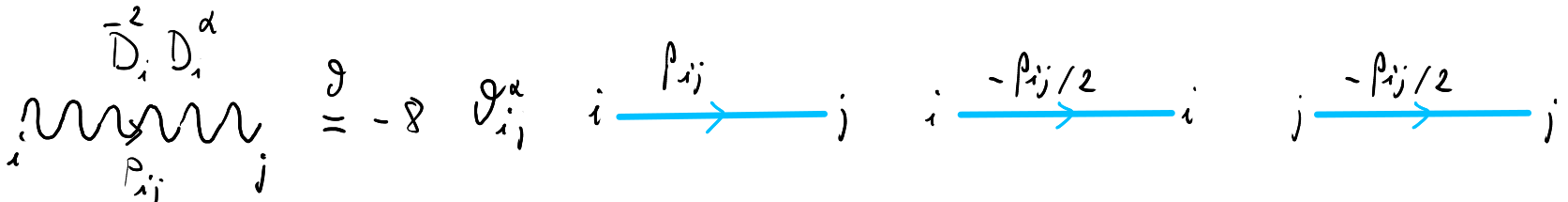}}~,\\
		\label{vl3b}	
		& \parbox[c]{0.82\textwidth}{\includegraphics[width = 0.82\textwidth]{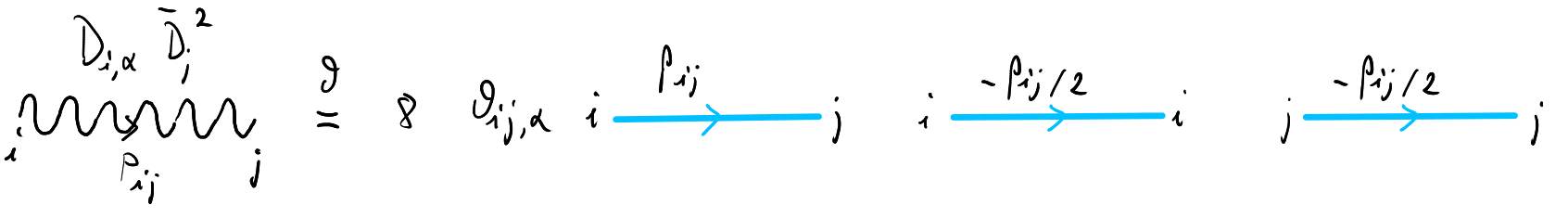}}~,\\
	\end{align}	
	two with four spinor derivatives:
	\begin{align}
		\label{vl4a}
		& \parbox[c]{0.65\textwidth}{\includegraphics[width = 0.65\textwidth]{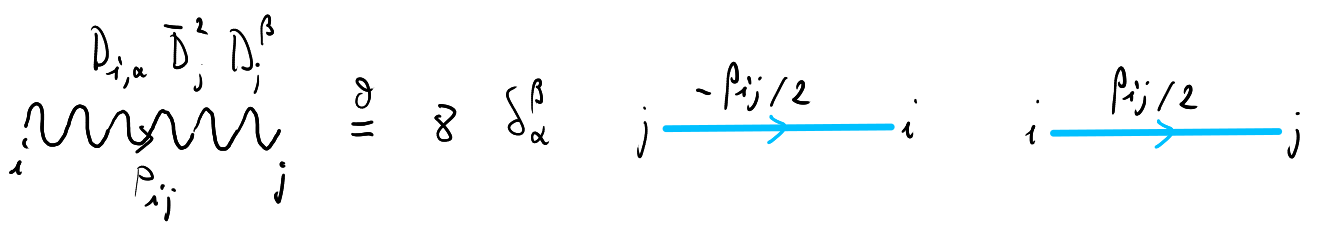}}~,\\
		\label{vl4b}	
		& \parbox[c]{0.24\textwidth}{\includegraphics[width = 0.24\textwidth]{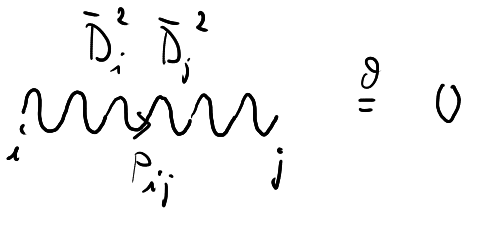}}~,\\
	\end{align}	
	one with five spinor derivatives:
	\begin{equation}
		\label{vl5}
		\parbox[c]{0.24\textwidth}{\includegraphics[width = 0.24\textwidth]{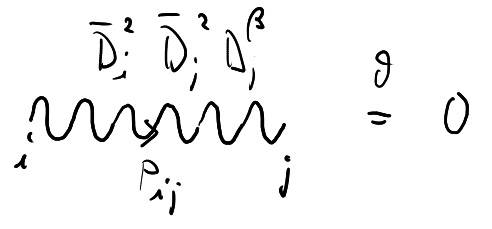}}
	\end{equation}
	and finally a single one with six spinor derivatives:
	\begin{equation}
		\label{vl6}
		\parbox[c]{0.7\textwidth}{\includegraphics[width = 0.7\textwidth]{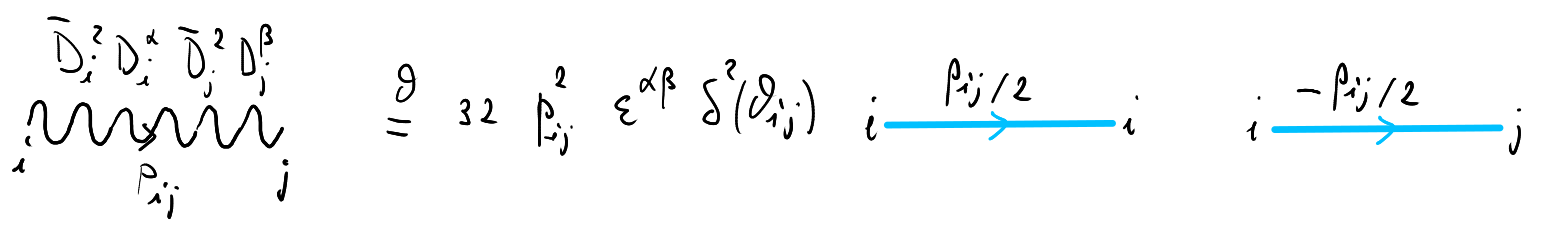}}
	\end{equation}
	
	Once the spinorial covariant derivatives have been assigned to the vector lines, the ``bare'' gluon vertices only contain the information about the integration; thus for instance
	\begin{align}
		\label{td30}
		\parbox[c]{0.3\textwidth}{\includegraphics[width = 0.3\textwidth]{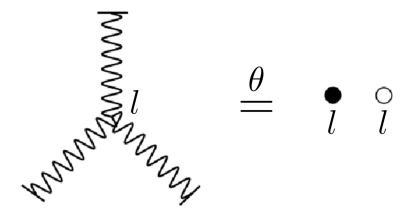}}~.
	\end{align}
	
	Consider the situation in which the ``decorated'' vector lines of figure \ref{fig:vecvert} are inserted inside superdiagrams with scalar external states, which have an overall trivial Lorentz structure. The various spinor quantities and tensors appearing above,  like $\theta_{ij}^\alpha$, $\epsilon_{\alpha\beta}$, and $\delta^\alpha_\beta$, must conspire to produce scalar structure. Thus they can only give overall numerical factors or Grassmannian scalar products of the type  $\theta_{ij}\cdot\theta_{kl}$. Remembering that $\theta_{ij} = \theta_i - \theta_j$ and that $\delta^2(\theta)=\theta^2$, one has:
	\begin{align}
		\label{e:thetatodeltas1}
		&\theta_{ij}\cdot\theta_{kl}=\frac{1}{2}\left(\delta^2(\theta_{ij})+\delta^2(\theta_{kl})-\delta^2(\theta_i-(\theta_j+\theta_k-\theta_l))\right)~.
	\end{align}
	Thus, the Grassmann structure of all superdiagrams formed with the elements we considered so far can be recast as linear combinations of integrals of exponentials and Grassmann $\delta$ functions which we know how to deal with using the $\theta$-diagrammatic technique.
	
	In other theories based on $\cN=1$ superfields, other combinations of spinorial covariant derivatives, beside those appearing in eq.s (\ref{vl1}-\ref{vl6}), might appear. They could be treated by applying the same principle. Furthermore
	the number of possible different combinations is finite and one could write down the analogues of the rules in eq.s (\ref{vl1}-\ref{vl6}) for all of them. Thus for a generic theory with chiral/antichiral and vector $\cN=1$ superfields, with spinorial covariant derivatives appearing on the vectors only, all the Grassmann integrals in any superdiagram can always be expressed in terms of $\theta$-diagrams.   
	
	\subsection{Some $\theta$-diagrammatical properties}
	\label{subsec:thetaprop}
	Let us pause here to recapitulate some properties of our $\theta$-diagrammatic notation, which have used somehow implicitly above. 
	
	First of all, the order in which white and black dots appear in the graphical representation of the $\theta$-diagram is not relevant: they represent integrals over certain Grassmann variables, and all the occurrences of these variables in other elements of the diagram must be integrated over. Thus, for example,
	\begin{align}
		\label{prop1}
		\parbox[c]{0.15\textwidth}{\includegraphics[width = 0.15\textwidth]{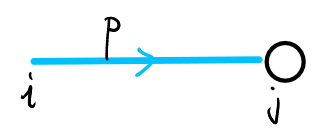}}
		= \int d^2\bar{\theta}_j\, \rme^{2(\theta_i p \bar{\theta}_j)}~,
	\end{align}
	even if the white dot appears rightmost in the drawing.   
	
	Another graphical property that we have been using is the fact that endpoints of solid or dotted $\theta$ lines  get attached to dots carrying the same index, remembering that the lines go from a black to a white dot: thus, for instance,  
	\begin{align}
		\label{prop2}
		\parbox[c]{0.9\textwidth}{\includegraphics[width = 0.9\textwidth]{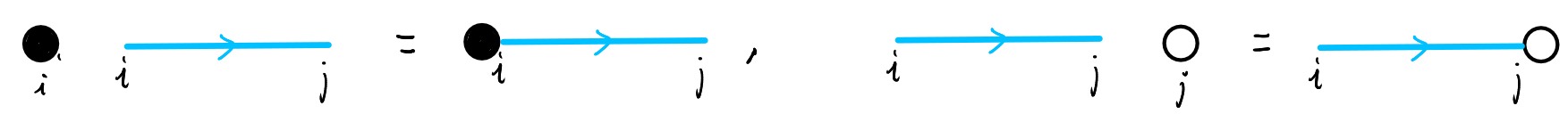}}
	\end{align}
	as well as
	\begin{align}
		\label{prop3}
		\parbox[c]{0.9\textwidth}{\includegraphics[width = 0.9\textwidth]{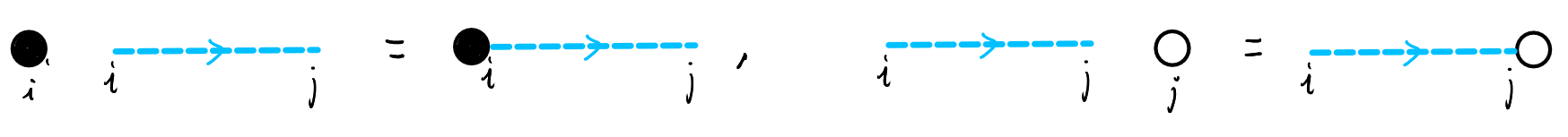}}~.
	\end{align}
	This also works when one end is already attached to a dot, or when a dot has already other lines attached to it.
	
	In the $\theta$ diagram, the lines can be constrained as consequence of the integration of some Grassmannian delta function. These delta function might set some variable to zero, as in the case of the $\delta^2(\theta)$ or $\delta^2(\bar{\theta})$ present in the hyper-multiplet vertices, or identify two different variables, as in the case of the $\delta^2(\theta_{ij})=\delta^2(\theta_i-\theta_j)$ and the $\delta^2(\bar{\theta}_{ij})=\delta^2(\bar{\theta}_i-\bar{\theta}_j)$ present in vectorial superpropagators, see eq. (\ref{td21}). 
	
	It is easy to see that such constraints amount to the following rules:
	\begin{align}
		\label{const1}
		\parbox[c]{0.84\textwidth}{\includegraphics[width = 0.84\textwidth]{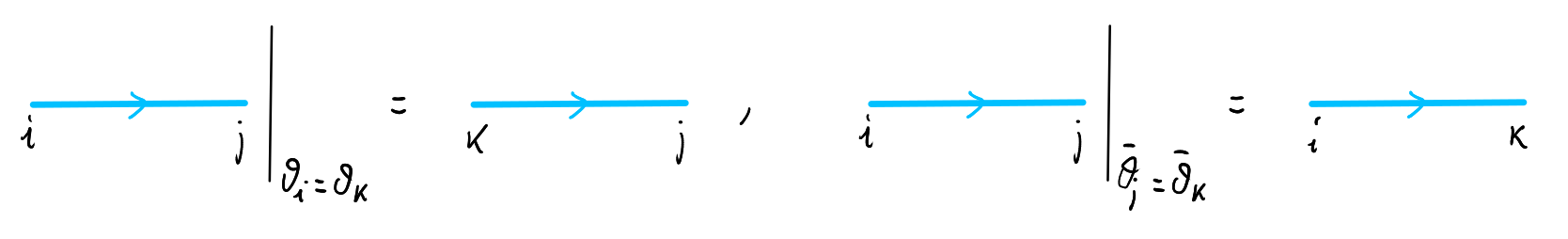}}
	\end{align}
	and
	\begin{align}
		\label{const2}
		\parbox[c]{0.84\textwidth}{\includegraphics[width = 0.84\textwidth]{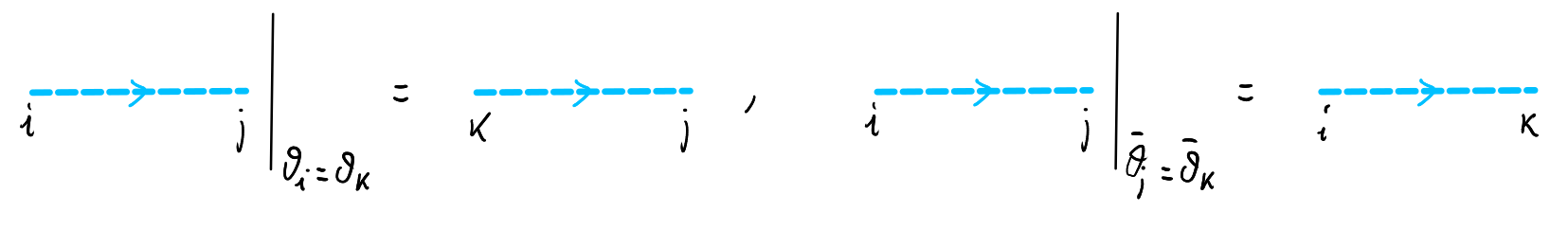}}
	\end{align}
	as well as
	\begin{align}
		\label{const3}
		\parbox[c]{0.48\textwidth}{\includegraphics[width = 0.48\textwidth]{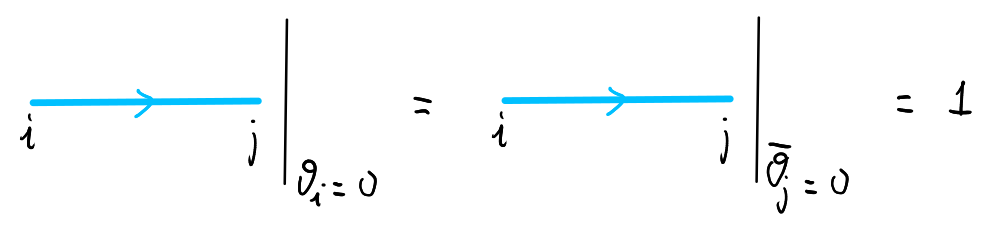}}
	\end{align}
	and
	\begin{align}
		\label{const4}
		\parbox[c]{0.48\textwidth}{\includegraphics[width = 0.48\textwidth]{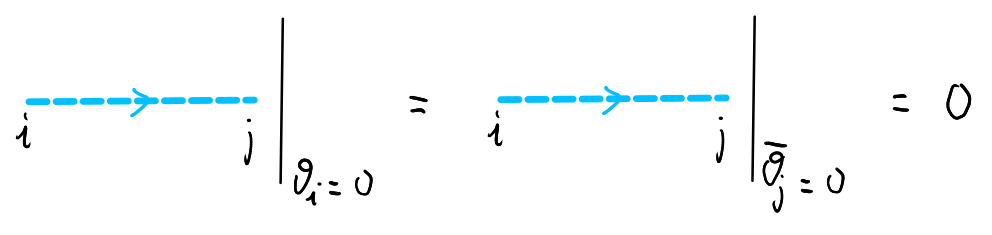}}~.
	\end{align}
	
	Another important property is the fact that we can combine the solid lines with the same endpoints, summing their momenta, as described in eq. (\ref{td6}):
	\begin{equation}
		\parbox[c]{.44\textwidth}{\includegraphics[width = .44\textwidth]{td6.png}}~.
	\end{equation}
	
	Finally we would like to point out the fact that solid $\mathit{\theta}$-lines can be \emph{stripped off} from a $\theta$-diagram; we will show this through an example. Let's consider a portion of $\theta$-diagram with the following form:
	\begin{align}
		\parbox[c]{0.16\textwidth}{\includegraphics[width = 0.16\textwidth]{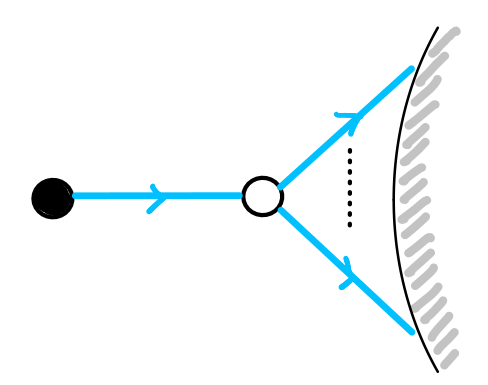}}
	\end{align}
	and the decomposition of its external line:
	\begin{align}
		\parbox[c]{0.76\textwidth}{\includegraphics[width = 0.76\textwidth]{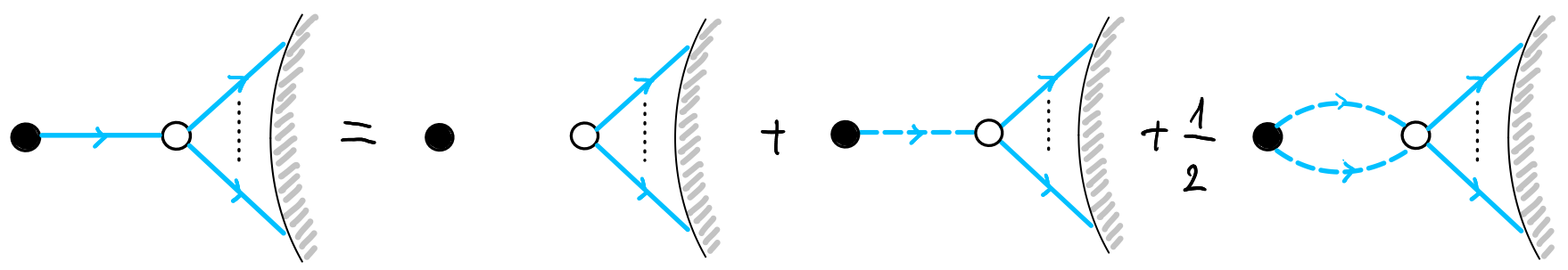}}~.
	\end{align}
	The only contribution that we can accept is the one where two dashed $\theta$-lines connect to the external vertex; thus we remain with:
	\begin{align}
		\parbox[c]{0.6\textwidth}{\includegraphics[width = 0.6\textwidth]{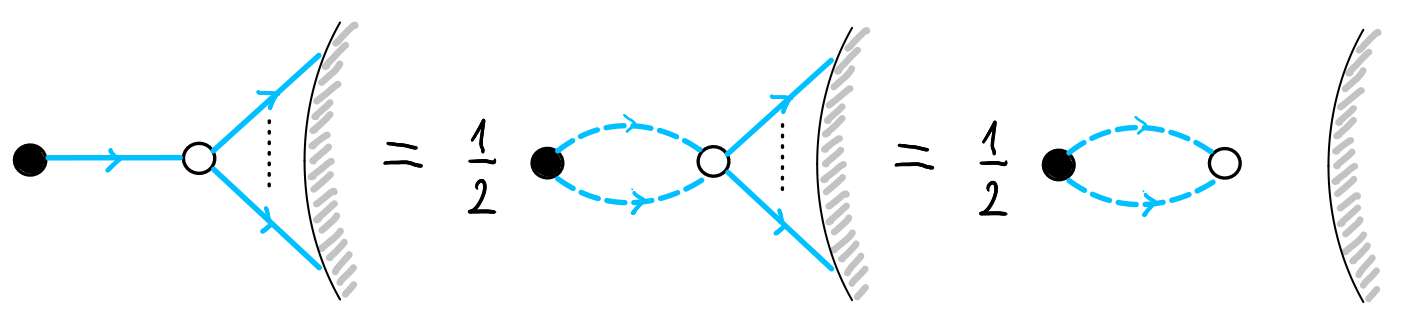}}~,
	\end{align}
	where in the last step we took into account that, since two dashed lines are already attached to the white $\theta$-dot, no further lines can be attached to it. 
	Using eq. (\ref{td8}), we can summarize this property as follows:
	\begin{align}\label{estrip}
		\parbox[c]{0.36\textwidth}{\includegraphics[width = 0.36\textwidth]{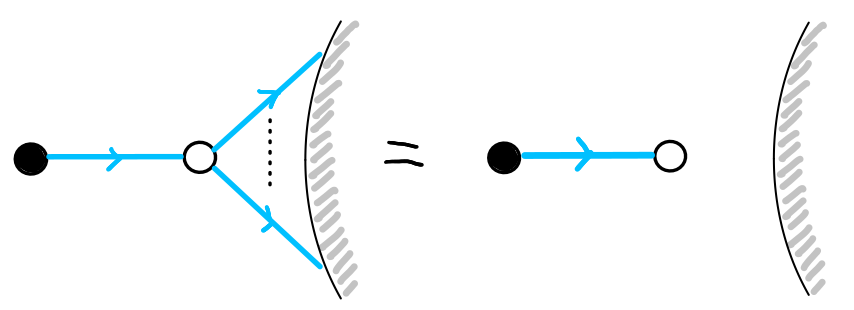}}~.
	\end{align}
	
	This property obviously holds independently of the color of the most external $\theta$-dot.

\section{Analyizing an example}
\label{sec:example}
In this section we illustrate the $\theta$-diagram methodology by means of an example which contains all the ingredients discussed above.

Let us consider a superdiagram that contributes at the three-loop order to the propagator of the adjoint scalar in an $\cN=2$ SYM theory with matter, given by
\begin{align}
	\label{vvv}
	\cW_{ab} = \parbox[c]{.4\textwidth}{\includegraphics[width = .4\textwidth]{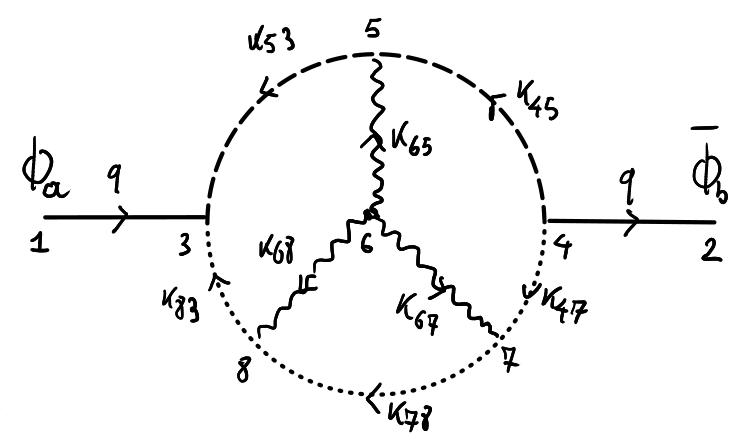}}~.
\end{align}
We have explicitly written here our rather obvious convention for the names of the momenta running in the various propagators; in the following we will sometimes just assume them.

To the ``external'' ring of chiral propagators and to all vertices but the central one we can assign $\theta$-diagrammatic elements according to the rules discussed above, in particular those in eq.s (\ref{td10},\ref{td11},\ref{td12},\ref{td22}). These elements are
\begin{align}
	\label{s4_2}
	\parbox[c]{.9\textwidth}{\includegraphics[width = .9\textwidth]{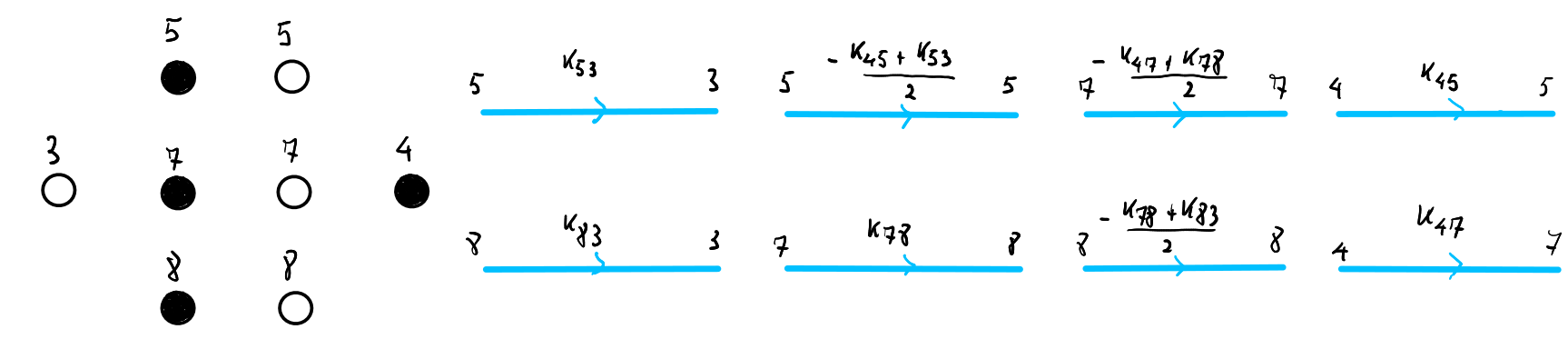}}~.
\end{align}		

Building up the Feynman diagram, we actually have to take into account six contributions corresponding to the different ways in which the triple and the single covariant derivative structures that appear originally in the three-gluon vertex -- see figure \ref{fig:vecvert} -- can be assigned to the vector propagators. In other words, the gluon part of this superdiagram is given by 
\begin{align}
	\label{td27}
	\parbox[c]{.75\textwidth}{\includegraphics[width = .75\textwidth]{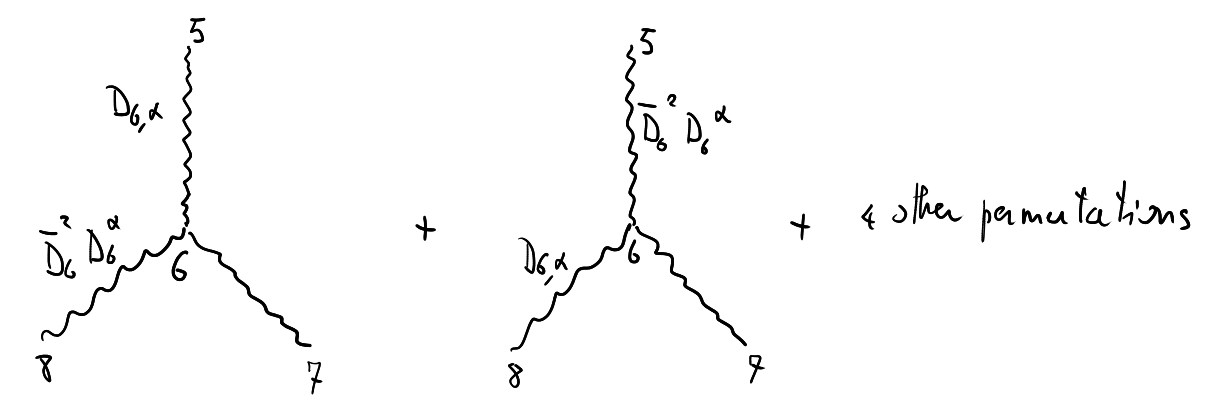}}~.
\end{align}		

We denote by $\cZ_{ijk}$, with $i,j,k$ in the range $5,7,8$, the Grassmann part of contribution to the diagram $\cW_{ab}$ in which $\bar D_6 D_6^\alpha$ acts on the vector propagator from the node $6$ to the node $i$ and
$D_{6,\alpha}$ acts on the propagator from $6$ to $k$. Thus, for instance,
\begin{align}
	\label{s4_3}
	\parbox[c]{.4\textwidth}{\includegraphics[width = .4\textwidth]{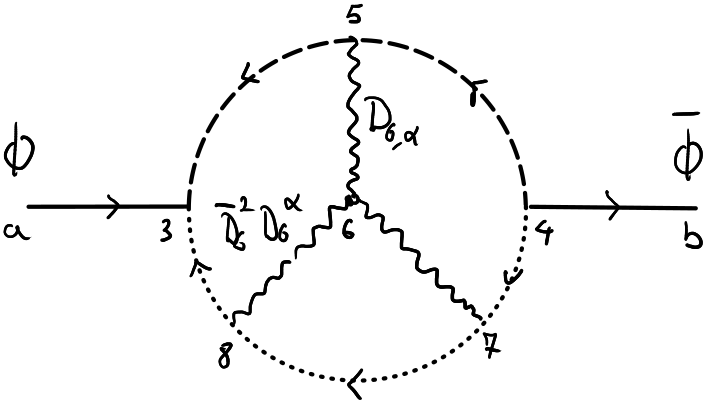}} \et \cZ_{875},
\end{align}
where for simplicity we did not write the momenta associated to the various propagators. Altogether we have
\begin{align}
	\label{td30bis}
	\cW_{ab} \et \cZ_{875} + \cZ_{578} + \cZ_{758} + \cZ_{857} + \cZ_{587} + \cZ_{785}~.
\end{align} 

The term $\cZ_{ijk}$ contains the following ingredients:
\begin{align}
	\label{td31}
	\parbox[c]{.6\textwidth}{\includegraphics[width = .6\textwidth]{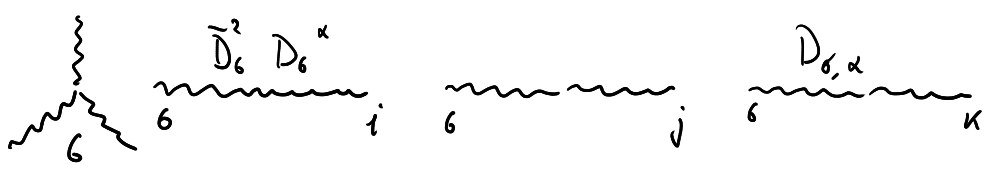}} 
\end{align}
which, according to the rules in eq.s (\ref{td30},\ref{vl3a},\ref{td21}) and (\ref{vl1}), correspond to 
\begin{align}
	\label{s4_4}
	\parbox[c]{.8\textwidth}{\includegraphics[width = .8\textwidth]{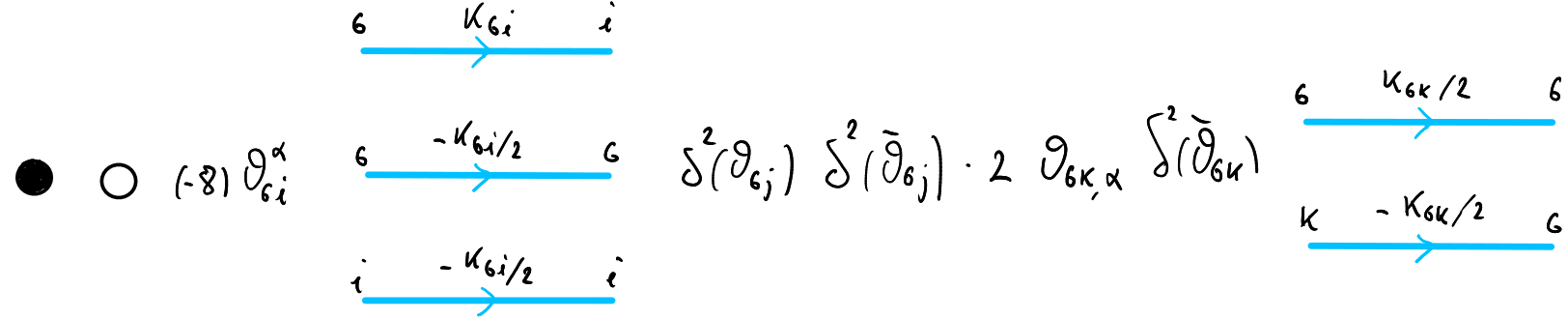}}~. 
\end{align}
The contraction $\theta_{6i}\cdot\theta_{6k}$ appears, for which we can use the property (\ref{e:thetatodeltas1}) obtaining
\begin{align}
	\label{contr6}
	\theta_{6i}\cdot\theta_{6k} = \frac 12 \left[-\delta^2(\theta_{ki}) + \delta^2(\theta_{6i}) + \delta^2(\theta_{6k}) \right]~.
\end{align} 
Moreover, we can integrate over $\theta_6$ and $\bar{\theta}_6$ variables, which appear only inside this portion of the diagram, using  the $\delta$ functions $\delta^2(\theta_{6j})$ and $\delta^2(\bar\theta_{6j})$. We rewrite thus the factors in eq. (\ref{td31}) as
\begin{align}
	\label{s4_5}
	\parbox[c]{.6\textwidth}{\includegraphics[width = .6\textwidth]{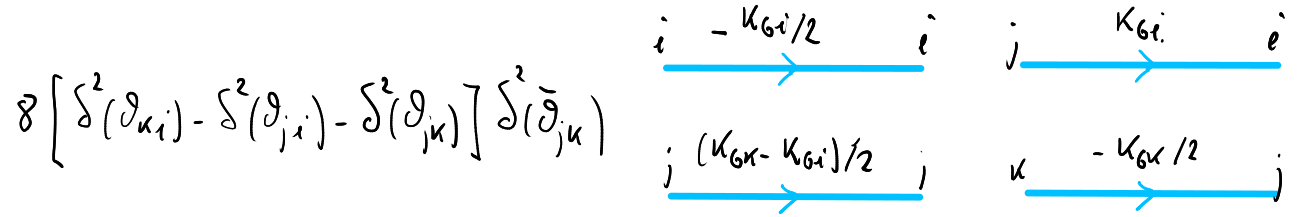}}~. 
\end{align}
Here we have used the property (\ref{td6}) for the solid lines from $j$ to $j$; note that these elements were written in eq. (\ref{s4_4}) as solid lines from $6$ to $6$, which is the same in presence of $\delta^2(\theta_{6j})\delta^2(\bar\theta_{6j})$.  Eq. (\ref{s4_5}) contains the sum of three terms, containing different $\delta$ functions. Correspondingly, we write 
\begin{align}
	\label{ZetaA123}
	\cZ_{ijk} = 8 \left(\cA^{(1)}_{ijk} - \cA^{(2)}_{ijk} - \cA^{(3)}_{ijk}\right)~,
\end{align}
where $\cA^{(1)}_{ijk}$ contains $\delta^2(\theta_{ki})$, $\cA^{(2)}_{ijk}$ contains $\delta^2(\theta_{ji})$ and $\cA^{(3)}_{ijk}$ contains $\delta^2(\theta_{jk})$.

We will now describe in some detail the evaluation of  $\cZ_{875}$ and $\cZ_{578}$; the other terms can be treated analogously and we well just quote the result. 

\paragraph{Evaluation of $\cZ_{875}$}
We start from $\cA^{(1)}_{875}$. We put together the contributions in eq. (\ref{s4_2}) with the first addend in (\ref {s4_5}) with $i,j,k= 8,7,5$, we take into account the factor of $\delta^2(\theta_{58})\delta^2(\bar{\theta}_{75})$ and we collect the lines between the same nodes according to the property (\ref{td6}). Finally, we exploit momentum conservation at each node of the diagram, see eq. (\ref{vvv}). In this way we obtain a $\theta$-diagram that we can easily evaluate using the rules discussed in sections \ref{subsec:intrthetad} and \ref{subsec:dcac}: 
\begin{align}
	\label{s4_6}
	\cA_{875}^{(1)} = \parbox[c]{.6\textwidth}{\includegraphics[width = .6\textwidth]{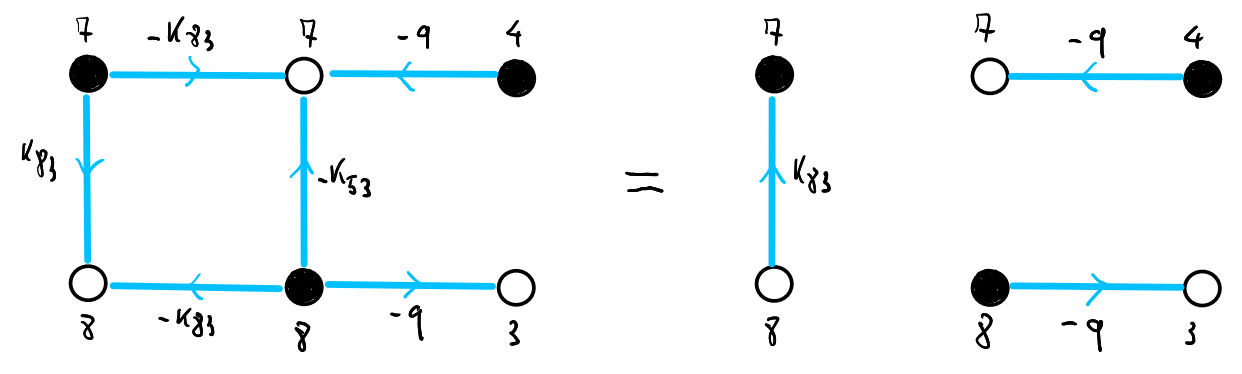}} 
\end{align}
In the second step we simply stripped off the external lines, according to the procedure discussed around eq. (\ref{estrip}). This simplification is implemented in the code that we will discuss in section \ref{sec:description}. Now we can apply to the r.h.s. the results (\ref{td8},\ref{l2int}) obtaining
\begin{align}
	\label{a1875}
	\cA_{875}^{(1)} = - k_{83}^2 \, (q^2)^2~. 
\end{align}

Let us now consider $\cA^{(2)}_{875}$. We combine the contributions in eq. (\ref{s4_2}) with the second addend in (\ref {s4_5}) with $i,j,k= 8,7,5$. We have now a factor of $\delta^2(\theta_{78})\delta^2(\bar{\theta}_{75})$;
collecting the lines and using momentum conservation we arrive at the following $\theta$-diagram: 
\begin{align}
	\label{s4_7}
	\cA_{875}^{(2)} = \parbox[c]{.75\textwidth}{\includegraphics[width = .75\textwidth]{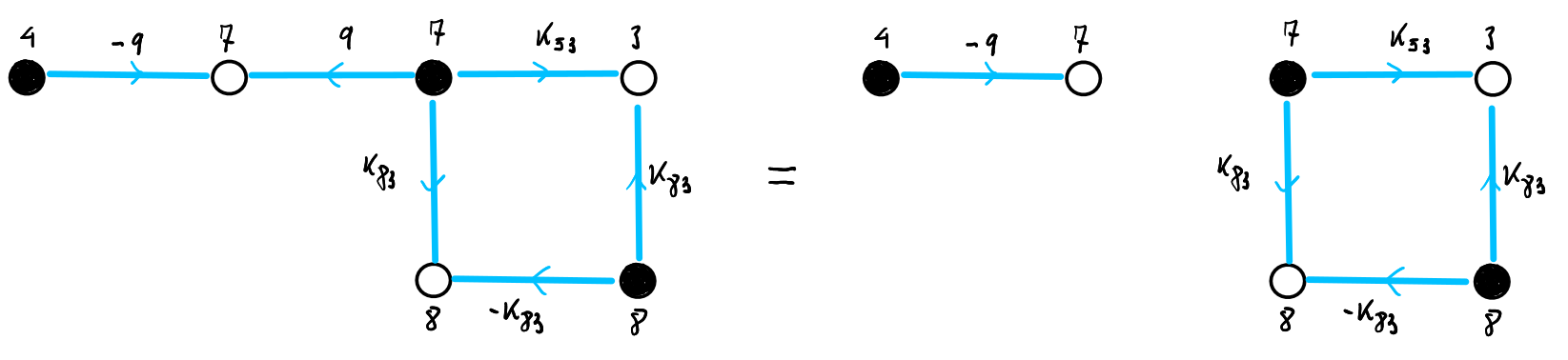}}~. 
\end{align}
In the second step we have stripped off the line $\tline{47}$.
The resulting expression can be evaluated using the results for a single line and the one for a four-nodes loop. The second one was given in eq. (\ref{z4res}), in which we now have to set $p_1=k_{53}$, $p_2=-p_3= p_4= k_{83}$: taking into account the momentum conservation relation $k_{53}+ k_{83} = -q$, the result is simply $q^2\, k_{83}^2$. Altogether we get thus
\begin{align}
	\label{a2875}
	\cA_{875}^{(2)} = - k_{83}^2  (q^2)^2~. 
\end{align}

For what concerns $\cA_{875}^{(3)}$, the terms in eq. (\ref{s4_2}) have to be combined with the third addend in 
(\ref{s4_5}) with $i,j,k= 8,7,5$, which contains a factor of $\delta^2(\theta_{75})\delta^2(\bar{\theta}_{75})$. Collecting the lines and using momentum conservation, we get this time 
\begin{align}
	\label{td38}
	\cA_{875}^{(3)} = \parbox[c]{.33\textwidth}{\includegraphics[width = .33\textwidth]{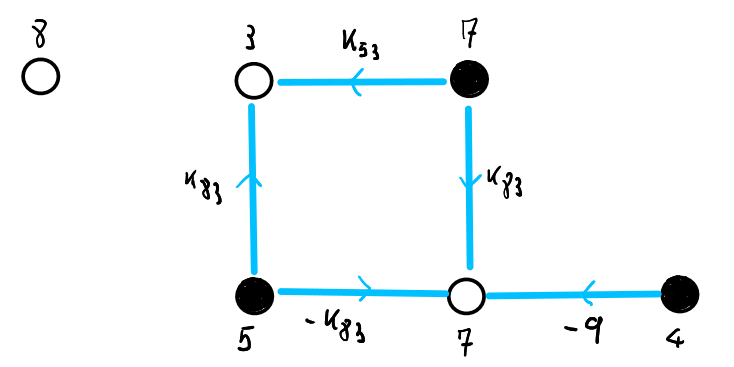}}~. 
\end{align}
This diagram is obviously vanishing because the integration over the Grassmann variables of the isolated white dot cannot be saturated by any line. Thus
\begin{align}
	\label{a3875}
	\cA_{875}^{(3)} = 0~. 
\end{align}
Inserting the results (\ref{a1875},\ref{a2875},\ref{a3875}) into eq. (\ref{ZetaA123}) we get
\begin{align}
	\label{Z875res} \cZ_{875} = 8 \left(\cA_{875}^{(1)} - \cA_{875}^{(2)} - \cA_{875}^{(3)}\right) = 0~.
\end{align}

\paragraph{Evaluation of $\cZ_{578}$}
This contribution is obtained by combining the elements in eq. (\ref{s4_2}) with those in eq. (\ref{s4_5}) with $i,j,k = 5,7,8$. We split it into the three terms $\cA_{578}^{(a)}$, with $a=1,2,3$, according to eq. (\ref{ZetaA123}). Thus $\cA_{578}^{(1)}$ corresponds to the first addend in (\ref{s4_5}), and contains the factor $\delta^2(\theta_{85}) \delta^2(\bar{\theta}_{78})$. Collecting the lines and using momentum conservation we obtain
\begin{align}
	\label{td3940}
	\cA_{578}^{(1)} = \parbox[c]{.28\textwidth}{\includegraphics[width = .28\textwidth]{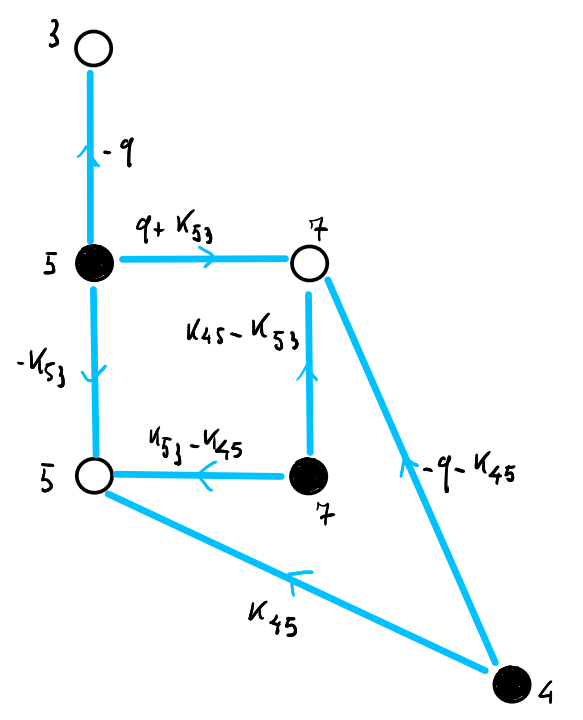}}
	= \parbox[c]{.32\textwidth}{\includegraphics[width = .32\textwidth]{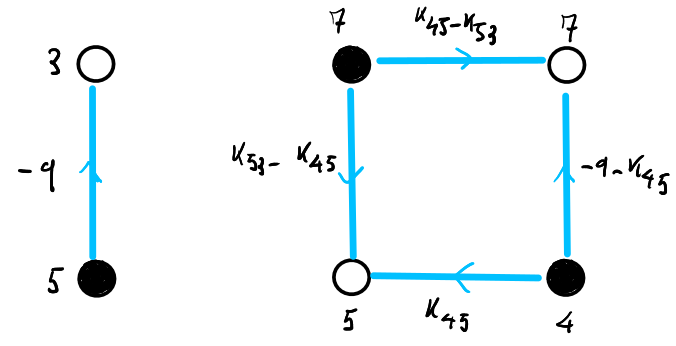}}
	~, 
\end{align}
where in the second step we have simplified the $\theta$-diagram by
stripping off the line $\tline{35}$. We can now evaluate the single line and the four-nodes loop using eq.s (\ref{td8},\ref{l2int}) and (\ref{z4res}). We obtain in the end
\begin{align}
	\label{a1578}
	\cA_{578}^{(1)} = - (q^2)^2 (k_{45} - k_{53})^2~.
\end{align}

The term $\cA_{578}^{(2)}$ corresponds to the second addend in eq. (\ref{s4_5}), and contains the factor $\delta^2(\theta_{75}) \delta^2(\bar{\theta}_{78})$. Proceeding in the by now usual way we arrive at the following $\theta$-diagram: 
\begin{align}
	\label{td41}
	\cA_{578}^{(2)} = \parbox[c]{.3\textwidth}{\includegraphics[width = .3\textwidth]{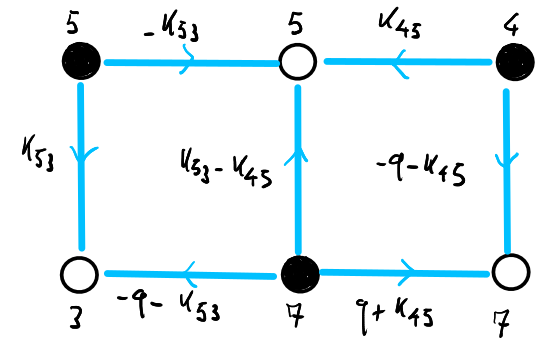}}~.
\end{align}	
The decomposition of such a double-box diagram was exhibited in figure \ref{fig:doublesquare}. Each of the terms in the decomposition is either a single line or a dashed loop whose value is obtained from eq.s (\ref{td20}) and (\ref{trtoa}-\ref{recab}). After a bit of algebra, the final outcome is 
\begin{align}
	\label{a2578}
	\cA_{578}^{(2)} = - (q+k_{45})^2 \, k_{53}^2\, (q+ k_{53})^2~.
\end{align}	

The term $\cA_{578}^{(3)}$ comes from the third addend in (\ref{s4_5}). It is proportional to $\delta^2(\theta_{78}) \delta^2(\bar{\theta}_{78})$ and the associated $\theta$-diagram is found to be the following:
\begin{align}
	\label{td42}
	\cA_{578}^{(3)} = \parbox[c]{.3\textwidth}{\includegraphics[width = .3\textwidth]{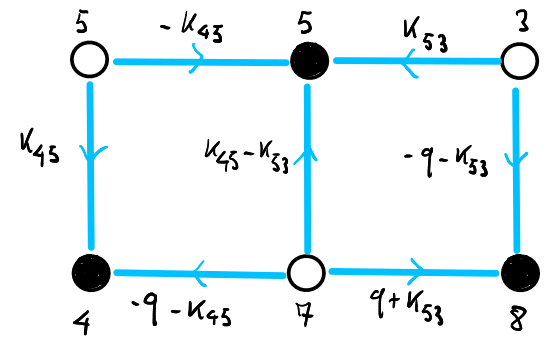}}~.
\end{align}	
This double box diagram can be evaluated analogously to the one of eq. (\ref{td41}) and the result is
\begin{align}
	\label{a3578}
	\cA_{578}^{(3)} = - q^2\, k_{45}^2 (q+k_{53})^2~.
\end{align}	

It is now straightforward to obtain 
\begin{align}
	\label{Z578res}
	\cZ_{578} = 8 \left(A^{(1)}_{578}-A^{(2)}_{578}-A^{(3)}_{578}\right)
	= 16q^2\,(k_{45}^2\,k^2_{53} + q^2\, k_{45}\cdot k_{53} 
	+ k_{45}^2\, q\cdot k_{53} + k_{53}^2\, q\cdot k_{45})~.
\end{align}

Similarly one can compute all the terms $\cZ_{ijk}$ in the Grassmann part of the diagram $\cW_{ab}$. The total result agrees with \cite{Billo:2019}, where the computation was performed only partly by means of $\theta$-diagrammatic techniques as the covariant spinorial derivatives were treated by brute force. 

\subsection{Ordering issues}
\label{subsec:ord}
The techniques introduced here allow to compute the Grassmann part of more complicated diagrams that would really be hard to deal with other methods. For diagrams which contain more than one vertex with spinor covariant derivatives, one has to pay attention to 
ordering issues which are relevant for determining the overall sign of each contribution. 

\begin{figure}
	\includegraphics[width= 0.9\textwidth]{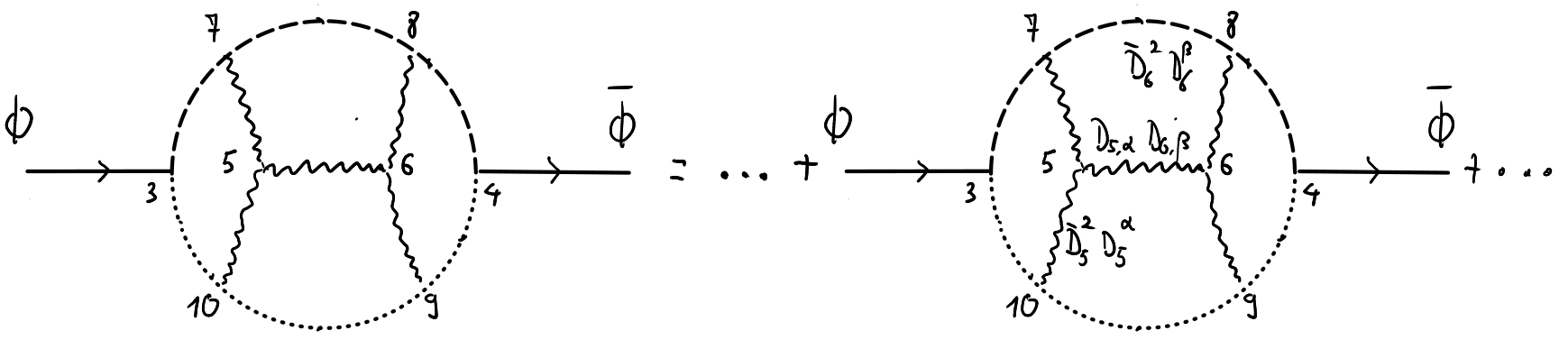}
	\caption{An example of diagram in which ordering issues must be taken into account when applying the $\theta$-diagrammatic rules.}
	\label{fig:double3vertex}
\end{figure}

\begin{figure}
	\includegraphics[width=0.9 \textwidth]{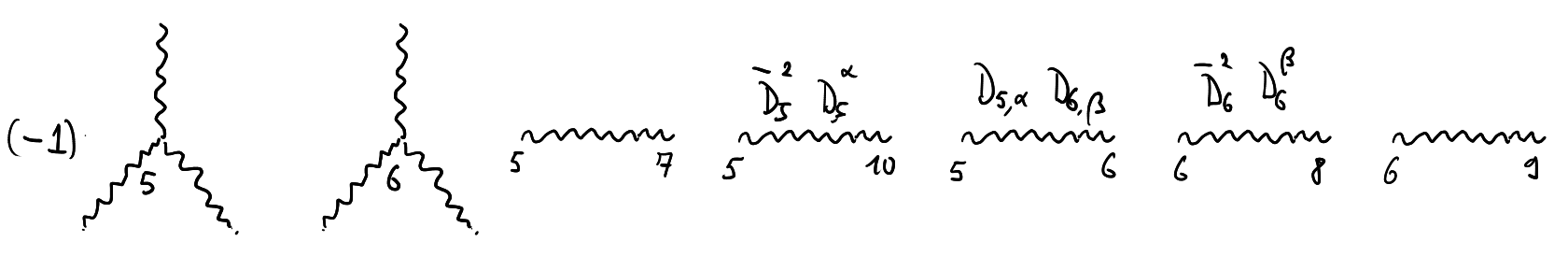}
	\caption{An example of a decomposition where the overall sign is non trivial.}
	\label{fig:double3vertexdecomposition}
\end{figure}

Suppose for instance having to compute the diagram in figure \ref{fig:double3vertex}.
Using the condensed notation $V_i=V(x_i,\theta_i,\bar{\theta}_i)$, the contractions that lead to the term singled out in the right 
hand side of figure \ref{fig:double3vertex} are the following:
\begin{gather}
	\bcontraction{...}{V_7}{...V_8...(\bar{D}_5)^2[D_5^\alpha[V_5]]}{V_5}
	\acontraction[2ex]{...V_7...V_8...(\bar{D}_5)^2[D_5^\alpha[}{V_5}{]]V_5 D_{5,\alpha}[V_5](\bar{D}_6)^2[D_6^\alpha[V_6]]V_6 D_{6,\alpha}[V_6]...V_9...}{V_{10}}
	\bcontraction{...V_7...V_8...(\bar{D}_5)^2[D_5^\alpha[V_5]]V_5 D_{5,\alpha}[}{V_5}{](\bar{D}_6)^2[D_6^\alpha[V_6]]V_6 D_{6,\alpha}[}{V_6}
	\bcontraction[2ex]{...V_7...V_8...(\bar{D}_5)^2[D_5^\alpha[V_5]]V_5 D_{5,\alpha}[V_5](\bar{D}_6)^2[D_6^\alpha[V_6]]}{V_6}{D_{6,\alpha}[V_6]...}{V_9}
	\acontraction{...V_7...}{V_8}{...(\bar{D}_5)^2[D_5^\alpha[V_5]]V_5 D_{5,\alpha}[V_5](\bar{D}_6)^2[D_6^\alpha[}{V_6}
	...V_7...V_8...(\bar{D}_5)^2[D_5^\alpha[V_5]]V_5 D_{5,\alpha}[V_5](\bar{D}_6)^2[D_6^\alpha[V_6]]V_6 D_{6,\alpha}[V_6]...V_9...V_{10}...
	\label{e:generalWickscontractions}
\end{gather}
Our approach is to separate this superdiagram in its elementary constituents, to which we assign their $\theta$-diagrammatical counterparts. In doing this we have to take into account that, while the vectorial superfield is a commuting object, each covariant spinor derivative flips its statistics.  
We must thus pay attention to the signs that arise if some exchanges of such elements in the Wick contractions. These signs are accounted for in the code
that accompanies this paper. 
In the case at hand, the decomposition of the particular contribution we are considering is given in figure \ref{fig:double3vertexdecomposition} and includes a global $(-1)$ factor.

\section{Superdiagrams with vector or spinor external states}
\label{sec:vecspin}
We will now discuss how to handle the Grassmann integrations in $n$-point functions whose external states are 
spinorial or vectorial. To extend the $\theta$-diagrammatic method to these cases it suffices to understand what Grassmannian factors the non-scalar external lines introduce. The necessary ingredients have already been written down in section \ref{subsec:ext_lines}; here we will recast them in a $\theta$-diagrammatic form.

We shall begin with vector external states, which are quite easy to handle.

\subsection{External vectors}
\label{subsec:extvec}
The Grassmann part of a vector line attached on one end to an external vector state $v^\mu$ has been given in eq. (\ref{e:vmue}). We can introduce a $\theta$-diagrammatical notation for it as follows:
\begin{align}
	\label{e1}
	\parbox[c]{.7\textwidth}{\includegraphics[width = .7\textwidth]{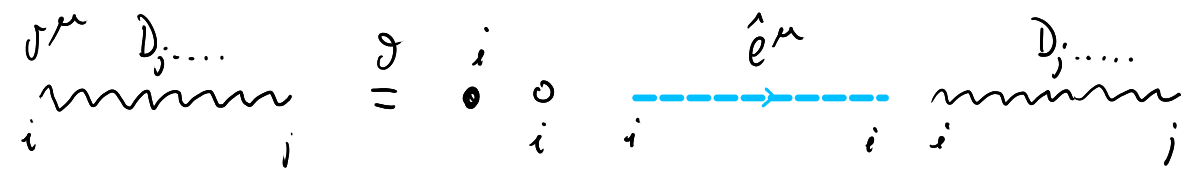}}~,
\end{align}	
where the element
\begin{align}
	\label{e2}
	\parbox[c]{.35\textwidth}{\includegraphics[width = .35\textwidth]{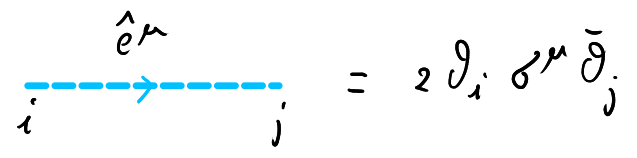}}~.
\end{align}	
is nothing else that the dashed line introduced in eq. (\ref{td1}) with the momentum $p$ replaced by the versor $\hat e^\mu$ in the direction $\mu$. This versor has components $\hat e^\mu_\nu = \delta^\mu_\nu$, so that its scalar product with any vector $a$ yields
\begin{align}
	\label{emuprop}
	\hat e^\mu \cdot a = a^\mu~.
\end{align}	  
Thus in particular $\hat e^\mu \cdot \sigma = \sigma^\mu$ and $\hat e^\mu\cdot \bar{\sigma} = \bar{\sigma}^\mu$.

Let us note that, in the case in which the external vector line carries no spinorial covariant derivatives, its expression simplifies into 
\begin{align}
	\label{e3}
	\parbox[c]{.35\textwidth}{\includegraphics[width = .35\textwidth]{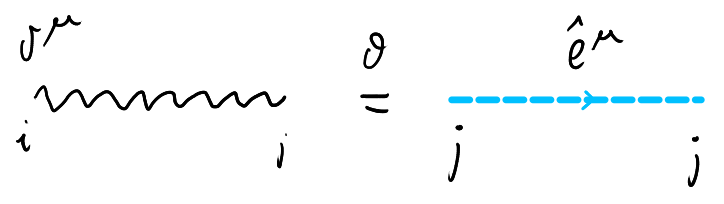}}
\end{align}	
since the supervector propagator is given by $\delta^2(\theta_{ij}) \delta^2(\bar\theta_{ij})$ and can be used to integrate the $\theta_i$ and $\bar{\theta}_i$ variables, which only occur in this part of the diagram, according to (\ref{e:vmue}).

Consider a very simple example, akin to the case with scalar external states considered in section \ref{subsec:simpleexample}, namely the one-loop superdiagram  
\begin{align}
	\label{e4}
	\cW_{ab}^{\mu\nu} = \parbox[c]{.3\textwidth}{\includegraphics[width = .3\textwidth]{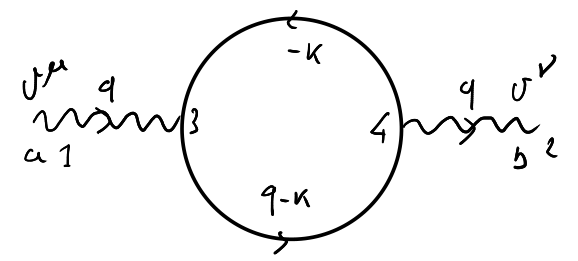}}~.
\end{align}	
Its Grassmann part $\cZ^{\mu\nu} \et \cW_{ab}^{\mu\nu}$ is given, according to the rule just introduced and to eq.s (\ref{td10},\ref{td22}), by
\begin{align}
	\label{e5}
	\cZ^{\mu\nu} = \parbox[c]{.25\textwidth}{\includegraphics[width = .25\textwidth]{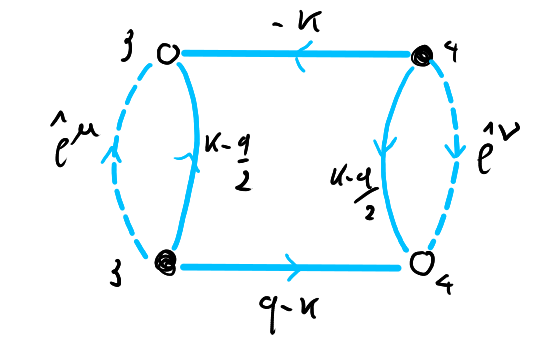}}~.
\end{align}	
This $\theta$-diagram can be expanded as usual into cycles of dotted lines, but such cycles are bound to contain the two $\hat e^\mu$ and $\hat e^\nu$ dotted lines. Therefore one gets
\begin{align}
	\label{e6}
	\cZ^{\mu\nu} = \parbox[c]{.5\textwidth}{\includegraphics[width = .5\textwidth]{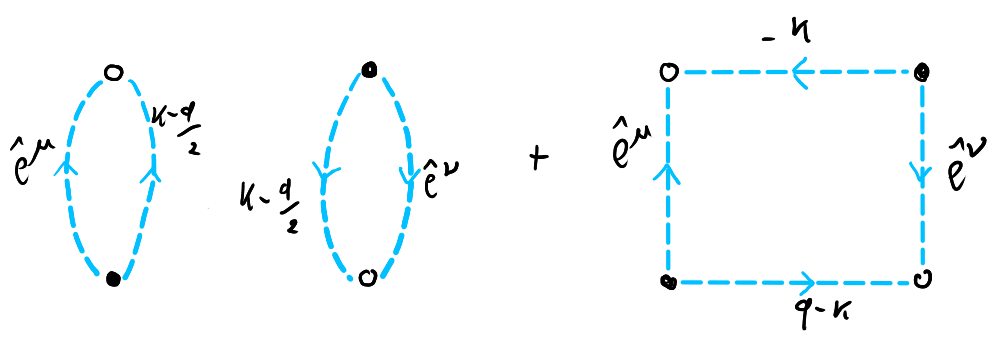}}~.
\end{align}
Both terms can now be computed using the trace rule of eq. (\ref{td20}) and the formul\ae\, in Appendix \ref{subapp:spinconv}, treating the versors $\hat e^{\mu}$ and $\hat e^\nu$ as any other quadri-momentum and then taking into account eq. (\ref{emuprop}). 

As another example, let us consider the following superdiagram, corresponding to  a one loop cubic vertex correction:
\begin{align}
	\label{e4bis}
	\cW^{\mu}_{abc} = \parbox[c]{.3\textwidth}{\includegraphics[width = .3\textwidth]{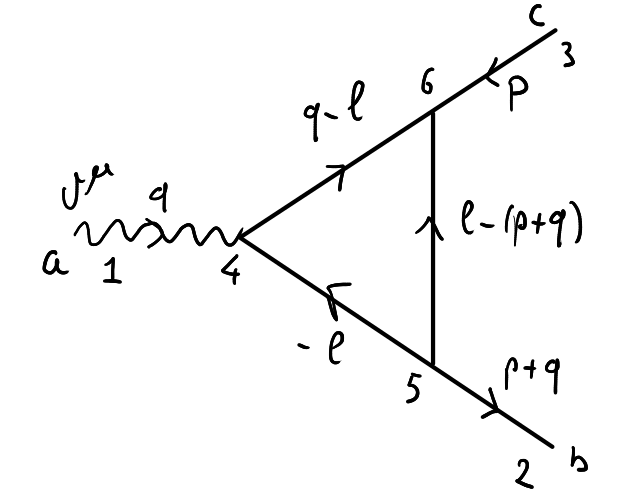}}~.
\end{align}	
Its Grassmann part $\cZ^{\mu} \et \cW_{abc}^{\mu}$ is given, using the rules introduced above, by
\begin{align}
	\label{e4tris}
	\cZ^{\mu} = \parbox[c]{.25\textwidth}{\includegraphics[width = .25\textwidth]{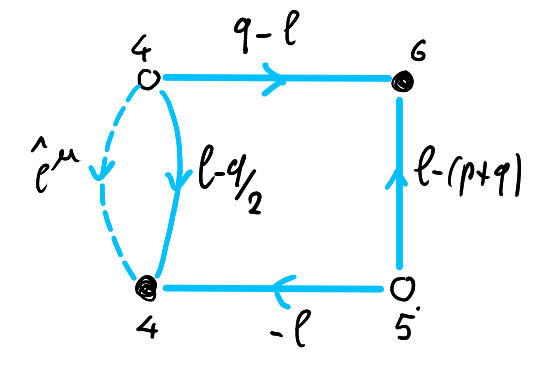}}~.
\end{align}
Expanding this $\theta$-diagram into cycles of dotted lines, bound to contain the $\hat e^\mu$ dotted line, we get
\begin{align}
	\label{e4quad}
	\cZ^{\mu} = \parbox[c]{.5\textwidth}{\includegraphics[width = .5\textwidth]{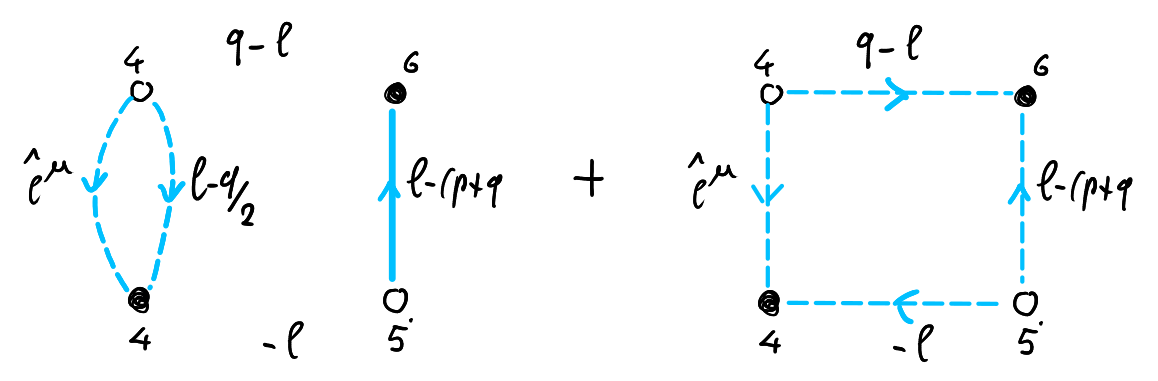}}~.
\end{align}
Again, both terms can be computed using eq. (\ref{td20}) and Appendix \ref{subapp:spinconv}.

\subsection{Spinors from (anti-)chiral multiplets as external states}
\label{subsec:chirext}
Let us now consider superdiagrams whose external states are spinors sitting in $\cN=1$ chiral or anti-chiral multiplets%
\footnote{Note that, from the $\cN=2$ point of view, if the $\cN=1$ chiral or antichiral multiplet is the adjoint one which, together with the $\cN=1$ vector multiplet, makes up the $\cN=2$ gauge multiplet, then these spinors are among the gauginos of the theory.}.

The Grassmann part of a chiral line attached at one end to a chiralino or to an anti-chiralino has been given in
eq.s (\ref{e:sce2},\ref{e:asce2}). In a (partially) 
$\theta$-diagrammatic notation it can be expressed as follows:
\begin{align}
	\label{e7}
	\parbox[c]{.5\textwidth}{\includegraphics[width = .5\textwidth]{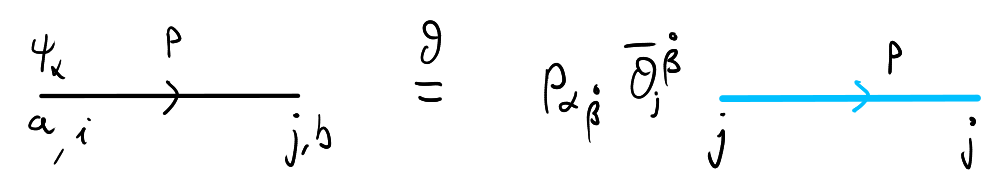}}
\end{align}
and
\begin{align}
	\label{e8}
	\parbox[c]{.5\textwidth}{\includegraphics[width = .5\textwidth]{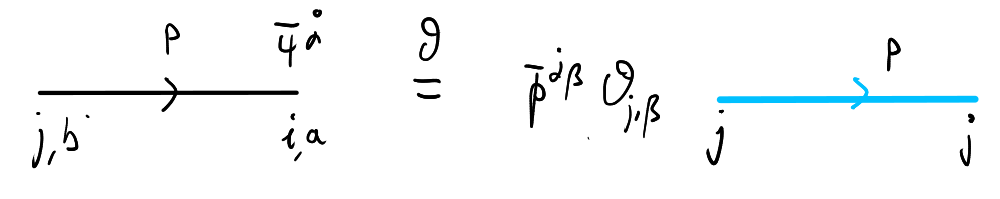}}~.
\end{align}

The novelty is the presence of ``bare'' $\theta$-variables whose spinorial indices are not contracted with the spinor indices carried by other $\theta$ variables. This leads, when these external lines are combined with the inner vertices, to the appearance of new $\theta$-diagrammatic elements. Indeed, consider for instance an external line of the type (\ref{e7}) attached to a chiral cubic vertex, whose Grassmann part was given in eq. (\ref{td12}). We have
\begin{align}
	\label{e9}
	\parbox[c]{.75\textwidth}{\includegraphics[width = .75\textwidth]{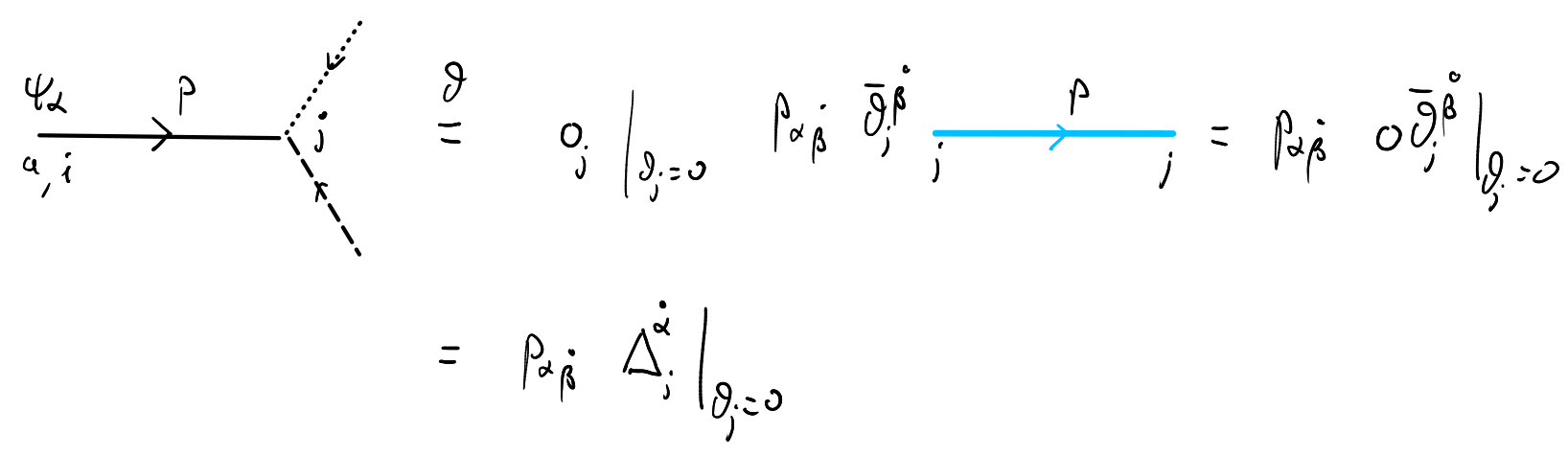}}
\end{align}
and similarly
\begin{align}
	\label{e9bis}
	\parbox[c]{.45\textwidth}{\includegraphics[width = .45\textwidth]{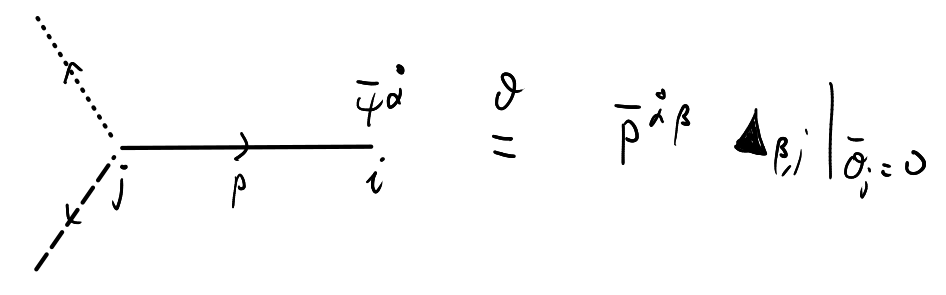}}~,
\end{align}
where we introduced the notation
\begin{align}
	\label{e10}
	\parbox[c]{.7\textwidth}{\includegraphics[width = .7\textwidth]{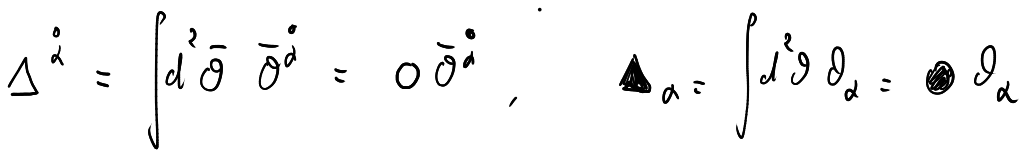}}~.
\end{align}

Consider for instance the following simple one-loop example:
\begin{align}
	\label{e11}
	\cW_{ab,\alpha}^{\dot\alpha} = \parbox[c]{.4\textwidth}{\includegraphics[width = .4\textwidth]{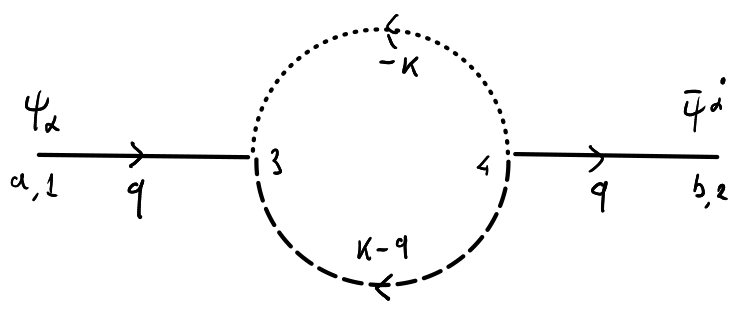}}~.
\end{align}	
Using the rules just stated, its Grassmann part $\cZ_{~\alpha}^{\dot\alpha} \et \cW_{ab,\alpha}^{\dot\alpha} $ is given by
\begin{align}
	\label{e12}
	\cZ_{~\alpha}^{\dot\alpha} = \parbox[c]{.75\textwidth}{\includegraphics[width = .75\textwidth]{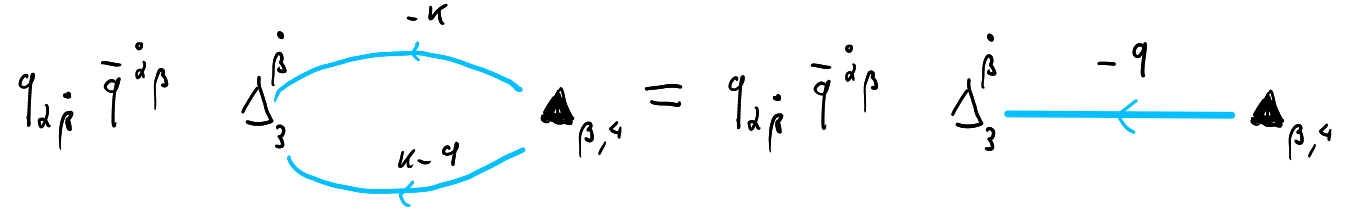}}~.
\end{align}	

Let us note that the external spinors appear in a specific order, dictated by the Wick contractions that originate the diagram one is considering, which implies a specific ordering of the ``bare'' $\theta$-variables in the $\theta$-diagram that carry the spinor indices of the external spinors. If this order is altered, we have to keep track of the ensuing overall sign - again, this is automatic in the code.   

The crucial property of these triangular symbols is that, in the expansion of a $\theta$-diagram, only the terms where precisely one dotted line is attached to each triangular vertex survive since these represent an integration already containing a fermionic coordinate. This is analogous to the fact that exactly two dotted lines must be attached to each black or white dot, as stated after eq. (\ref{td7}). Thus the possible decompositions of the $\theta$-diagram must contain open trajectories -- which cannot intersect or overlap -- starting on a triangle end ending on another one. In the case of eq. (\ref{e12}) there is just one possibility:
\begin{align}
	\label{e13}
	\cZ_{~\alpha}^{\dot\alpha} = \parbox[c]{.33\textwidth}{\includegraphics[width = .33\textwidth]{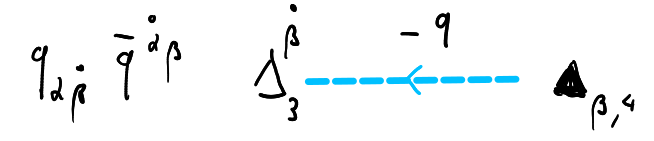}}~.
\end{align}	
Let us consider a more elaborate example, namely the following two-loop quartic diagram, which we indicate only schematically, without labeling the momenta and the nodes:
\begin{align}
	\label{e13bis}
	\parbox[c]{.3\textwidth}{\includegraphics[width = .3\textwidth]{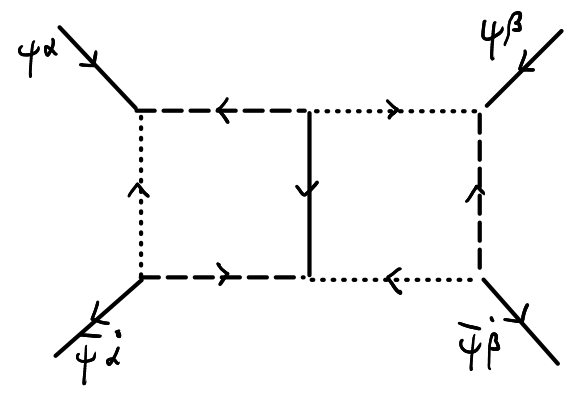}}~.
\end{align}	
In this case there are many more possibilities to draw the open lines and the decomposition of the $\theta$-diagram contains the following structures (we omit writing the factors of the external momenta attached to the triangles):
\begin{align}
	\label{e13tris}
	\parbox[c]{.8\textwidth}{\includegraphics[width = .8\textwidth]{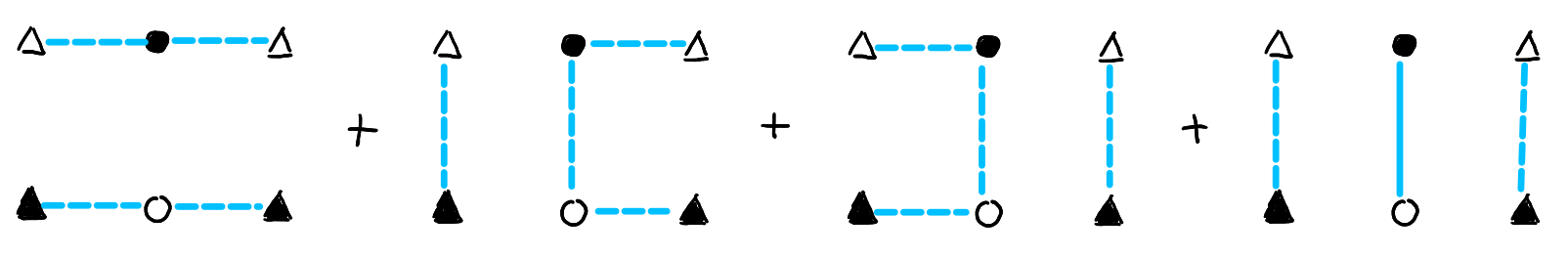}}~.
\end{align}	

The open paths connecting the triangles represent Grassmannian integrations; for instance
\begin{align}
	\label{e14}
	\parbox[c]{.35\textwidth}{\includegraphics[width = .35\textwidth]{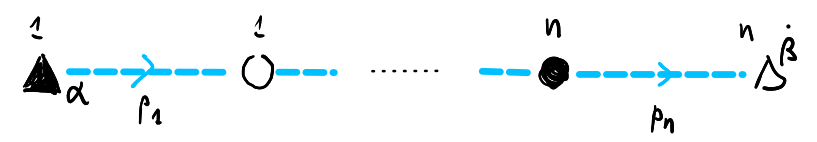}}
	= \int d\Theta\, \theta_{1,\alpha} 2(\theta_1 p_1 \bar{\theta}_1)\ldots 2(\theta_n p_n \bar{\theta}_n)
	\bar{\theta}_n^{\dot{\beta}}~,
\end{align}	
where by $d\Theta$ we indicate the integration over all the involved Grassmann variables.
These expressions can be evaluated by the same Fierz techniques utilized in eq. (\ref{td20}) for the closed paths. The result is
\begin{align}
	\label{e14bis}
	\parbox[c]{.35\textwidth}{\includegraphics[width = .35\textwidth]{e14}}
	= \frac{(-1)^{n+1}}{2}(p_1 \bar q_1 \ldots \bar{q}_{n-1} p_n)_{\alpha}^{\dot{\gamma}}~.
\end{align}	
Similarly, one has
\begin{align}
	\label{e15}
	\parbox[c]{.35\textwidth}{\includegraphics[width = .35\textwidth]{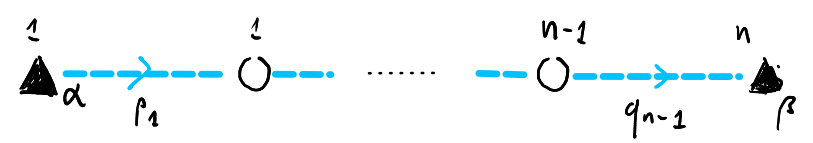}}
	= \frac{(-1)^{n}}{2} (p_1 \bar q_1 \ldots  p_{n-1} \bar{q}_{n-1})_{\alpha\beta}
\end{align}	
and
\begin{align}
	\label{e16}
	\parbox[c]{.35\textwidth}{\includegraphics[width = .35\textwidth]{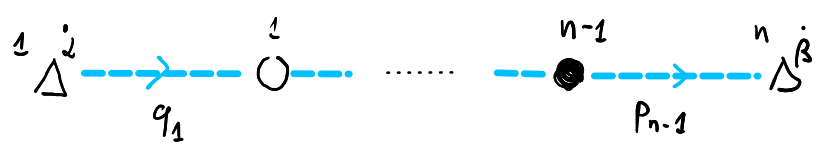}}
	= \frac{(-1)^{n}}{2} (\bar q_1 p_1\ldots  \bar{q}_{n-1} p_{n-1})^{\dot\alpha\dot\beta}
\end{align}	
All of these products of $\sigma^\mu$-matrices
can be treated algorithmically as described in Appendix (\ref{subapp:spinconv}). 

For instance, in the first two cases we have
\begin{align}
	\label{e16bis}
	\parbox[c]{.2\textwidth}{\includegraphics[width = .2\textwidth]{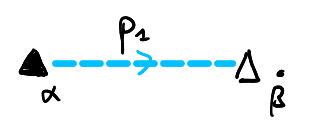}}
	= \frac 12 (p_1)_\alpha^{~\dot{\beta}} = \frac 12 (p_1)_\mu \,(\sigma^\mu)_\alpha^{~\dot{\beta}}
\end{align}	
and
\begin{align}
	\label{e16tris}
	\parbox[c]{.3\textwidth}{\includegraphics[width = .3\textwidth]{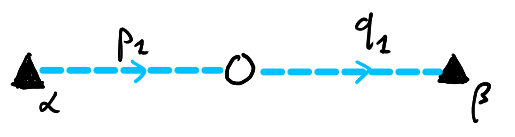}}
	= \frac 12 (p_1)_\mu \, (q_1)_\nu\, (\sigma^{\mu}\bar{\sigma}^{\nu})_{\alpha\beta}
	= -\frac 12 p_1 \cdot q_1 \, \epsilon_{\alpha\beta} + \frac 12 (p_1)_\mu \, (q_1)_\nu\, 
	(\sigma^{\mu\nu})_{\alpha\beta}~. 
\end{align}	
Note that the open chains with two black triangles, i.e. with two external chiral spinor indices, transform in the $(2,1) \otimes (2,1)$ representation of the spin group, which is reducible into $(1,1) \oplus (3,1)$, i.e., into an antisymmetric and a symmetric part in these spinor indices. 

Coming back to the simple case of eq. (\ref{e13}), the final result is 
\begin{align}
	\label{e17}
	\cZ_{~\alpha}^{\dot{\alpha}} = (-1)\bar q^{\dot{\alpha}\beta}\times \frac{(-1)}{2} (-q)_{\beta\dot{\gamma}} \epsilon^{\dot{\gamma}\dot{\beta}}q_{\alpha\dot\beta}
	= -\frac 12 (\bar q q \bar q)^{\dot{\alpha}}_{\ \alpha} = \frac{q^2}{2} (\bar q)^{\dot{\alpha}}_{~\alpha}=\frac{q^2}{2}q_\mu(\sigma^\mu)_\alpha^{\ \dot{\alpha}}~.
\end{align}	

The first $(-1)$ factor comes from exchanging the theta-variables of the triangular $\theta$-vertices in eq. (\ref{e13}) to match the one in the formula (\ref{e14}).

\subsection{Gauginos as external states}
\label{subsec:gaugext}
Let us consider now  the case in which we have a spinor from the $\cN=1$ vector multiplet in an external state. The  Grassmann parts of a vector line attached at one end to an external gaugino or anti-gaugino were given in eq.s (\ref{e:ge}) and (\ref{e:age}). We can describe them in $\theta$-diagrammatic notation as follows: 
\begin{align}
	\label{e18}
	\parbox[c]{.65\textwidth}{\includegraphics[width = .65\textwidth]{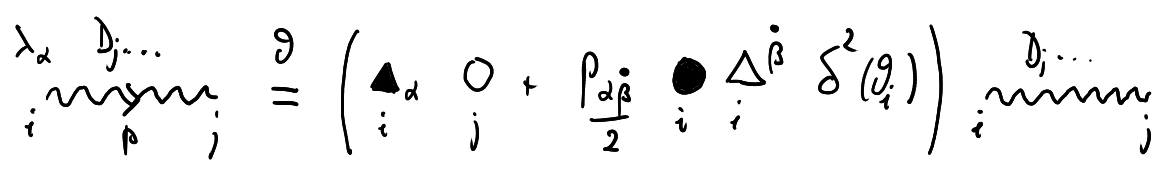}}
\end{align}
and
\begin{align}
	\label{e19}
	\parbox[c]{.65\textwidth}{\includegraphics[width = .65\textwidth]{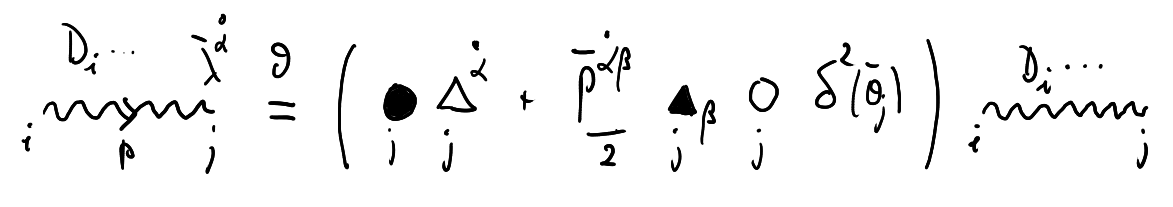}}~.
\end{align}
Thus each gaugino external line leads to the sum of two diagrammatic elements; this is a consequence of the fact that we chose not to impose the Wess-Zumino gauge. The $\theta$-diagrams with gauginos as external states will therefore comprise multiple terms already before effecting their decomposition. 

Let us note that if the external gaugino line carries no spinorial derivatives its contributions simplify because they involve the vector line with no spinor derivatives, which is just $\delta^2(\theta_{ij}) \delta^2(\bar{\theta}_{ij})$. This allows to integrate out the $\theta$ variables in the external node. Thus, for instance, if the external gaugino is attached to a cubic adjoint vertex, see figure \ref{fig:vecvert}, we have
\begin{align}
	\label{e20}
	\parbox[c]{.48\textwidth}{\includegraphics[width = .48\textwidth]{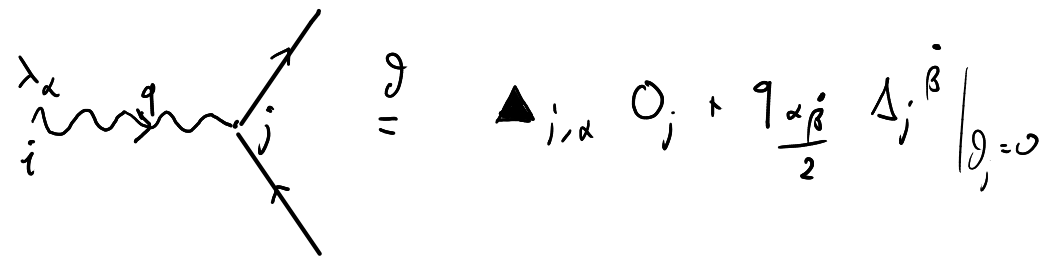}}
\end{align}
and
\begin{align}
	\label{e21}
	\parbox[c]{.48\textwidth}{\includegraphics[width = .48\textwidth]{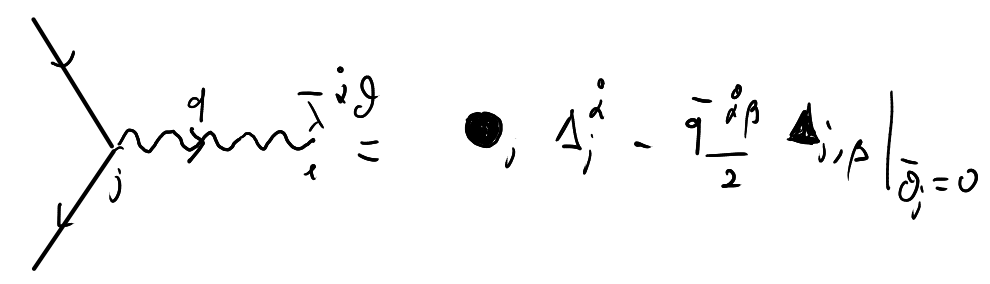}}~.
\end{align}

Let us consider, for instance, the following one-loop superdiagram:
\begin{align}
	\label{e21bis}
	\cW_{ab,\alpha}^{\dot{\alpha}} = 
	\parbox[c]{.4\textwidth}{\includegraphics[width = .4\textwidth]{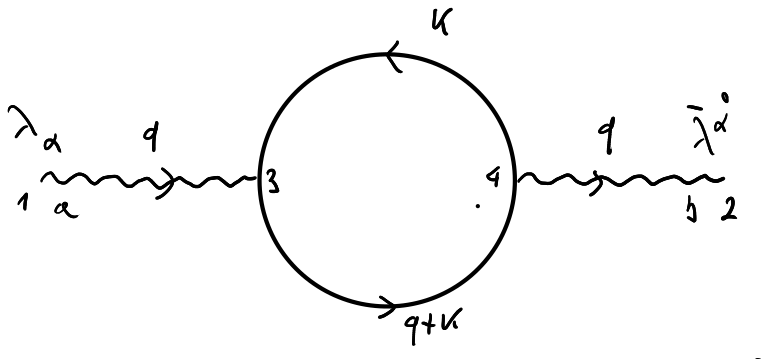}}~.
\end{align}
According to the rules we just described, its Grassmann part $\cZ_{~\alpha}^{\dot{\alpha}}\et \cW_{ab,\alpha}^{\dot{\alpha}}$ is given in $\theta$-diagrammatical terms by 
\begin{align}
	\label{e22}
	\cZ_{~\alpha}^{\dot{\alpha}} = 
	\parbox[c]{.8\textwidth}{\includegraphics[width = .8\textwidth]{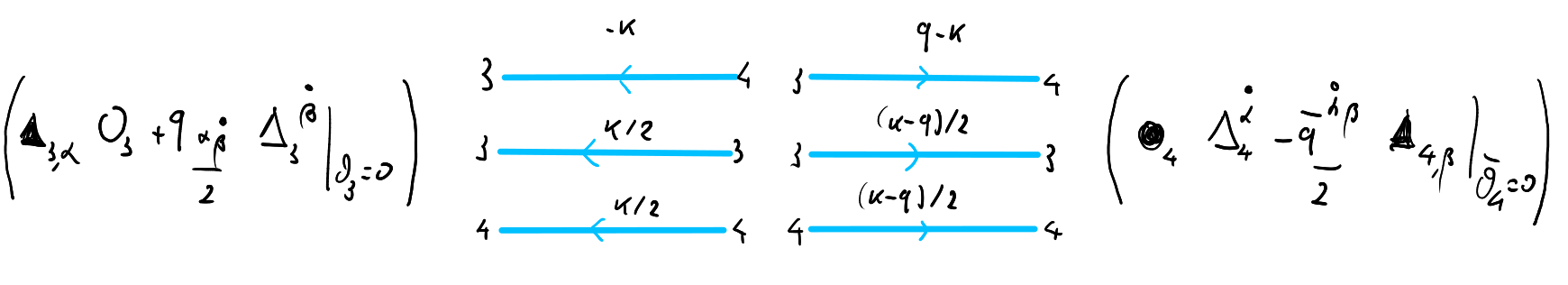}}~.
\end{align}
It comprises therefore four $\theta$-diagrams: we write
$\cZ_{~\alpha}^{\dot{\alpha}} = \cZ_{(1)\alpha}^{\dot{\alpha}} + \cZ_{(2)\alpha}^{\dot{\alpha}} + \cZ_{(3)\alpha}^{\dot{\alpha}} + \cZ_{(4)\alpha}^{\dot{\alpha}}$ and we have
\begin{align}
	\label{e23}
	\cZ_{(1)\alpha}^{\dot{\alpha}} = 
	\parbox[c]{.5\textwidth}{\includegraphics[width = .5\textwidth]{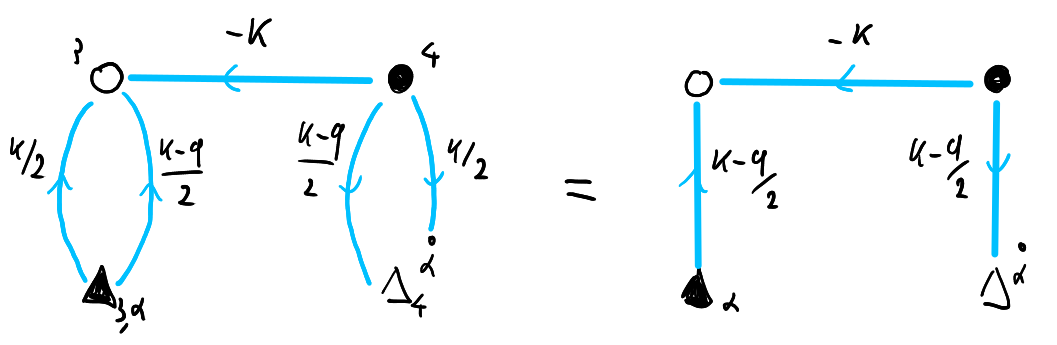}}~.
\end{align}
In the first step we kept the labels on the nodes to facilitate the comparison with eq. (\ref{e22}); in the second step we omitted such labels and we joined the lines between the same nodes. In the following terms we will write directly the analogue of this second expression, namely
\begin{align}
	\label{e24e25}
	\cZ_{(2)\alpha}^{\dot{\alpha}} = 
	\parbox[c]{.25\textwidth}{\includegraphics[width = .25\textwidth]{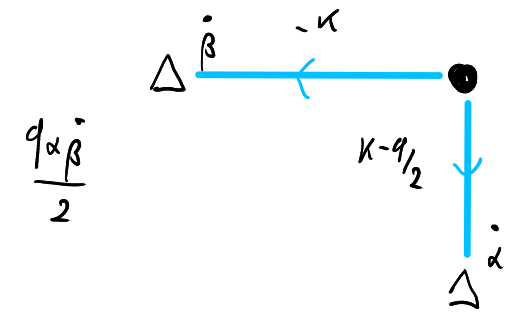}}~,~~~
	\cZ_{(3)\alpha}^{\dot{\alpha}} = 
	\parbox[c]{.25\textwidth}{\includegraphics[width = .25\textwidth]{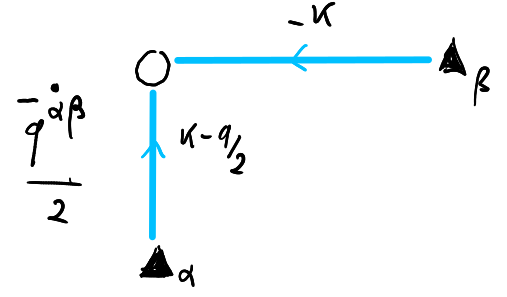}}
\end{align}
and finally
\begin{align}
	\label{e26}
	\cZ_{(4)\alpha}^{\dot{\alpha}} = 
	\parbox[c]{.34\textwidth}{\includegraphics[width = .34\textwidth]{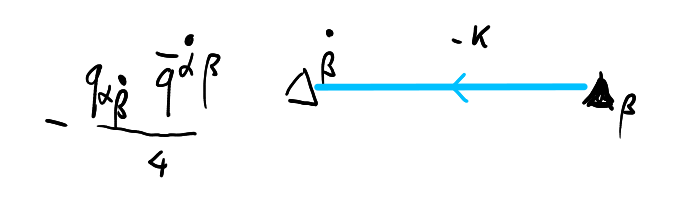}}~.
\end{align}
The decomposition of these $\theta$-diagrams goes as follows. We find
\begin{align}
	\label{e27}
	\cZ_{(1)\alpha}^{\dot{\alpha}} = 
	\parbox[c]{.65\textwidth}{\includegraphics[width = .65\textwidth]{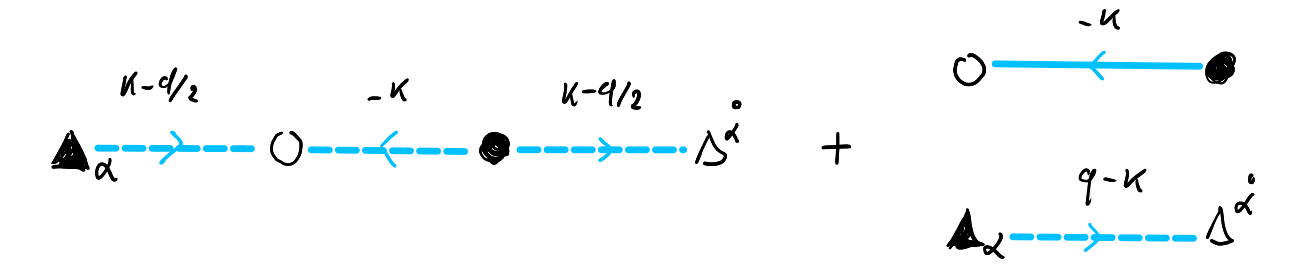}}
\end{align}
as well as
\begin{align}
	\label{e28e29}
	\cZ_{(2)\alpha}^{\dot{\alpha}} = 
	\parbox[c]{.32\textwidth}{\includegraphics[width = .32\textwidth]{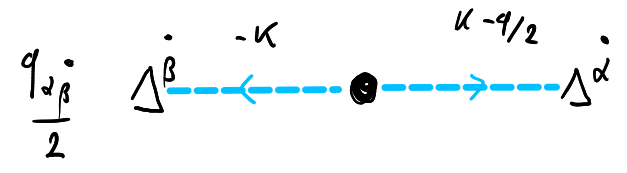}}~,~~~
	\cZ_{(3)\alpha}^{\dot{\alpha}} = 
	\parbox[c]{.32\textwidth}{\includegraphics[width = .32\textwidth]{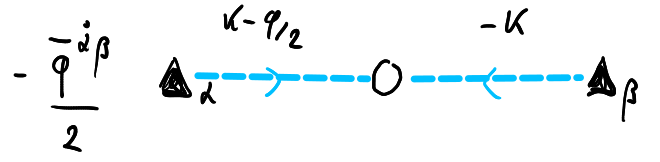}}
\end{align}
and 
\begin{align}
	\label{e30}
	\cZ_{(4)\alpha}^{\dot{\alpha}} = 
	\parbox[c]{.27\textwidth}{\includegraphics[width = .27\textwidth]{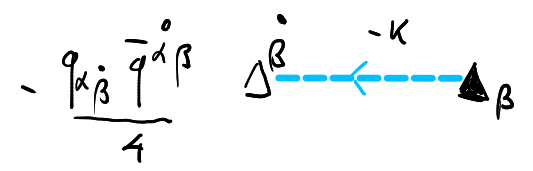}}~.
\end{align}

\section{Use of the Grassmann integration algorithm}
\label{sec:description}
We have insisted that the approach to the Grassmann integration in superdiagrams we described in the previous sections is algorithmic and can therefore be implemented in a computer code. We have chosen to realize this possibility as a Mathematica notebook\footnote{We decided to not build a package because our coding requires to  make Mathematica's \q{\ttt{.}} operation orderless. We chose to localize this choice in a single notebook to avoid disrupting other Mathematica computations by the user.} which is made available together with this paper.
The code, as it stands,
allows to deal with the type of matter content and interactions considered in the bulk of the paper. It can also be applied to other theories whose interaction vertices contain only products of superfields with some combinations of spinorial covariant derivatives on vector superfields. It is straightforward to modify the program in order to include massive chiral multiplets, in which case chiral superpropagators terms like $\langle \Phi \Phi\rangle$ or $\langle \bar{\Phi} \bar{\Phi}\rangle$ appear. Furthermore it is in principle possible to extend the program to handle vertices with all possible combinations of spinorial derivatives an all superfields. It is true that the code can certainly be optimized by more expert programmers but yet it proves our point: the procedure described in the text can be implemented on a computer.  

The notebook is called \texttt{Theta-diagrams.nb}. In the following, we assume that this notebook has been opened in \texttt{Mathematica} and evaluated. 
Beside the notebook \texttt{Theta-diagrams.nb}, there is a folder, called \texttt{Theta-programs}, which contains the breakup of the code into smaller notebooks, each containing some basic step of the procedure, accompanied by a few more examples.  

In this section we illustrate how to use this software to obtain from a superdiagram the momenta polynomial resulting from  the Grassmann integration. We proceed by examples to illustrate how the superdiagram must be specified and how the output has to be interpreted. 

\paragraph{Diagrams with (anti)chiral superfields}
Let us begin with a very simple example, namely the superdiagram in eq. (\ref{1loopQ}), which we redraw here for commodity of the reader:
\begin{align}
	\label{1loopQbis}
	\parbox[c]{.35\textwidth}{\includegraphics[width = .35\textwidth]{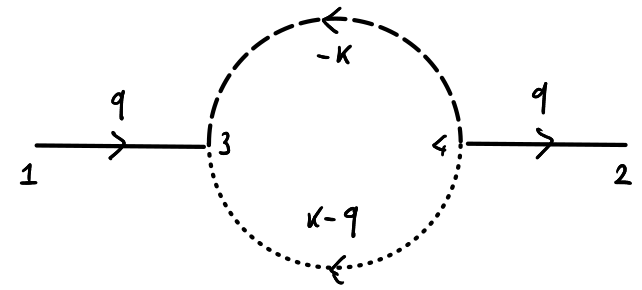}}~.
\end{align}
To evaluate the corresponding Grassmann integral, namely the function $\cZ$ of eq. (\ref{Z1loopQ}), the following command must be issued:
\begin{align*}
	\parbox[c]{\textwidth}{\includegraphics[width = \textwidth]{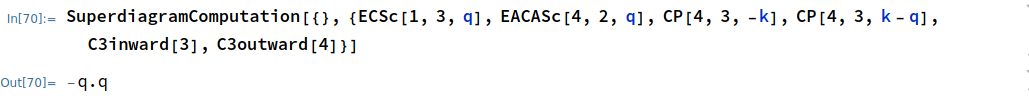}}
\end{align*}
The first argument is empty because it must be used only when gauginos appear in the external states. The other arguments describe the elements of the diagram. 
\begin{itemize}
	\item \texttt{ECSc[1,3,q]} is the external line of the chiral scalar going from the vertex labeled as 1 to the vertex labeled as 3. Similarly, \texttt{EACAS[4,2,q]} is the external line or the anti-chiral scalar from vertex 4 to vertex 2.
	\item \texttt{CP[4,3,-k]} stands for the chiral superfield propagator that carries momentum $-k$ and goes from the vertex 4 to the vertex 3. \texttt{CP[4,3,k-q]} is the other chiral superfield propagator of the diagram.
	\item \texttt{C3inward[3]} and \texttt{C3outward[4]} signal the presence of three-point chiral supervertices with incoming lines in position 3, and outgoing lines in position 4, see eq. (\ref{det2}). 
\end{itemize}
The command returns the output \texttt{-q$\cdot$q}, namely $-q^2$, in agreement with eq. (\ref{Z1loopQbis}).

Let us consider a more complex example, namely the three-loop superdiagram of eq. (\ref{td15}). We implement momentum conservation at each vertex and choose three independent loop momenta $r,s,t$ as follows:
\begin{align}
	\label{1loopQtris}
	\parbox[c]{.4\textwidth}{\includegraphics[width = .4\textwidth]{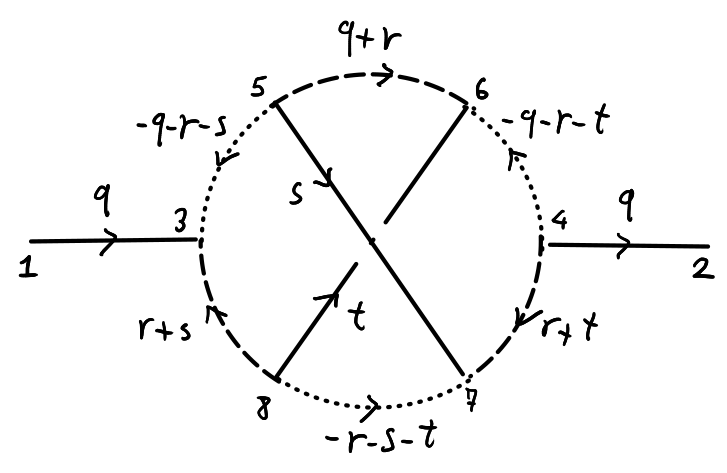}}~.
\end{align}
The corresponding Grassmann integral is computed by the following command:
\begin{align*}
	\parbox[c]{\textwidth}{\includegraphics[width = \textwidth]{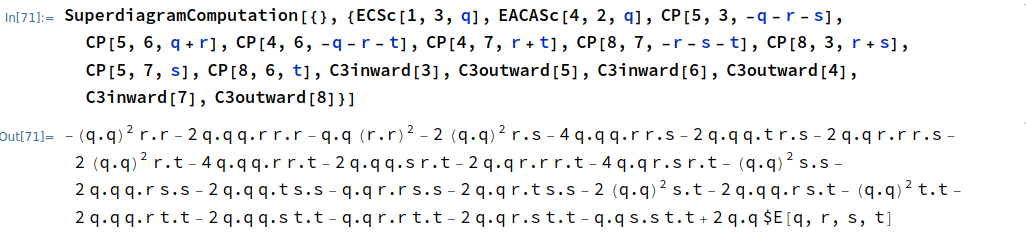}}
\end{align*}
The result, which is proportional to $q^2$, is written almost completely in terms of scalar products of momenta. The only other type of expression that appears is \texttt{\$E[q,r,s,t]}, which stands for $\epsilon_{\mu\rho\sigma\tau}q^\mu r^\rho s^\sigma t^\tau$.   

\paragraph{Diagrams with internal vector superfields}
We consider now the three loops example with an internal three-vector vertex of eq. (\ref{s4_2}). 
Implementing momentum conservation and choosing three independent loop momenta $r,s,t$ we write it as
\begin{align}
	\label{sec603}
	\parbox[c]{.4\textwidth}{\includegraphics[width = .4\textwidth]{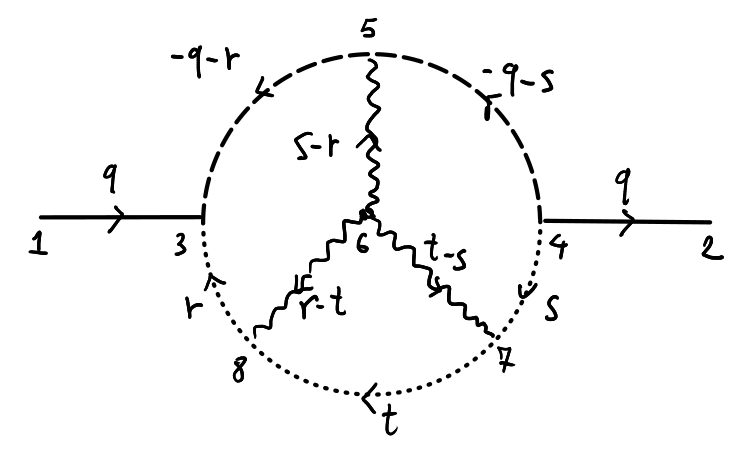}}~.
\end{align}
To evaluate the Grassmann integrals for this superdiagram, we have to explicitly define the momenta variables for the gluon propagators. We choose
\begin{align*}
	\parbox[c]{\textwidth}{\includegraphics[width = \textwidth]{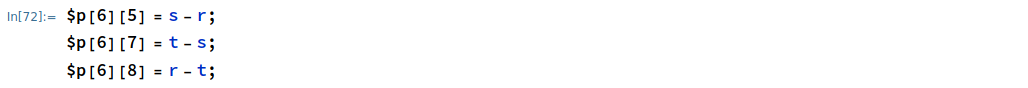}}
\end{align*}
and then compute 
\begin{align*}
	\parbox[c]{\textwidth}{\includegraphics[width = \textwidth]{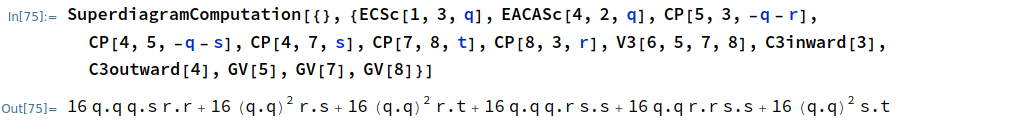}}
\end{align*}
Among the arguments we have some new ingredients. 
\begin{itemize}
	\item \texttt{V3[6,5,7,8]} represents the 3-vector vertex in $6$, from which three vector superpropagators point toward the vertices $5,7,8$. 
	\item In case one needs to consider a diagram in which a propagator connects two legs of the vertex \texttt{V3} the input should be of the form \texttt{V3[o,a,a,b]}. Note that the index \texttt{a} to be used has to be different from all other indices used to specify vertices or external lines. A similar remark applies also to four-vector vertices, which in the program are denoted in general as \texttt{V4[o,a,b,c,d]}. 
	\item
	Note that the vector propagators departing from the 3-vector vertex  must not appear separately in the input, and the program assign automatically a name to their momenta: \texttt{\$p[6][5]} is the momentum flowing in the propagator from $6$ to $5$, and so on. This is why we had to give the instructions \texttt{\$p[6][5] = s-r}, to enforce our choice of names before calling the program\footnote{The default choice of the momenta for these vertexes is flowing outward, when two of these are connected the program defines \texttt{\$p[i][j]} with \texttt{i<j}.}.
	\item \texttt{GV[i]} indicates that in position $i$ there is a ``generic vertex'' whose Grassmann contribution is just $\int\!d^4\theta_i$. This is the case, see eq. (\ref{det4}), of the vector-chiral-chiral vertices in positions  5,7 and 8.
\end{itemize}
The result, which is proportional to $16 q^2$, is very simple although the computation, 
described in section \ref{sec:example}, is quite involved.  

\paragraph{Diagrams with vector external lines.}
The program also handles superdiagrams with external vector lines. Let us consider for instance the example given in eq. (\ref{e4}), that we redraw here for ease of comparison to the input that must be given to the program:
\begin{align}
	\label{e4-bis}
	\parbox[c]{.35\textwidth}{\includegraphics[width = .35\textwidth]{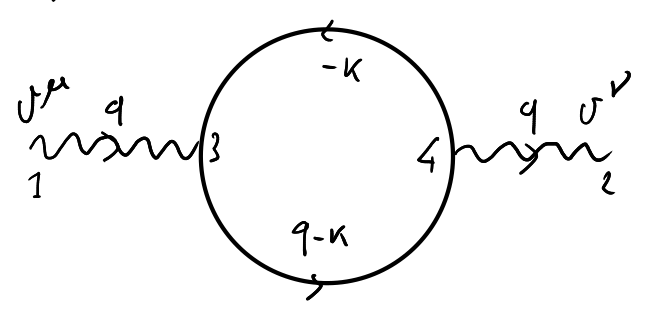}}~.
\end{align}
The corresponding Grassmann integrations are evaluated by the following command:
\begin{align*}
	\parbox[c]{\textwidth}{\includegraphics[width = \textwidth]{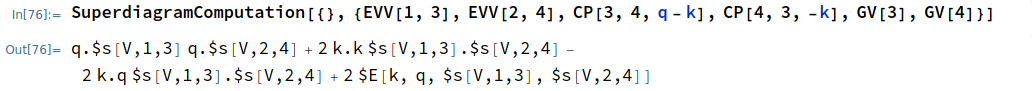}}
\end{align*}
Note the following points: 
\begin{itemize}
	\item \texttt{EVV[i,j]} stands for a vector external lines attached to a vertex in position $j$. 
	\item The external indices $\mu$ and $\nu$ are not specified explicitly in the input. The program associates automatically to each external vector line \texttt{EVV[i,j]} a unit vector \texttt{\$s[V,i,j]} that appears in the output. With the conventions taken in (\ref{e4-bis}) we should therefore read \texttt{\$s[V,1,3]} as $\hat{e}^\mu$ and \texttt{\$s[V,4,2]} as $\hat{e}^\nu$.
\end{itemize}
Using the properties of the unit vectors $\hat{e}^\mu$, see the discussion from eq. (\ref{e1}) to eq. (\ref{emuprop}), the output of the program corresponds therefore to 
\begin{align}
	\label{e:vectoroutput}
	& q\cdot \hat{e}^\mu\ q\cdot \hat{e}^\nu +2k\cdot k\ \hat{e}^\mu\cdot \hat{e}^\nu -2k\cdot q\ \hat{e}^\mu\cdot \hat{e}^\nu+2k_\rho q_\sigma 	\hat{e}^\mu_\lambda \hat{e}^\nu_\eta\epsilon^{\rho\sigma\lambda\eta}=\notag\\
	=q^\mu q^\nu+2k\cdot k\delta^{\mu\nu}-2k\cdot q\delta^{\mu\nu}+2k_\rho q_\sigma 	\epsilon^{\rho\sigma\mu\nu}.
\end{align}

Consider now the diagram of eq. (\ref{e4bis}), namely 
\begin{align}
	\label{e4bisbis}
	\parbox[c]{.3\textwidth}{\includegraphics[width = .3\textwidth]{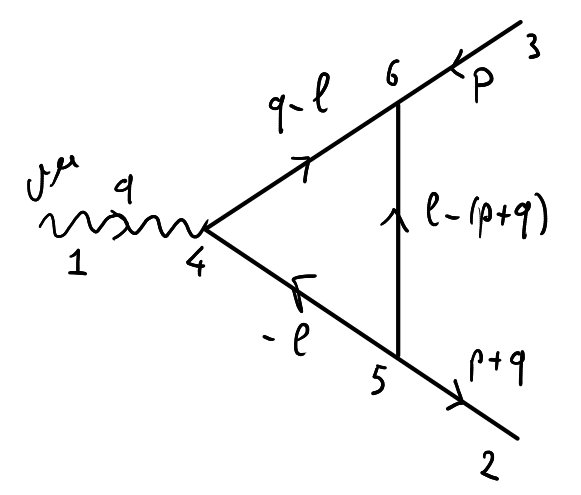}}~.
\end{align}	
The corresponding Grassmann integration can be performed with the command
\begin{align*}
	\parbox[c]{\textwidth}{\includegraphics[width = \textwidth]{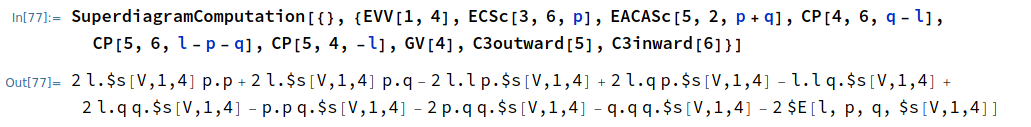}}
\end{align*}
According to the conventions already discussed for the diagram (\ref{e4-bis}), \texttt{\$s[V,1,3]} stands for the unit vector $\hat e^\mu$ and, after simple manipulations, the result has the form
\begin{align}
	\label{e4bisbisres}
	2\ell^\mu (p^2 + p\cdot q) - 2 p^\mu (\ell^2-\ell\cdot q) - q^\mu\left(\ell^2 - 2 \ell\cdot q + (p+q)^2\right) - 2 \epsilon^{\mu\nu\rho\sigma} \ell_\nu p_\rho q_\sigma~.
\end{align}

\paragraph{Diagrams with chiralino external lines}
Let us discuss now the the case in which some of the external states are spinors, focusing as usual on an example. We consider the superdiagram of eq. (\ref{e21bis}), namely
\begin{align}
	\label{e11bis}
	\parbox[c]{.35\textwidth}{\includegraphics[width = .35\textwidth]{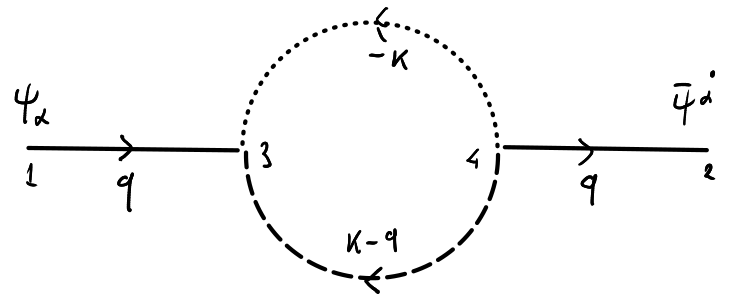}}~.
\end{align}
The corresponding Grassmann integrals are computed using the following call:
\begin{align*}
	\parbox[c]{\textwidth}{\includegraphics[width = \textwidth]{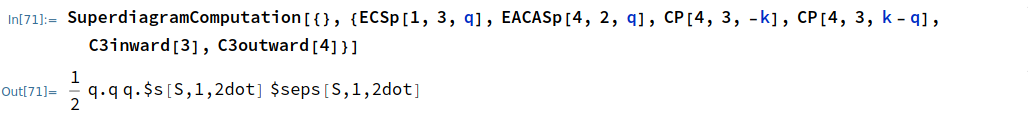}}
\end{align*}
In this expression, 
\texttt{\$seps[S,1,2dot]} stands for $(\sigma^\mu)_{\alpha}^{~\dot{\alpha}}$ while \texttt{\$s[S,1,2dot]} corresponds to $\hat{e}_\mu$. Indeed \texttt{\$seps} indicates a $\sigma^\mu$ matrix with an index saturated via a $\e$ (the \texttt{S} simply reminds us that that term comes from a spinorial contribution) whose spinorial indices are those associated to the external points $1$ and $2$ in the diagram, namely $\alpha$ and $\dot{\alpha}$. The index $\mu$ is dummy, and it has to be saturated with a structure labeled by the same arguments, namely \texttt{\$s[S,1,2dot]}. As already explained after eq. (\ref{e4-bis}),  \texttt{\$s} indicates a unit vector, which here has to be $\hat e_\mu$. Thus the above results amounts to
\begin{align}
	\label{resspin1}
	\frac 12 q^2\, q_\mu \left(\sigma^\mu\right)_\alpha^{~\dot\alpha} =  \frac 12 q^2\, q_\alpha^{~\dot\alpha}~.
\end{align}	

Let us now consider the more complicated case introduced  in eq. (\ref{e13bis}) and discussed thereafter. In the decomposition of the corresponding Grassmannian integrals more tensorial structures appear, which are codified in a non completely obvious way in the output of the program. Let us redraw the diagram in question here: 
\begin{align}
	\label{e4spinors}
	\parbox[c]{.4\textwidth}{\includegraphics[width = .4\textwidth]{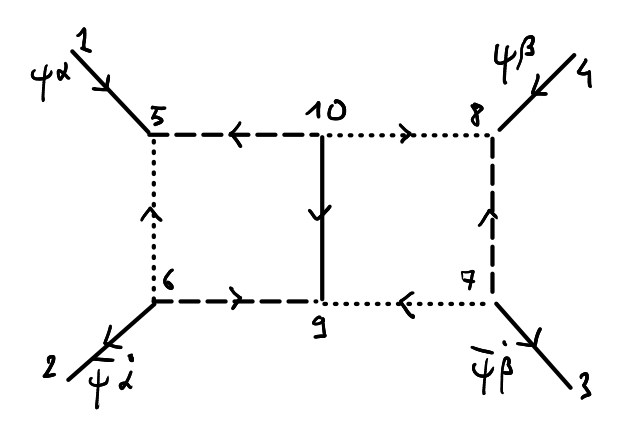}}~.
\end{align}	
Here we labeled the nodes so as to facilitate the comparison with the input that has to be given to the program in order to compute the corresponding Grassmann integrals, namely 
\begin{align*}
	\parbox[c]{\textwidth}{\includegraphics[width = \textwidth]{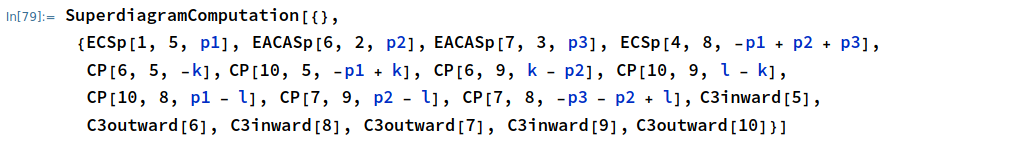}}~.
\end{align*}
We used momentum conservation to express the external momentum $p_4$ in terms of $p_1$, $p_2$ and $p_3$ and we denoted by $k$ and $l$ the two independent loop momenta running in the two boxes - the assignments of momenta to the various lines can be understood from the call itself, whose syntax should by now be clear.

The resulting expression carries two chiral and two anti-chiral indices, namely it transforms in the $(2,1)\otimes(2,1)\otimes(1,2)\otimes(1,2)$ representation, which is reducible into $(3,3)\oplus (3,1)\oplus(1,3)\oplus(1,1)$. The program, however, which implements the decomposition into open paths described in eq. (\ref{e13tris}), does not completely carry out the reduction. It does so for the terms arising from the first term in the decomposition (\ref{e13tris}), but not or the others%
\footnote{In the first decomposition of (\ref{e13tris}), the tensor product is performed in two steps as $[(2,1)\otimes(2,1)]\otimes [(1,2)\otimes (1,2)]$; in the other cases it is organized as $[(2,1)\otimes(1,2)]\otimes [(2,1)\otimes (1,2)]$ and the tensor product between the two brackets is left indicated.}. As a consequence, the output of the program is not optimized and is very long, so we do not report it here. It contains various new structures. For instance, the following term appears in the output of the program call above:
\begin{align*}
	\parbox[c]{0.55\textwidth}{\includegraphics[width = 0.55\textwidth]{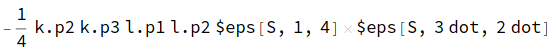}}~.
\end{align*}
Here  \texttt{\$eps[S,1,4]} stands for $\epsilon_{\alpha\beta}$: indeed, \texttt{\$eps} denotes an $\epsilon$ tensor, the argument \texttt{S} indicates that its indices are spinorial while \texttt{1,4} states that  these spinorial indices are those associated with the lines denoted as $1$ and $4$ in the input for this diagram. With analogous logic, \texttt{\$eps[S,3dot,2dot]} stands for $\epsilon^{\dot\beta\dot\alpha}$. Altogether this term reads therefore
\begin{align}
	\label{math-8res}
	-\frac{1}{4}(k\cdot p_2)(k\cdot p_3)(\ell\cdot p_1)(\ell\cdot p_2)\epsilon_{\alpha\beta} \,\epsilon^{\dot\beta\dot\alpha}~.
\end{align}

Another term appearing in the output is the following:
\begin{align*}
	\parbox[c]{\textwidth}{\includegraphics[width = \textwidth]{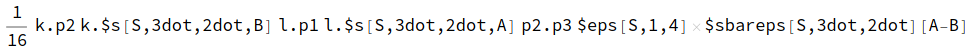}}~.
\end{align*}
Here \texttt{\$sbareps[S,1
	3dot,2dot][A-B]} stands for $\left(\bar\sigma^{\mu\nu}\right)^{\dot\beta \dot\alpha}$. Indeed \texttt{\$sbareps[\ldots][A-B]} indicates a $\bar\sigma$ matrix with two vector indices with raised spinor indices -- see the remark after eq. \ref{trsmnsrs}. The arguments \texttt{S,3dot,2dot} mean that the spinor indices are those associated with the external lines numbered as $3$ and $2$ in the input. 
The quantity \texttt{\$s[S,3dot,2dot,A]} has to be interpreted as the unit vector $\hat e_\mu$. In fact, \texttt{\$s} indicates a unit vector, while the arguments \texttt{S,1,4,A} implicate that the vector index of this unit vector is the first one (this is the meaning of \texttt{A}) of the indices of the sigma matrix whose spinor indices are associated to the lines $3$ and $2$. Similarly, \texttt{\$s[S,3dot,2dot,B]} means $\hat e_\nu$. Overall the  output term above is
\begin{align}
	\label{math-9-res}
	\frac{1}{16}(k\cdot p_2)(k\cdot\hat{e}_\nu)(\ell\cdot p_1)(\ell\cdot\hat{e}_\mu)(p_2\cdot p_3)\epsilon_{\alpha\beta} (\bar\sigma^{\mu\nu})^{\dot\beta\dot{\alpha}}
	= \frac{1}{16}(k\cdot p_2)(\ell\cdot p_1)(p_2\cdot p_3) \ell_\mu k_\nu (\bar\sigma^{\mu\nu})^{\dot\beta\dot{\alpha}}\, \epsilon_{\alpha\beta}~. 
\end{align} 

Of course, there is an analogous notation for $\left(\sigma^{\mu\nu}\right)_{\a\b}$ matrices, which are called \texttt{\$seps}.

\paragraph{Diagrams with gaugino external lines}
The external gaugino (\texttt{EG}) and anti-gaugino (\texttt{EAG}) lines have to be called within the first argument of the \texttt{SuperdiagramComputation} command.  For instance, consider the diagram of figure eq. (\ref{e21bis}), which we redraw here for ease of comparison:  
\begin{align}
	\label{e21bisbis}
	\parbox[c]{.4\textwidth}{\includegraphics[width = .4\textwidth]{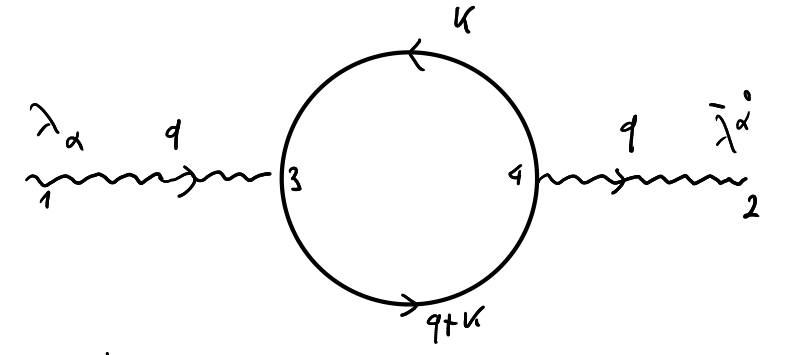}}~.
\end{align}
Its Grassmann part is computed via the following call\footnote{Note that \ttt{EG} and \ttt{EAG}, differently from the other superdiagrammatic elements, are not functions but just place-holders that the command \ttt{SuperdiagramComputation} recognizes and elaborates properly.}:
\begin{align*}
	\parbox[c]{\textwidth}{\includegraphics[width = \textwidth]{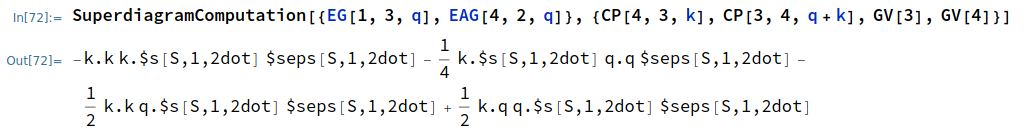}}
\end{align*}
There is no new ingredient in this result which, with manipulations already used previously, reads explicitly
\begin{align}
	\label{math-10-res}
	-\left(k^2 + \frac{q^2}{4}\right)\, k_\alpha^{~\dot{\alpha}}
	- \frac 12 \left(k^2 - (k\cdot q)\right) \, q_\alpha^{~\dot{\alpha}}~. 
\end{align}	

\paragraph{Ordering issues}
When there are spinorial external lines, which have a fermionic statistics, the order in which they are written in the correlator is relevant to determine the overall sign. The program returns the sign which is appropriate for the correlator in which the external spinors are ordered exactly in which they appear as arguments of the  \texttt{SuperdiagramComputation} call. Our convention is thus that the gauginos appear as the leftmost elements. 

\vskip 1cm
\noindent {\large {\bf Acknowledgments}}
\vskip 0.2cm
We would like to thank A. Armato, L. Bianchi, M. Frau, F. Galvagno, A. Lerda, S. Penati, I. Pesando and M. Preti for useful discussions.
This research is partially supported  by the MUR PRIN contract 2020KR4KN2``String Theory as a
bridge between Gauge Theories and Quantum Gravity'' and by the INFN project ST\&FI
``String Theory \& Fundamental Interactions''. 
\vskip 1cm
\begin{appendix}

\section{Notations and conventions}
\label{app:notations}
\subsection{General conventions}\label{s:compendiumconvenctions}

We will work in the Euclidean formulation of the theory. The Minkowskian theory can be recovered by an (inverse) Wick rotation. Tensors will be (anti)-symmetrized with strength one, namely we will define  
\begin{align}
	\label{symmdef}
	A_{(\mu_1...\mu_n)}&=\frac{1}{n!}\sum_{\sigma\in\S_n}A_{\mu_{\sigma(1)}...\mu_{\sigma(n)}}~\text{,}\\
	A_{[\mu_1...\mu_n]}&=\frac{1}{n!}\sum_{\sigma\in\S_n}\text{sgn}(\sigma)A_{\mu_{\sigma(1)}...\mu_{\sigma(n)}}~\text{.}
\end{align}
We will use commas to avoid symmetrization/anti-symmetrization of certain indexes as in the example that follows:
\begin{align}
	A^{[\mu,\ \nu,\ \rho]}=\frac{1}{2}(A^{\mu\nu\rho}-A^{\rho\nu\mu})~\text{.}
\end{align}
We will also adopt the following as generalized delta:
\begin{align}
	\delta^{\mu\nu}_{\rho\sigma}=\delta^\mu_\rho\delta^\nu_\sigma-\delta^\mu_\sigma\delta^\nu_\rho\text{.}
\end{align}
We will usually consider correlation functions in momentum space. Our conventions are such that, fo instance, the (super)propagators we have 
\begin{align}\label{e:propagatorconventions}
	&\braket{\Phi (x_i,\theta_i,\bar{\theta}_i)\bar{\Phi}(x_j,\theta_j,\bar{\theta}_j)}=\int\! \frac{d^4 p}{(2\pi)^4}\braket{\Phi (\theta_i,\bar{\theta}_i)\bar{\Phi}(\theta_j,\bar{\theta}_j)}(p)\,
	\rme^{\ii p\cdot (x_i-x_j)}~.
\end{align}
and in (super)-vertices the momenta associated to the various legs are naturally considered to be directed outward.

\subsection{Spinorial and grassmannian conventions}
\label{subapp:spinconv}
We denote by $\psi$ a chiral spinor of components $\psi_\alpha$ with $\alpha=1,2$, which transforms in the $(2,1)$ representation of $\mathrm{SO}(4)\sim \mathrm{SU}(2)_L \times \mathrm{SU}(2)_R$
and by $\bar \psi$ an anti-chiral one of 
components $\bar{\psi}^{\dot{\alpha}}$, with $\dot{\alpha}=1,2$, which transforms in the $(1,2)$.
Scalar spinor bilinears are denoted as follows:
\begin{align}
	\label{contractions}
	(\psi\chi) &\equiv 
	\epsilon^{\alpha\beta}\,\psi_\beta\,\chi_\alpha 
	~,\\[1mm]
	(\bar\psi\bar\chi) &\equiv \epsilon_{\dot{\alpha}\dot{\beta}}\,\bar\psi^{\dot{\beta}}\,\bar\chi^{\dot{\alpha}}
\end{align}
where
\begin{equation}
	\label{epsilons}
	\epsilon^{12} = \epsilon^{\dot{1}\dot{2}} =\epsilon_{21} = \epsilon_{\dot{2}\dot{1}} = 1~.
\end{equation}
Since the spinors are of Grassmannian nature, these scalar products are symmetric. 
For the ``square'' of spinors, we use -- when there is no risk of confusion with the component $\alpha=2$ of the spinor itself, the notation
\begin{align}
	\label{t2tb2}
	\psi^2 \equiv (\psi\psi)~,~~~
	\bar{\psi}^2 \equiv (\bar{\psi}\bar{\psi})~.
\end{align}

The spinor indices are raised and lowered with the following rules:
\begin{align}
	\label{raislow}
	\psi^\alpha = \epsilon^{\alpha\beta}\,\psi_\beta~,~~~
	\psi_\alpha = \epsilon_{\alpha\beta}\,\psi^\beta~,~~~
	\bar\psi^{\dot{\alpha}} =\epsilon^{\dot{\alpha}\dot{\beta}}\,\bar\psi_{\dot{\beta}}~,~~~
	\bar\psi_{\dot{\alpha}} =\epsilon_{\dot{\alpha}\dot{\beta}}\,\bar\psi^{\dot{\beta}}~,
\end{align}
so  that the bilinears can also be written as
\begin{align}
	\label{scalcontr}
	(\psi \chi)=\psi^\alpha\chi_\alpha,~~~ (\bar{\psi}\bar{\chi})=\bar{\psi}_{\dot{\alpha}}\bar{\chi}^{\dot{\alpha}}~.
\end{align}

\paragraph{Integration formul\ae}
The basic integration rules are
\begin{align}
	\label{intcomp}
	\int d\theta^\alpha\,\theta_\beta = \delta^\alpha_\beta~,~~~
	\int d\bar\theta_{\dot\alpha}\,\bar\theta^{\dot\beta} = \delta^{\dot\beta}_{\dot\alpha}~.
\end{align}
One defines then
\begin{gather}
	\label{intt1}
	\int d^2\theta =-\frac 14 \epsilon_{\alpha\beta}\int d\theta^\alpha d\theta^\beta~,~~~
	\int d^2\bar{\theta} =-\frac 14 \epsilon^{\dot{\alpha}\dot{\beta}}\int d\bar{\theta}_{\dot{\alpha}} d\bar{\theta}_{\dot{\beta}}~,
\end{gather}
so that 
\begin{align}
	\label{intt2}
	\int d^2\theta\, \theta^2 = 1~,~~~
	\int d^2\bar{\theta}\, \bar{\theta}^2 = 1~.
\end{align} 
Thus $\theta^2$ and $\bar{\theta}^2$ act as fermionic $\delta$-functions; 
more in general, if we write $\theta_{ij} = \theta_i - \theta_j$, we have
\begin{align}
	\label{deltat}
	\theta_{ij}^2 = \delta^2(\theta_{ij})~,~~~
	\bar\theta_{ij}^{\,2} = \delta^2(\bar\theta_{ij})~.
\end{align} 
We also use the notation
\begin{align}
	\label{delta4t}
	\theta_{ij}^2\, \bar\theta_{ij}^{\,2} = \delta^4(\theta_{ij})~.
\end{align}

\paragraph{Clifford algebra}	
We realize the Euclidean Clifford algebra
\begin{equation}
	\label{cliff4}
	\sigma_\mu\bar\sigma_\nu + \sigma_\nu\bar\sigma_\mu =
	-2\,\delta_{\mu\nu}\,\mathbf{1}
\end{equation}
by means of the matrices $(\sigma^\mu)_{\alpha\dot\beta}$ and
$(\bar\sigma^{\mu})^{\dot\alpha\beta}$ that we take to be 
\begin{equation}
	\label{sigmas}
	\sigma^\mu =
	(\vec\tau,-\ii\mathbf{1})~,\qquad
	\bar\sigma^\mu =
	-\sigma_\mu^\dagger = (-\vec\tau,-\ii\mathbf{1})~,
\end{equation}
where $\vec\tau$ are the ordinary Pauli matrices. They are such that
\begin{equation}
	\label{traspsigma}
	(\bar\sigma^{\mu})^{\dot\alpha\alpha}=\epsilon^{\alpha\beta}\,\epsilon^{\dot{\alpha}\dot{\beta}}(\sigma^\mu)_{\beta\dot\beta}~.
\end{equation}

With these matrices we can write the four-vectors as bispinors:
\begin{equation}
	\label{bispinork}
	k_{\alpha\dot{\beta}} = k_\mu \, (\sigma^\mu)_{\alpha\dot{\beta}}~,~~~
	\bar k^{\alpha\dot\beta} = k^\mu \, (\bar\sigma_\mu)^{\dot\alpha\beta}~.		
\end{equation}
We will often use the notations $k$ and $\bar{k}$ to indicate the matrices $k_{\alpha\dot{\beta}}$ and  $\bar k^{\alpha\dot\beta}$. One can then form spinor bilinears of the type
\begin{equation}
	\label{tptb}
	(\theta k \bar{\theta}) = \theta^\alpha\, k_{\alpha\dot{\beta}} \,\bar{\theta}^{\dot{\beta}}~.
\end{equation}
Because of the property (\ref{traspsigma}) we have 
\begin{align}
	(\theta k\bar{\theta})=-(\bar{\theta}\bar{k}\theta)~.
\end{align}

Let us introduce the matrices $(\sigma^{\mu\nu})_\alpha^{~\beta}$ and $(\bar\sigma^{\mu\nu})^{\dot{\alpha}}_{~\dot{\beta}}$, defined as 
\begin{align}
	\label{sigmamunu}
	(\sigma^{\mu\nu})_\alpha^{~\beta} = \frac 12 \left(\sigma^\mu\bar{\sigma}^\nu - \sigma^\nu\bar{\sigma}^\mu\right)_\alpha^{~\beta}~,~~~
	(\bar\sigma^{\mu\nu})^{\dot{\alpha}}_{~\dot{\beta}} = \frac 12 \left(\bar\sigma^\mu\sigma^\nu - \bar\sigma^\nu\sigma^\mu\right)^{\dot{\alpha}}_{~\dot{\beta}}~.
\end{align} 
They are (anti)self-dual:
\begin{align}
	\sigma^{\mu\nu} =\frac{1}{2}\epsilon^{\mu\nu\rho\sigma}\sigma_{\rho\sigma}~,~~~
	\bar{\sigma}^{\mu\nu} =-\frac{1}{2}\epsilon^{\mu\nu\rho\sigma}\bar{\sigma}_{\rho\sigma}~,
\end{align}
and are proportional to the generators of $\mathrm{SO}(4)$ Lorentz rotations on the chiral and antichiral spinors $\psi_\beta$ and $\bar\psi^{\dot{\beta}}$, as it follows by computing their commutators. In fact, none has the following product formul\ae:
\begin{align}
	\label{prodss}
	\sigma^{\mu\nu}\sigma^{\rho\sigma} = \left(\delta^{\mu\sigma}\delta^{\nu\rho} - \delta^{\mu\rho}\delta^{\nu\sigma} - \epsilon^{\mu\nu\rho\sigma}\right) \, \mathbf{1} 
	+ \left(-\delta^{\rho\mu}\sigma^{\nu\sigma}+\delta^{\rho\nu}\sigma^{\mu\sigma}-\delta^{\sigma\nu}\sigma^{\mu\rho}+\delta^{\sigma\mu}\sigma^{\nu\rho}\right)~,
\end{align}
and
\begin{align}
	\label{prodsb}
	\bar\sigma^{\mu\nu}\bar\sigma^{\rho\sigma} & = \left(\delta^{\mu\sigma}\delta^{\nu\rho} - \delta^{\mu\rho}\delta^{\nu\sigma} + \epsilon^{\mu\nu\rho\sigma}\right) \, \mathbf{1} 
	+ \left(-\delta^{\rho\mu}\bar{\sigma}^{\nu\sigma}+\delta^{\rho\nu}\bar{\sigma}^{\mu\sigma}-\delta^{\sigma\nu}\bar{\sigma}^{\mu\rho}+\delta^{\sigma\mu}\bar{\sigma}^{\nu\rho}\right)~.
\end{align}
Taking the trace, we have
\begin{align}
	\label{trsmnsrs}
	\tr \left(\sigma^{\mu\nu}\sigma^{\rho\sigma}\right) 
	& = 2 \left(\delta^{\mu\sigma}\delta^{\nu\rho} - \delta^{\mu\rho}\delta^{\nu\sigma} 
	- \epsilon^{\mu\nu\rho\sigma}\right)~,
	\notag\\
	\tr \left(\bar\sigma^{\mu\nu}\bar\sigma^{\rho\sigma}\right)
	& = 2 \left(\delta^{\mu\sigma}\delta^{\nu\rho} - \delta^{\mu\rho}\delta^{\nu\sigma} 
	+ \epsilon^{\mu\nu\rho\sigma}\right)~. 
\end{align}		
Let us also notice that the tensors  $(\sigma^{\mu\nu})_{\alpha\beta}$ and $(\bar\sigma^{\mu\nu})^{\dot{\alpha}\dot{\beta}}$ obtained by lowering or raising one spinor index are symmetric in $(\alpha,\beta)$ or $(\dot{\alpha},\dot{\beta})$.

\paragraph{Decomposition of bispinors}
Let us consider a bispinor $A_{\alpha\beta}$. It transforms in the representation $(2,1)\otimes (2,1)$ which is reducible into $(1,1) \oplus (3,1)$, i.e. into a singlet and a symmetric triplet, as follows:
\begin{align}
	\label{decA}
	A_{\alpha\beta} = - a \, \epsilon_{\alpha\beta} + \frac 12 a_{\mu\nu}\, (\sigma^{\mu\nu})_{\alpha\beta}~.
\end{align}
If we consider the matrix $A$ of components $A_\alpha^{~\beta}$, the above decomposition reads
\begin{align}
	\label{decAbis}
	A = a \, \mathbf{1} + \frac 12 a_{\mu\nu}\, \sigma^{\mu\nu}~,
\end{align}
and, using the trace relation (\ref{trsmnsrs}), we can determine the coefficients of the decomposition as follows:
\begin{align}
	\label{coeffdecA}
	a = \frac 12 \tr A~,~~~
	a_{\mu\nu} = -\frac 14 \tr (A\sigma_{\mu\nu})~.
\end{align}
Analogously, an anti-chiral bispinor ${\bar C}^{\dot\alpha}_{~\dot{\beta}}$ an be decomposed in matrix terms as follows:
\begin{align}
	\label{decCbis}
	\bar C 
	= \frac 12 \tr \bar C \, \mathbf{1} - \frac 18 \tr (\bar C\bar\sigma_{\mu\nu})\, \bar\sigma^{\mu\nu}~.		
\end{align} 

In particular, when the bispinor is factorized, i.e. when $A_{\alpha\beta} = \psi_\alpha\chi_\beta$, we have the Fierz identity
\begin{align}
	\label{fierzchir}
	\psi_\alpha \chi_\beta = \frac 12 (\chi\psi)\, \epsilon_{\alpha\beta} + \frac 18 (\chi\sigma_{\mu\nu}\psi)\, \left(\sigma^{\mu\nu}\right)_{\alpha\beta}~.
\end{align} 
If the two spinors are identical, only the anti-symmetric structure survives:
\begin{align}
	\label{fierzchirug}
	\psi_\alpha \psi_\beta = \frac 12 (\psi\psi)\, \epsilon_{\alpha\beta} ~.
\end{align} 

Similarly in the anti-chiral case we have
\begin{align}
	\label{fierzachir}
	\bar\psi^{\dot\alpha} \bar\chi^{\dot\beta} = \frac 12 (\bar\chi\bar\psi)\, \epsilon^{\dot\alpha\dot\beta} + \frac 18 (\bar\chi\bar\sigma_{\mu\nu}\bar\psi)\, \left(\bar\sigma^{\mu\nu}\right)^{\dot\alpha\dot\beta}~.
\end{align}		
Again, if the two spinors are identical, only the anti-symmetric structure survives:
\begin{align}
	\label{fierzachirug}
	\bar\psi^{\dot\alpha} \bar\psi^{\dot\beta} = \frac 12 (\bar\psi\bar\psi)\, \epsilon^{\dot\alpha\dot\beta} ~.
\end{align} 

\paragraph{Strings of $\sigma$ matrices}
In the evaluatin of the superdiagrams one often has to deal with  bispinors  obtained by repeated products of $\sigma$ and $\bar{\sigma}$ matrices, of the form
\begin{align}
	\label{Asigma}	
	(\sigma^{\mu_1}\bar{\sigma}^{\nu_1}...\sigma^{\mu_n}\bar{\sigma}^{\nu_n})_\alpha^{\ \beta}
\end{align}	
or
\begin{align}
	\label{Csigma}
	(\bar{\sigma}^{\mu_1}\sigma^{\nu_1}...\bar{\sigma}^{\mu_n}\sigma^{\nu_n})^{\dot{\alpha}}_{\ \dot{\beta}}~.
\end{align}	
They can be decomposed according to eq.s (\ref{decAbis}) and (\ref{decCbis}), and the tensorial coefficients of their decomposition can be determined recursively. Let us consider the chiral case (\ref{Asigma}), for the antichiral case the procedure is perfectly analogous. We write
\begin{align}
	\label{Atdec}
	\sigma^{\mu_1}\bar{\sigma}^{\nu_1}...\sigma^{\mu_n}\bar{\sigma}^{\nu_n}
	= a^{\mu_1\ldots\nu_n }\, \mathbf{1} + \frac{1}{2}b^{\mu_1 \ldots \nu_n}_{\rho\sigma}\, \sigma^{\rho\sigma}
\end{align}	
and we want to determine the numerical tensorial coefficients $a$ and $b$. Let us note that the coefficients $a$ correspond to the traces of these products of matrices:
\begin{align}
	\label{trtoa}
	\tr\left(\sigma^{\mu_1}\bar{\sigma}^{\nu_1}...\sigma^{\mu_n}\bar{\sigma}^{\nu_n}\right) 
	= 2 a^{\mu_1\ldots\nu_n }~. 
\end{align}

In the case $n=1$, we have
\begin{align}
	\label{inconddec}
	a^{\mu_1\nu_1} = -\delta^{\mu_1\nu_1}~,~~~
	b^{\mu_1\nu_1}_{\rho\sigma} =\frac{1}{2}( \delta^{\mu_1\nu_1}_{\rho\sigma}+\epsilon^{\mu_1\nu_1}_{\ \ \ \ \ \rho\sigma})~,
\end{align}
where in the second equality the double delta is anti-symmetrized with strength one. These relations represent the initial conditions for a recursive procedure in $n$. In fact, on the one side 
\begin{align}
	\label{Atdecnp1}
	\sigma^{\mu_1}\bar{\sigma}^{\nu_1}...\sigma^{\mu_{n+1}}\bar{\sigma}^{\nu_{n+1}}
	= a^{\mu_1\ldots\nu_{n+1} }\, \mathbf{1} + \frac{1}{2}b^{\mu_1 \ldots \nu_{n+1}}_{\rho\sigma}\, \sigma^{\rho\sigma}~.
\end{align}	
On the other side, using eq.s (\ref{Atdec}) and (\ref{inconddec}), we have
\begin{align}
	\label{Atdecrec}
	\sigma^{\mu_1}\bar{\sigma}^{\nu_1}...\sigma^{\mu_{n+1}}\bar{\sigma}^{\nu_{n+1}}
	= \left(a^{\mu_1\ldots\nu_{n} }\, \mathbf{1} + \frac{1}{2}b^{\mu_1 \ldots \nu_{n}}_{\rho\sigma}\, \sigma^{\rho\sigma}\right)\left(-\delta^{\mu_{n+1}\nu_{n+1}}\, \mathbf{1} + \sigma^{\mu_{n+1}\nu_{n+1}}\right)~.
\end{align}	
We can expand the products using eq. (\ref{prodss}) and compare the result with eq. (\ref{Atdecnp1}). In this way we find, after some algebric manipulations,  the following recursion relations:
\begin{align}
	\label{recab}
	a^{\mu_1\ldots\nu_{n+1} }  =& -a^{\mu_1\ldots\nu_n }\, \delta^{\mu_{n+1}\nu_{n+1}}  
	- 2  b^{\mu_1 \ldots \nu_{n},\mu_{n+1}\nu_{n+1}} 
	~,
	\notag\\
	b^{\mu_1 \ldots \nu_{n+1}}_{\rho\sigma} 
	& =  -b^{\mu_1 \ldots \nu_{n}}_{\rho\sigma} \, \delta^{\mu_{n+1}\nu_{n+1}}
	+ \frac{1}{2}a^{\mu_1\ldots\nu_n }( \delta^{\mu_{n+1}\nu_{n+1}}_{\rho\sigma}+\epsilon^{\mu_{n+1}\nu_{n+1}}_{\ \ \ \ \ \ \ \ \ \ \rho\sigma})+4b^{\mu_1...\nu_{n+1},\ \ \![\mu_{n+1}}_{\ \ \ \ \ \ \ \ \ \ [\mu}\delta^{\nu_{n+1}]}_{\nu]}~. 
\end{align} 
Starting from the initial condition (\ref{inconddec}) we can in this way obtain the coefficients $a$ and $b$ of the decomposition for any $n$. For instance, in the the $n=2$ case we find 
\begin{align}
	\label{abn2}
	\sigma^{\mu_1} \bar\sigma^{\nu_1} \sigma^{\mu_2} \bar\sigma^{\nu_2}  =&
	\left(\delta^{\mu_1\nu_1} \delta^{\mu_2\nu_2} - \delta^{\mu_1\mu_2} \delta^{\nu_1\nu_2}
	+ \delta^{\mu_1\nu_2} \delta^{\mu_2\nu_1} -
	\epsilon^{\mu_1\nu_1\mu_2\nu_2}\right)\mathbf{1}+\notag\\
	&-\delta^{\mu_1\nu_1}\sigma^{\mu_2\nu_2}-\delta^{\mu_2\nu_2}\sigma^{\mu_1\nu_1}-\delta^{\mu_1\mu_2}\sigma^{\nu_1\nu_2}-\delta^{\nu_1\nu_2}\sigma^{\mu_1\mu_2}+\delta^{\mu_1\nu_2}\sigma^{\nu_1\mu_2}+\delta^{\nu_1\mu_2}\sigma^{\mu_1\nu_2}~,
\end{align}  
which implies, according to eq. (\ref{trtoa}), that 
\begin{align}
	\label{trn2}
	\tr \left(\sigma^{\mu_1} \bar\sigma^{\nu_1} \sigma^{\mu_2} \bar\sigma^{\nu_2}\right)
	= 2 \left(\delta^{\mu_1\nu_1} \delta^{\mu_2\nu_2} - \delta^{\mu_1\mu_2} \delta^{\nu_1\nu_2}
	+ \delta^{\mu_1\nu_2} \delta^{\mu_2\nu_1} -
	\epsilon^{\mu_1\nu_1\mu_2\nu_2}\right)~.  
\end{align}

In a completely analogous way we can determine recursively the decomposition of the antichiral bispinors given by products of $\sigma$ matrices as in eq. (\ref{Csigma}). The expressions have the same form but for the Levi-Civita tensors appearing with the opposite sign. 

This recursive procedure to determine the products of $\sigma$ matrices is implemented in the Mathematica code distributed with this paper.   

\paragraph{Spinorial derivatives}
We use the symbols
\begin{align}\label{e:raiselowerder1}
	\partial_\alpha=\frac{\partial}{\partial \theta^\alpha}~,~~~
	\bar{\partial}_{\dot{\alpha}}=\frac{\partial}{\partial \bar{\theta}^{\dot{\alpha}}}
\end{align}
and
\begin{align}
	\label{dertup}	
	\partial^\alpha= \frac{\partial}{\partial \theta_\alpha} ~,~~~
	\bar{\partial}^{\dot{\alpha}}=\frac{\partial}{\partial \bar{\theta}_{\dot{\alpha}}}~,
\end{align}
in such a way that $\partial_\alpha \theta^\beta = \partial^\beta \theta_\alpha = \delta^\beta_\alpha$ and
$\partial_{\dot\alpha} \theta^{\dot\beta} = \partial^{\dot\beta} \theta_{\dot\alpha} = \delta^{\dot\beta}_{\dot\alpha}$. Note that in this way 
\begin{align} 
	\label{raisder}
	\partial^\alpha = -\epsilon^{\alpha\beta}\partial_\beta~,~~~
	\bar{\partial}^{\dot{\alpha}}=-\epsilon^{\dot{\alpha}\dot{\beta}}\partial_{\dot{\beta}}~,
\end{align}
with a sign difference with respect to the general rule (\ref{raislow}).

\paragraph{Spinor covariant derivatives}  	
Covariant derivatives are defined as:
\begin{align}
	D_\alpha=\partial_\alpha+i\sigma^\mu_{\alpha\dot{\alpha}}\bar{\theta}^{\dot{\alpha}}\partial_\mu~,~~~ \bar{D}_{\dot{\alpha}}=-\bar{\partial}_{\dot{\alpha}}-i\theta^{\alpha}\sigma^\mu_{\alpha\dot{\alpha}}\partial_\mu~.
	\label{e:spacetimederivatives}
\end{align}   	
The spinor indices of the covariant derivatives are raised according to the convention in eq. (\ref{raisder}). 
In momentum space, we have
\begin{align}
	\label{Dmom}
	D_{\alpha}=\partial_{\alpha}- p_{\alpha\dot{\alpha}}\bar{\theta}^{\dot{\alpha}}~,~~~
	\bar{D}_{\dot{\alpha}}=-\bar{\partial}_{\dot{\alpha}}+\theta^\alpha p_{\alpha\dot{\alpha}}~,
\end{align}
where $p$ il the momentum out-flowing from the node in which sits the superfield to which the covariant derivative is applied. When it is not obvious which momentum should appear, we specify it using the notations $D_{\alpha}(p)$, and $\bar{D}_{\dot{\alpha}}(p)$. The squared covariant derivatives are defined as follows:
\begin{align*}
	D^2=\epsilon^{\alpha\beta}D_{\alpha}D_{\beta}\ \ \ \ \ \ \ \ \ \bar{D}^2=\epsilon_{\dot{\alpha}\dot{\beta}}\bar{D}^{\bar{\alpha}}\bar{D}^{\bar{\beta}}
\end{align*}

\paragraph{Grassmannian integration by parts}\label{s:integrationbyparts}
Let $B_i$ be bosonic quantities, and $F_i$ a fermionic elements. Then one has the following  integration by parts rules for the covariant derivative  $D_{\alpha}(p)$: 
\begin{align}
	\int\! d^2\theta\ \left[D_\alpha(p)F\right]B&=\int\! d^2\theta\ F \left[D_\alpha(-p)B\right]~,\\
	\int\! d^2\theta\ \left[D_\alpha(p)B_1\right]B_2&=-\int\! d^2\theta\ B_1 \left[D_\alpha(-p)B_2\right]~,\\
	\int\! d^2\theta\ \left[D_\alpha(p)F_1\right]F_2&=\int\! d^2\theta\ F_1 \left[D_\alpha(-p) F_2\right]~.
\end{align}
The rules for $\bar{D}^{\dot{\alpha}}(p)$ are analogous. 
One also has
\begin{align}
	\int\! d^2\theta\ \left[D^2(p)F\right]B
	&=-\int\! d^2\theta\ F \left[D^2(-p)B\right]~,\\ \label{e:bbintegrationbyparts}
	\int\! d^2\theta\ \left[D^2(p)B_1\right]B_2&=\int\! d^2\theta\ B_1\left[D^2(-p)B_2\right]~,\\
	\int\! d^2\theta\ \left[D^2(p)F_1\right]F_2&=\int\! d^2\theta\ F_1 \left[D^2(-p)F_2\right]~.
\end{align}
\end{appendix}

\providecommand{\href}[2]{#2}

\begingroup\raggedright

\endgroup

\end{document}